\documentclass[10pt]{iopart}
\pdfoutput=1

\expandafter\let\csname equation*\endcsname=\relax
\expandafter\let\csname endequation*\endcsname=\relax
\usepackage{color}
\usepackage{amsmath}
\usepackage{amssymb}
\usepackage{enumerate}
\usepackage{graphicx}
\usepackage{mathbbol}
\usepackage{amsfonts}
\usepackage{url}

\newcommand{\nn}{\nonumber}

\newcommand{\bea}{\begin{eqnarray}}
\newcommand{\eea}{\end{eqnarray}}
\newcommand{\beq}{\begin{equation}}
\newcommand{\eeq}{\end{equation}}

\def\XXint#1#2#3{{\setbox0=\hbox{$#1{#2#3}{\int}$}
 \vcenter{\hbox{$#2#3$}}\kern-.5\wd0}}

\usepackage[utf8]{inputenc}
\usepackage{amsmath}
\usepackage{hyperref}
\usepackage{graphicx}
\usepackage{amsfonts}
\usepackage{amsthm}
\usepackage{cases}
\usepackage{bm}
\usepackage{amssymb}

\usepackage{color}
\definecolor{Blue}{rgb}{0.00, 0.00, 1.00}
\definecolor{Red}{rgb}{1.00, 0.00, 0.00}

\hypersetup{
    colorlinks=true,       
    linkcolor=red,          
    citecolor=blue,        
    filecolor=magenta,      
    urlcolor=cyan           
}

\newcommand{\be}{\begin{equation}}
\newcommand{\ee}{\end{equation}}

\newcommand{\beqn}{\begin{eqnarray}}
\newcommand{\eeqn}{\end{eqnarray}}

\DeclareMathOperator{\Ai}{Ai}
\DeclareMathOperator{\Bi}{Bi}
\DeclareMathOperator{\Li}{Li}

\DeclareMathOperator{\J}{J}

\def\q{\frac{\hbar^2}{2m}}

\newcommand{\moy}[1]{\ensuremath{\langle #1 \rangle}}

\makeatletter
\renewcommand\@appendixstar{\@@par
 \ifnumbysec 
 \@addtoreset{table}{section}
 \@addtoreset{figure}{section}\fi
 \setcounter{section}{0}
 \setcounter{subsection}{0}
 \setcounter{subsubsection}{0}
 \setcounter{equation}{0}
 \setcounter{figure}{0}
 \setcounter{table}{0}
 \def\thesection{\Alph{section}} 
 \def\theequation{\ifnumbysec
      \Alph{section}.\arabic{equation}\else
      \Alph{section}\arabic{equation}\fi}
 \def\thetable{\ifnumbysec
      \Alph{section}\arabic{table}\else
      A\arabic{table}\fi}
 \def\thefigure{\ifnumbysec
      \Alph{section}\arabic{figure}\else
      A\arabic{figure}\fi}}
\makeatother

\begin{document}
\title[]{Non-interacting fermions in hard-edge potentials}

\author{Bertrand Lacroix-A-Chez-Toine}
\address{LPTMS, CNRS, Univ. Paris-Sud, Universit\'e Paris-Saclay, 91405 Orsay, France}

\author{Pierre Le Doussal}
\address{CNRS-Laboratoire de Physique Th\'eorique de l'Ecole Normale Sup\'erieure, 24 rue Lhomond, 75231 Paris Cedex, France}

\author{Satya N. Majumdar}
\address{LPTMS, CNRS, Univ. Paris-Sud, Universit\'e Paris-Saclay, 91405 Orsay, France}

\author{Gr\'egory Schehr}
\address{LPTMS, CNRS, Univ. Paris-Sud, Universit\'e Paris-Saclay, 91405 Orsay, France}

\begin{abstract}
We consider the spatial quantum and thermal fluctuations of non-interacting Fermi gases of $N$ particles confined in $d$-dimensional non-smooth potentials. We first present a thorough study of the spherically symmetric pure hard-box potential, with vanishing potential inside the box, both at $T=0$ and $T>0$. We find that the correlations near the wall are described by a ``hard edge'' kernel, which depend both on $d$ and $T$, and which is different from the ``soft edge'' Airy kernel, and its higher $d$ generalizations, found for smooth potentials. We extend these results to the case where the potential is non-uniform inside the box, and find that there exists a family of kernels which interpolate between the above ``hard edge'' kernel and the ``soft edge'' kernels. Finally, we consider one-dimensional singular potentials of the form $V(x)\sim x^{-\gamma}$ with $\gamma>0$. We show that the correlations close to the singularity at $x=0$ are described by this ``hard edge'' kernel for $1\leq\gamma<2$ while they are described by a broader family of ``hard edge'' kernels known
as the Bessel kernel for $\gamma=2$ and, finally by the Airy kernel for $\gamma>2$. These one-dimensional kernels also appear in random matrix theory, and we provide here the mapping between the $1d$ fermion
models and the corresponding random matrix ensembles.
Part of these results were announced in a recent Letter, EPL {\bf 120}, 10006 (2017). 
\end{abstract}

\maketitle

\tableofcontents

\newpage

\section{Introduction and main results}

\subsection{General introduction and motivations}

Recent experimental developments in ultra-cold gases of fermions or bosons \cite{BDZ08} has generated 
a lot of theoretical interest for quantum many-body systems~\cite{GPS08}. In particular, even in the absence 
of interactions, a situation that can be accessed experimentally \cite{BDZ08,GPS08}, such systems of quantum particles 
display very rich behaviours, arising purely form the quantum statistics. For noninteracting Fermions, which
we focus on here, nontrivial spatial correlations naturally emerge from 
the Pauli exclusion principle. These correlations can be probed experimentally, 
using the recently developed Fermi quantum microscopes~\cite{Fermicro1,Fermicro2,Fermicro3}. This
certainly calls for a detailed characterisation of the spatial fluctuations in such noninteracting Fermi gas.  

Most of the current experiments are actually performed in the presence of a trapping potential, which creates
an edge {in space}, beyond which the density of fermions vanishes.  
While the correlations in the bulk, i.e. far from the edge, are well described by standard approaches such as the {local density approximation} (LDA) \cite{castin}, these approximations break down close to the edge of the Fermi gas \cite{Kohn}. It was recently shown that Random Matrix Theory (RMT), and related determinantal point processes (DPP), provide very powerful tools to describe the statistical fluctuations of Fermi gases, both at zero and finite temperature \cite{us_finiteT,DPMS:2015,fermions_review}. These techniques not only reproduce the LDA results in the bulk in a very controlled way but also allow to describe the edge properties of the Fermi gas \cite{us_finiteT,DPMS:2015,fermions_review}. 

\begin{figure}[h]
\centering
\includegraphics[width=1.\linewidth]{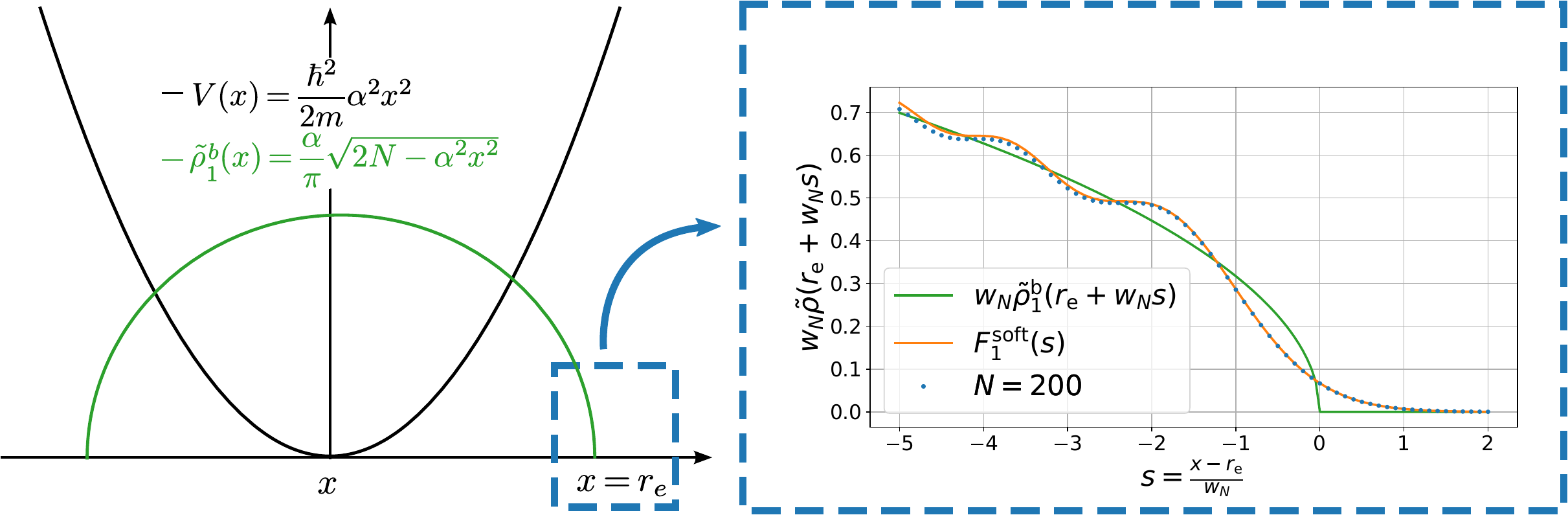}
\caption{Sketch of a one-dimensional harmonic potential (black line) and the corresponding bulk density given in Eq. \eqref{dens_oh_1d} (green line). Close to the edge $|x-r_{\rm e}|\sim w_N$, the rescaled finite $N$ density (blue dots) presents oscillations that are not described by the bulk density. On the other hand, the rescaled density shows a good agreement with the scaling form for large $N$ given in Eq. \eqref{dens_oh_edge} (orange line).}\label{Fig_airy_LDA}
\end{figure}

To illustrate the connections between trapped fermions and RMT, let us consider $N$ noninteracting fermions in a one-dimensional harmonic potential $V(x)=\frac{1}{2}m \omega x^2$. At $T=0$ the system is in its ground state and the associated many-body wave function $\Psi_0(x_1,\cdots,x_N)$ can be computed explicitly. One finds that the quantum joint probability density function (PDF) of the positions is given by \cite{marino_prl}
\be
|\Psi_0(x_1,\cdots,x_N)|^2=\frac{1}{Z_N}\prod_{i<j}^N |x_i-x_j|^2 e^{-\sum_{k=1}^N \alpha^2 x_k^2 }\label{GUE_fermions}\;,
\ee
where $Z_N$ is a normalisation constant and $\alpha = \sqrt{m \omega/\hbar}$ is a characteristic inverse length scale. This expression (\ref{GUE_fermions}) thus establishes a one-to-one mapping between the scaled fermion's positions $\alpha\,x_i$'s and the eigenvalues $\lambda_i$'s 
of random matrices belonging to the so-called Gaussian Unitary Ensemble (GUE) \cite{mehta,For10}. 
%
%
From this mapping, one immediately obtains that, for large $N$, the density of fermions $\tilde \rho(x)$ has a finite support $[-r_e,+r_e]$ with $r_e = \sqrt{2N}/\alpha$, and is given by the Wigner semi-circle  
\be
\tilde \rho(x)\approx\tilde \rho_1^{\rm b}(x)=\frac{\alpha}{\pi}\sqrt{2N-\alpha^2 x^2}\;,\label{dens_oh_1d}
\ee
where the superscript '${\rm b}$' stands for bulk and the subscript $1$ for one dimension,
{and here and below the density $\tilde \rho$ is normalized to $N$, $\int dx \tilde \rho(x)=N$.}
In fact, the statistics of any observable at $T=0$ can be obtained from the determinantal structure of the correlations of the $x_i$'s~\cite{us_finiteT,DPMS:2015,fermions_review,marino_prl,Eis2013,CLM15}. Indeed, from the Wick theorem, any $p$-point correlation function of the positions $R_p(x_1, \cdots, x_p)$ can be written as a determinant of a $p \times p$ matrix whose entries are given by the so-called kernel $K_\mu(x,y)$. Here $\mu$ denotes the Fermi energy of the system, which is related to the number of particles $N$ in the system. In particular, the Fermion density $\tilde \rho(x)$ is given by $\tilde \rho(x) = K_\mu(x,x)$. 
{The correlation kernel for the fermions can be obtained from the known RMT results and displays two different scaling regimes:}
\begin{itemize}
\item A {\it bulk regime}, when both $x$ and $y$ are far from the edges at $\pm r_e = \pm \sqrt{2N}/\alpha$, where the density is finite. In this regime, for $x$ and $y$ close by with $x - y = O(\ell(x)) \ll 1$, with $\ell(x) = \pi/\tilde \rho_1^{\rm b}(x)$ being the typical inter-particle distance at point $x$, the correlations are translationally invariant and described by the so-called sine-kernel \cite{us_finiteT,fermions_review,Eis2013}
\be\label{sine_k}
K_\mu(x,y) \approx \frac{1}{\ell(x)} K_1^{\rm b}\left(\frac{x-y}{\ell(x)} \right) \;, \; K_1^{\rm b}(r)=\frac{\sin(r)}{\pi r}\;,
\ee
{which has been much studied} in RMT~\cite{mehta,For10}. 
\item An {\it edge regime}, when both $x$ and $y$ are located close to the edges of the spectrum at $\pm r_{\rm e}=\pm \sqrt{2N}/\alpha$, within a typical scale $w_N= 2^{-1/2}N^{-1/6}/\alpha$. At the edge (say $+r_e$), the kernel takes the scaling form
\be\label{scaling_edge_intro}
K_\mu(x,y) \approx \frac{1}{w_N} K_1^{\rm soft}\left(\frac{x-r_e}{w_N}, \frac{y-r_e}{w_N} \right)
\ee  
where $K_1^{\rm soft}(x,y)$ is the so-called Airy-kernel
 \cite{mehta,For10}
\be
K_{1}^{\rm soft}(x,y)=\int_0^{\infty}\Ai(u+x)\Ai(u+y)du=\frac{\Ai(x)\Ai'(y)-\Ai(y)\Ai'(x)}{x-y}\;,\label{airy_k}
\ee
where the superscript `${\rm soft}$' stands for the so-called soft edges, the standard denomination in RMT, and where $\Ai(x)$ is the Airy function. In particular, for large but finite $N$, the density $\tilde \rho(x)$ gets smoothened at the edge [see Fig. \ref{Fig_airy_LDA}], compared to the sharp $N \to \infty$ behaviour of the Wigner semi-circle in (\ref{dens_oh_1d}), and the edge density profile is given by 
\be
\tilde \rho(x)\approx\frac{1}{w_N}F_1^{\rm soft}\left(\frac{x-r_{\rm e}}{w_N}\right)\;\;{\rm with}\;\;F_1^{\rm soft}(s)=K_{1}^{\rm soft}(s,s)= \Ai'^2(s)-s\Ai^2(s)\;.\label{dens_oh_edge}
\ee
In Fig. \ref{Fig_airy_LDA}, we show a plot of the density near the edge. Another interesting consequence of this result (\ref{airy_k}) is that the typical fluctuations of the position of the rightmost fermion $x_{\max}=\max \{x_1, \cdots, x_N\}$ takes the scaling form, for large $N$,
\be\label{TW_distrib}
q_1(w)={\rm Prob} \left[x_{\max}\leq w\right]\approx {\cal F}_2\left(\frac{w-r_{\rm e}}{w_N}\right)\;,
\ee
where ${\cal F}_2(x)$ is the celebrated Tracy-Widom distribution for GUE \cite{TW}. 
This distribution is quite ubiquitous and has emerged in a variety of other systems (for a review see \cite{review_third_order}). 
It has also been observed experimentally, in nematic liquid crystals \cite{liquid_crystals} as well as in coupled optical fiber experiments
\cite{optical_fibers}. In fact, the system of noninteracting fermions in a one-dimensional harmonic potential  
is somehow the simplest system where the TW distribution could possibly be directly observed \cite{us_finiteT}.

\end{itemize} 

In the case of the harmonic potential, these results (\ref{sine_k}), (\ref{airy_k}) and (\ref{TW_distrib})
have been generalized to higher dimensions $d>1$ \cite{DPMS:2015,fermions_review,us_Wigner} [see also Eq. (\ref{K_LDA}, \ref{K_d_soft}) below], as well as to  finite temperature $T>0$ in any $d \geq 1$ \cite{us_finiteT,fermions_review,us_Wigner}, with a nontrivial dependence both on $d$ and $T$. The harmonic potential is very interesting because it is exactly solvable and it can thus bring a very useful insight. But since current experimental techniques allow to design confining potentials $V({\bf x})$ of different shapes \cite{Zwi2017,Hueck18}, it is natural and important to study how these results depend on $V({\bf x})$. Remarkably, it was shown that the results found for the harmonic well are actually universal
\cite{DPMS:2015,us_finiteT,Eis2013}, both in the bulk and at the edge, for a large class of {\it smoothly} varying spherically symmetric potentials, i.e. $V(|{\bf x}|)\sim |{\bf x}|^{p}$ with $p>0$ for instance (see \cite{fermions_review} for a more precise statement on the shape of the potential). It is thus natural to wonder what happens for non-smooth (or singular) potentials, which is precisely the question addressed in this paper.  

The most natural ``non-smooth'' example is the hard box potential, i.e. a potential $V_{\cal D}({\bf x})=0$ in a certain domain ${\cal D}$ and $V_{\cal D}({\bf x})=+\infty$ outside. A similar potential was recently designed experimentally in two dimensions \cite{Hueck18}. The hard box potential presents a strong discontinuity at the edge of the domain, where the single-particle wave functions must vanish. It is thus rather natural to expect that, close to the boundary of ${\cal D}$, which acts as a ``hard edge'' for the Fermi gas, the correlation kernel is different from the one found for smooth potentials [see Eqs. (\ref{airy_k}) and (\ref{K_d_soft})]. In a recent short Letter \cite{LLMS17}, we obtained analytical expressions for these new kernels for a pure hard box potential. The goal of the present paper is thus two-fold: (i) provide the details of the computations for the pure hard-box potential, which were only briefly sketched in \cite{LLMS17}, and (ii) analyse a much wider class of ``non-smooth'' potentials. The latter includes in particular non-uniform hard-box potentials as well as one-dimensional singular power-law potentials of the form $V(x)\sim |x|^{-\gamma}$ with $\gamma>0$. As we will see, such potentials lead to new universal kernels, which we analyze in detail.

Note that there has been interest in more mathematical literature on related topics in determinantal
processes and random matrices \cite{J07,BornemanFermions,MNS,Macedo,VerbaarschotGarcia1D} 
and, recently,
in explicit $1d$ noninteracting fermion models at zero \cite{Cunden1D} and finite temperature 
\cite{Liechty17,JohanssonLambert}.

\subsection{Main results}

\subsubsection{Hard box potential}\hfill\\


Let us first summarise our main results in the case of a spherical hard box potential, with a uniform potential inside the box
\be\label{hb_pot_d_result}
V_R({\bf x})=\begin{cases}
        & 0\;,\;\;|{\bf x}|\leq R\\
				&\\
        & \infty\;,\;\;|{\bf x}|>R\;.
       \end{cases}
\ee
\noindent {\bf Zero temperature ($T=0$):}  The eigenfunctions and the corresponding eigenvalues of the single particle Hamiltonian associated to $V_R({\bf x})$ can be computed exactly and the correlation kernel $K_\mu({\bf x}, {\bf y})$ can thus be obtained exactly for any finite number $N$ of fermions. By taking the large $N$ limit of this exact solution, we find that, at $T=0$, the correlations in the bulk, i.e. far from the boundary, are given by the usual sine-kernel (\ref{sine_k}) in $d=1$ or its generalisation in higher dimensions $d$ (\ref{K_LDA}). Indeed, one has \cite{DPMS:2015,fermions_review,torquato}
\be\label{bulk_kernel_d_intro}
K_\mu({\bf x}, {\bf y}) \approx k_F^d K_d^{\rm b}\left(k_F |{\bf x} - {\bf y}| \right) \;, \; K_d^{\rm b}(r) = \frac{{\rm J}_{d/2}(r)}{(2\pi r)^{d/2}} \;,
\ee 
where $k_F = \sqrt{2 m \mu}/\hbar$ and ${\rm J}_\nu(x)$ denotes the Bessel function of index $\nu$. In particular, the fermion density $\tilde \rho({\bf x}) = K_\mu({\bf x}, {\bf x})$ is thus uniform in the bulk $\tilde \rho(x) \approx \rho_0$ with $\rho_0 = k_F^d 2^{-d}/\gamma_d$ and $\gamma_d = \pi^{d/2} \Gamma(1+d/2)$. What about the correlations at the edge, i.e. close to the hard wall? 

\begin{figure}[t]
\centering
\includegraphics[width=\linewidth]{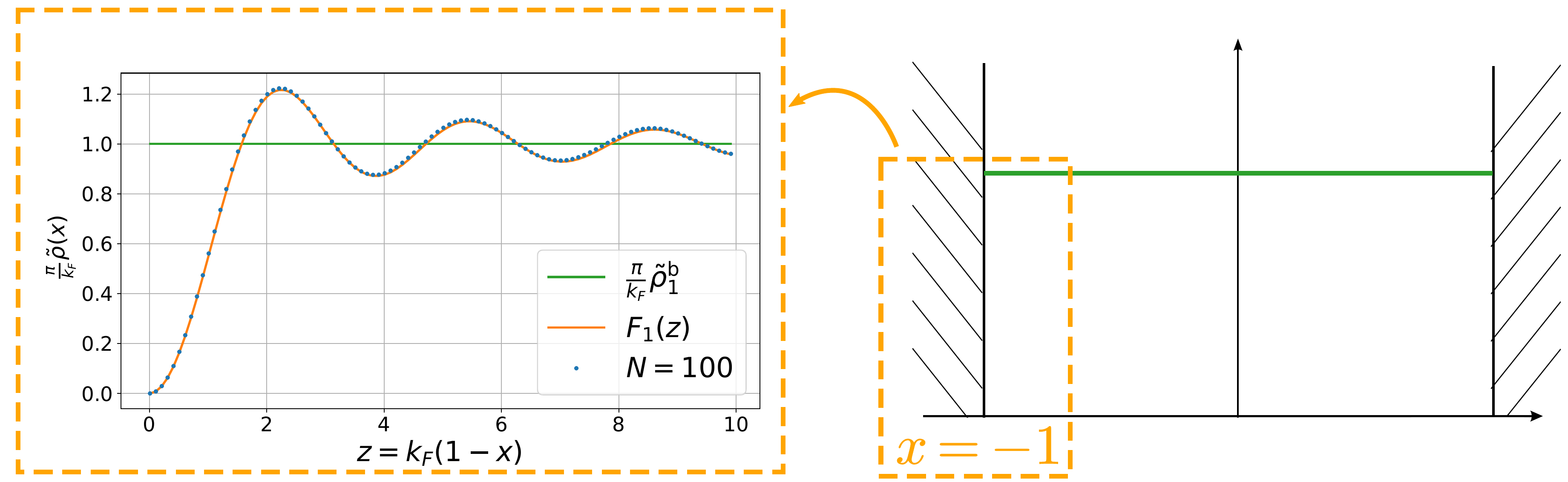}
\caption{Sketch of a hard box potential and the corresponding bulk density (in green). Close to the edge, the finite $N$ density takes the scaling form described in Eq. \eqref{edge_dens_hb_1d} while the scaling form for the bulk density in Eq. \eqref{k_F_hb} fails to describe the oscillations.}\label{Fig_box}
\end{figure}

To appreciate the difference with the case of smooth potentials, let us first consider the one-dimensional case. Indeed, in $d=1$, the quantum joint PDF of the positions of the fermions inside the box can be computed exactly, yielding
\be\label{joint_PDF_hb_1d_intro}
|\Psi_0({x}_1,\cdots,{x}_N)|^2=\frac{1}{Z_N}\prod_{k=1}^N \cos^2\left(\frac{\pi x_k}{2 R}\right)\prod_{i<j}^N \left|\sin\left(\frac{\pi x_i}{2 R}\right)-\sin\left(\frac{\pi x_j}{2 R}\right)\right|^2\;,
\ee
with $-R \leq x_i \leq R$ and where $Z_N$ is a normalisation constant. This result (\ref{joint_PDF_hb_1d_intro}) is quite different from the result found for the harmonic oscillator in Eq. (\ref{GUE_fermions}). While Eq. (\ref{GUE_fermions}) corresponds to the GUE, the joint PDF in (\ref{joint_PDF_hb_1d}) is actually related to the Jacobi Unitary Ensemble (JUE) of RMT. Indeed, defining $u_i=(1+\sin(\pi\,x_i/(2R)))/2$, the joint PDF for the $u_i$'s then reads (with $0\leq u_i \leq 1$, for all $i=1,\cdots, N$)
\be\label{JUE_intro}
P(u_1,\cdots,u_N)=\frac{1}{Z_N'}\prod_{k=1}^N \sqrt{u_k(1-u_k)}\prod_{i<j}^N \left|u_i-u_j\right|^2\;,
\ee
with $Z'_N$ another normalisation constant. This joint PDF in (\ref{JUE_intro}) is well known in RMT as the PDF of the eigenvalues of the JUE (with parameters $a = b = \frac{1}{2}^+$)~\cite{mehta,For10}. In Appendix \ref{App:Jacobi} we show that the JUE with arbitrary parameters $a$ and $b$ can be realized by a model of fermions in an appropriately chosen potential [see Eq. (\ref{potential_Jacobi})]. In RMT, it is well known that the edge behaviors in these two ensembles, GUE and JUE, are quite different (``soft edge'' for GUE versus ``hard edge'' for JUE). And indeed, in the large $N$ limit, we show that the kernel $K_\mu(x,y)$, for $x$ and $y$ both close to the edge at $x=R$ takes the scaling form  
\be\label{khb_1d_e_intro}
K_\mu({x}, {y})=k_F K_1^{\rm e}(k_F(R-x),k_F(R-y)),\;\;{\rm with}\;\;K_1^{\rm e}(u,v)=\frac{\sin(u-v)}{\pi(u-v)}-\frac{\sin(u+v)}{\pi(u+v)}\;,
\ee
where $k_F = \sqrt{2 m \mu}/\hbar$ -- we recall that $\mu$ is the Fermi energy. This limiting form of the kernel (\ref{khb_1d_e_intro}) is quite different from the edge kernel obtained for a soft potential, given in Eq. (\ref{airy_k}). From Eq. (\ref{khb_1d_e_intro}) we obtain the density profile at the edge, given by $\tilde \rho(x) = K_\mu(x,x)$, which reads in $d=1$
\be\label{density_1d_hwall}
\tilde \rho(x) \approx \rho_{0} F_1(k_F(R-x)) \; {\textrm{with}} \;\; F_1(z) = 1 - \frac{\sin 2z}{2z} \;,
\ee 
where $F_1(z)$ stands for $F_1^{\rm hard}(z)$ -- in contrast with $F^{\rm soft}_1(z)$ used for the soft edge (\ref{dens_oh_edge}). In the following, to simplify the notations, we will omit this superscript ``hard''. In Fig. \ref{Fig_box}, we show a plot of the density profile near the wall (\ref{density_1d_hwall}), which is quite different from
the density profile near the soft-edge corresponding to a smooth potential (\ref{dens_oh_edge}), plotted in Fig. \ref{Fig_airy_LDA}. 
Finally, we also obtain the generalisation of the kernel (\ref{khb_1d_e_intro}) in any higher dimensions $d \geq 1$, as given in Eq. \eqref{k_d_final}, which differs significantly from the edge kernel obtained for a soft potential in Eq. (\ref{airy_k}).  

\noindent{\bf Finite temperature $T>0$:} We next extend these results for the edge kernels in $d=1$ (\ref{khb_1d_e_intro}) and in $d \geq 1$ (\ref{k_d_final}) to finite temperature. 
The typical scale of fluctuations at finite temperature is set by the de Broglie wave length $\lambda_{T}=\sqrt{2\pi\hbar^2\beta /m}$, with $\beta=1/(k_B T)$ the inverse temperature. The kernel then takes at the edge the scaling form described in Eq. \eqref{k_edge_b_finite_2}.
These results obtained for a spherical boundary (and uniform potential) can be generalised to any smooth boundary \cite{LLMS17}. Note that non-smooth boundaries, like a wedge in $2d$, instead lead to different kernels which were calculated explicitly in \cite{LLMS17}.


\subsubsection{Non-uniform potential: from soft to hard edge}\hfill\\

The previous results show that the presence of a hard wall and uniform potential $V({\bf x})=0$ leads, in the large $N$ limit, to a correlation kernel, close to the wall at $|{\bf x}| = R$, which is quite different from the edge kernel near the ``soft edge'' at $|{\bf x}| = r_{\rm e}$ created by a smooth potential. It is thus natural to study a potential which interpolates between these two situations, namely 
\be\label{hb_pot_d_result}
V_R({\bf x})=\begin{cases}
        & V(|{\bf x}|)\;,\;\;|{\bf x}|\leq R\\
				&\\
        & \infty\;,\;\;\;\;\;\;\;\;|{\bf x}|>R\;,
       \end{cases}
\ee
where $V(|{\bf x}|)$ is non-uniform:  we call such a potential $V_R({\bf x})$ a ``truncated potential''. In the absence of a hard wall, the density of the Fermi gas displays an edge $r_{\rm e}$, such that $V(r_{\rm e}) = \mu$, with $\mu$ the Fermi energy, beyond which the density vanishes. Therefore two different situations may occur: (i) if $R \gg r_{\rm e}$ the edge kernel at $r_{\rm e}$ is given by the Airy kernel (\ref{airy_k}) (or its generalisation (\ref{K_d_soft}) for $d > 1$) since the effect of the wall is negligible, (ii) if $R \ll r_{\rm e}$ 
(i.e. $\frac{V(R)}{\mu} > 1$ at large $\mu$) the edge kernel at $R$ is given by the hard-edge kernel in (\ref{khb_1d_e_intro}) (or its generalisation (\ref{k_d_final}) for $d>1$). We show that when both $R$ and $r_{\rm e}$ are
of the same order, there is a kernel given in Eq. \eqref{K_1_l_final} for $d=1$ and in Eq. \eqref{K_ell_d} for $d>1$, parametrised by the parameter $\ell=(R-r_{\rm e})/w_N$, with {$w_N=\hbar^{2/3}/(2mV'(r_{\rm e}))^{1/3}$ the typical length scale at the soft edge,} that smoothly interpolates between the soft-edge and hard-edge kernel. Interestingly, the kernel that we found here in $d=1$, also appears in the study of the persistence properties of the Airy$_2$ process \cite{FerFri}.

\subsubsection{Singular potentials}\hfill\\

The hard box potential is very interesting from a theoretical point of view, but one may wonder
how one can realise experimentally such impenetrable barriers, say at the origin $x=0$. A natural way is to consider power-law repulsive potentials \cite{Andrews}
\be\label{singular_pot}
V(x)=\q \frac{\alpha(\alpha+1)}{|x|^{\gamma}}\;,\;\;|x|\neq 0\;\;{\rm with}\;\;\gamma>0 \,\; \;{\rm and} \; \alpha > 0\;.
\ee
In this paper, we show that the limiting kernel close to $x=0$ takes quite different scaling form depending on the value of $\gamma$, {namely at zero temperature:}
\begin{itemize}
\item for $0\leq \gamma<1$, the single particle wave-function $\phi_k(x)$ can be non-zero at $x=0$ and fermions can move from positive to negative value of $x$. The potential is not strong enough to confine the particles, hence we will not study this case here.
\item for $1\leq \gamma<2$, the single particle wave-function $\phi_k(x)$ has to vanish at $x=0$ and fermions can not leave the half-space. The correlation kernel then takes the same scaling form as for the hard box potential close to the origin
\be
K_{\mu}(x,y)\approx k_F K_{1}^{\rm e}(k_F x,k_F y)\;,\nn
\ee
where $K_{1}^{\rm e}(u,v)$ is given in Eq. (\ref{khb_1d_e_intro}). 
\item for $\gamma=2$, the correlation kernel depends continuously on $\alpha$ and is given by 
\begin{align}
&K_{\mu}(x,y)=2 k_F^2 \sqrt{x y} K_{\rm Be,\alpha+1/2}(k_F^2 x^2 , k_F^2 y^2)\;,\nn\\
&{\rm with}\;\;K_{\rm Be,\nu}(u,v)=\frac{ \sqrt{v}{\J}'_{\nu}(\sqrt{v}) \J_{\nu}(\sqrt{u}) 
- \sqrt{u} \J'_{\nu}(\sqrt{u}) \J_{\nu}(\sqrt{v}) }{2 (u-v)}\;, \label{K_bessel_intro}
\end{align}
which is the well known Bessel kernel, characteristic of hard edge scaling limits in random matrix theory \cite{For10}. {While for random matrices the Bessel kernel appears in terms of a scaled
coordinate near the edge in the large $N$ limit, it is noteworthy that for fermions in a pure $1/x^2$ potential
it is exact for any value of $\mu$.}

\item for $\gamma>2$ and independently of $\alpha>0$, there exists a finite smooth edge $r_{\rm e} \sim k_F^{-2/\gamma} \gg k_F^{-1}$ below which the density vanishes, i.e. it vanishes on $[0,r_e]$. Consequently, close to $r_e$, the limiting kernel is given by the Airy kernel 
\begin{align}
&K_{\mu}(x,y)\approx \frac{1}{w_N} K_{1}^{\rm soft}\left(\frac{r_{\rm e}-x}{w_N},\frac{r_{\rm e}-y}{w_N}\right)\;,\nn \\
&{\rm with}\;\;V(r_{\rm e}=\mu)\;\;{\rm and}\;\;w_N=\left(\frac{\hbar^2}{2m V'(r_{\rm e})}\right)^{\frac{1}{3}}\;. \label{K_airy_intro}
\end{align}
\end{itemize}

The case $\gamma=2$ in (\ref{singular_pot}) thus appears as the critical case, which interpolates between the `hard box' kernel (\ref{khb_1d_e_intro}), as $\alpha \to 0$ (using simply that $2\sqrt{u v} K_{{\rm Be},1/2}(u,v)=K_{1}^{\rm e}(u,v)$), and the Airy kernel (\ref{K_airy_intro}) as $\alpha \to \infty$ \cite{BFP98} [see also Eq. (\ref{convergence}) below]. {These results extend to finite temperature. The kernel for $\gamma=2$ 
at $T>0$ is presented in \eqref{finiteTBessel} and continously depends both on $\alpha$ and
$b$, a dimensionless inverse temperature parameter defined in \eqref{defbBessel}.
In principle, these results could also be extended to higher dimensions $d>1$, 
as we did for the hard box case, 
but this is left for future investigations.}

The paper is organized as follows. In Section \ref{Reminder_det} we recall the determinantal structure of the correlations, as well as the main results obtained for smooth trapping potentials. Section \ref{hb_1d_T0} is dedicated to the hard box potential at zero temperature and in one-dimension. We first discuss the case of a zero potential inside the box and then turn to the case of a non-uniform potential. In Section \ref{hb_d_T0}, we extend these results to the case of higher dimensions $d\geq 1$, still at $T=0$. In Section \ref{hb_d_T}, we treat the case of the hard box, with uniform potential inside, in dimension $d\geq 1$ and at finite temperature $T>0$. Finally in Section \ref{sing_pot}, we treat the case of different one dimensional singular power law potentials, i.e. $V(x)\sim |x|^{-\gamma}$ with $\gamma\geq 0$ and compute the correlation kernel close to the origin in $x=0$. {Appendix A contains the detailed derivation of the asymptotic behaviours of the crossover kernel between the hard and soft edges. Appendix B contains the derivation of the finite temperature bulk kernel, and Appendix C contains a study of fermions in the "Jacobi trap", related to the
general Jacobi ensemble of RMT.}


\section{Determinantal structure for a system of fermions at $T=0$}


We consider a $d$-dimensional system of $N$ spinless non-interacting fermions confined by an external potential $V({\bf x})$, where ${\bf x}$ is the position in $d$-dimension. The Hamiltonian of the system reads
\be\label{hamiltonian}
{\cal H}_N=\sum_{i=1}^N H_i,\;\;{\rm with}\;\;H_i=-\q\Delta_{{\bf x}_i }+V({\bf x}_i)\;.
\ee

\subsection{General framework}

At $T=0$, the system is in its ground state. Let us introduce the single particle eigenfunctions $\phi_{{\bf k}}({\bf x})$, which are  solutions of the Schr\"odinger equation
\be\label{spwf}
H\phi_{\bf k}({\bf x})=-\q\Delta_{{\bf x}}\phi_{\bf k}({\bf x})+V({\bf x})\phi_{\bf k}({\bf x})=\epsilon_{\bf k}\phi_{\bf k}({\bf x})\;.
\ee
These single particle eigenfunctions are orthonormal 
\be\label{ortho_wf}
\int d^d {\bf x} \, \phi_{\bf k}^*({\bf x})\phi_{\bf k'}({\bf x})=\delta_{{\bf k},{\bf k'}}\;.
\ee
We denote $\epsilon_1\leq\cdots\leq \epsilon_N$ the $N$ lowest energy levels and ${\bf k}_1,\cdots,{\bf k}_N$ their corresponding quantum numbers.
The last occupied level has an energy $\epsilon_N=\mu$ where $\mu$ is the Fermi energy. We consider the case where the many body ground state of energy $E_0=\sum_{k=1}^N \epsilon_k$ is non-degenerate. 
One can write this $N$-body ground state wave function as a Slater determinant built from the $N$ lowest level single-particle eigenfunctions
\be\label{GS_wf}
\Psi_0({\bf x}_1,\cdots,{\bf x}_N)=\frac{1}{\sqrt{N!}}\det_{1\leq i,j\leq N}\phi_{{\bf k}_i}({\bf x}_j)\;.
\ee
The quantum joint PDF for the positions ${\bf x}_1,\cdots,{\bf x}_N$ of the fermions is 
\be\label{GS_joint_1}
|\Psi_0({\bf x}_1,\cdots,{\bf x}_N)|^2=\frac{1}{N!}\det_{1\leq i,j\leq N}\phi_{{\bf k}_i}({\bf x}_j)\det_{1\leq l,m\leq N}\phi_{{\bf k}_l}^*({\bf x}_m)\;.
\ee
One can then use the identity $\det A \det B =\det AB$ to rewrite this joint PDF in Eq. \eqref{GS_joint_1} as
\be\label{GS_joint_2}
|\Psi_0({\bf x}_1,\cdots,{\bf x}_N)|^2=\frac{1}{N!}\det_{1\leq i,j\leq N}K_{\mu}({\bf x}_i,{\bf x}_j)\;.
\ee
The function $K_{\mu}({\bf x},{\bf y})$ is called the correlation kernel and is expressed from the $N$ lowest energy single-particle wave-functions, solutions of Eq. \eqref{spwf} as
\be\label{Kernel}
K_{\mu}({\bf x},{\bf y})=\sum_{i=1}^N\phi_{{\bf k}_i}^*({\bf x})\phi_{{\bf k}_i}({\bf y})=\sum_{\bf k}\phi_{\bf k}^*({\bf x})\phi_{\bf k}({\bf y})\Theta(\mu-\epsilon_{\bf k})\;,
\ee
with $\Theta(x)$ the Heaviside step-function.
The orthonormality of the wave functions \eqref{ortho_wf} implies the reproducibility property of the kernel
\be\label{reproducibility}
\int d^d{\bf y} \, K_{\mu}({\bf x},{\bf y})K_{\mu}({\bf y},{\bf z})=K_{\mu}({\bf x},{\bf z})\;.
\ee
The positions of the fermions $\lbrace {\bf x}_i,\;i=1,\cdots,N\rbrace$ form a $d$-dimensional determinantal point process \cite{J05,Bo11} of kernel $K_{\mu}({\bf x},{\bf y})$.
In particular, the $p$-point correlation function (for $1\leq p\leq N$) can be evaluated using Eq. \eqref{reproducibility} as
\be\label{correlation}
R_p({\bf x}_1,\cdots,{\bf x}_p)=\frac{N!}{(N-p)!}\int d^d{\bf x}_{p+1}\cdots d^d{\bf x}_N |\Psi_0({\bf x}_1,\cdots,{\bf x}_N)|^2=\det_{1\leq i,j\leq p}K_{\mu}({\bf x}_i,{\bf x}_j)\;,
\ee
where the kernel $K_{\mu}({\bf x},{\bf y})$ is given in Eq. \eqref{Kernel}. 
In particular, from Eq. \eqref{correlation}, the spatial density of the total system can be obtained as
\be\label{dens}
\tilde \rho({\bf x})=R_1({\bf x},{\bf x})=K_{\mu}({\bf x},{\bf x})=\sum_{\bf k}\left|\phi_{\bf k}({\bf x})\right|^2\Theta(\mu-\epsilon_{\bf k})\;.
\ee
{with $\int d^d {\bf x} \, \tilde \rho({\bf x})=N$.}

\subsection{A reminder on the case of smooth potentials at $T=0$}\label{Reminder_det}

Let us briefly recall the main {exact} results obtained in Refs. \cite{DPMS:2015,fermions_review} in the case of a smooth spherically symmetric potential $V({\bf x}) = V(|{\bf x}|)$. This will be useful in the following derivations and interpretations of our results. At $T=0$, the density in the bulk for large $N$, {which is correctly predicted by the LDA}, reads
\be
\tilde \rho({\bf x})\approx \tilde \rho_d^{\rm b}({\bf x})=\Omega_d\left(\frac{m\left[\mu-V(|{\bf x}|)\right]}{2\pi^2\hbar^2}\right)^{\frac{d}{2}}\;,\label{rho_LDA} 
\ee
which admits an edge at ${\bf x} = {\bf x}_{\rm e}$, such that $V(|{\bf x}_{\rm e}|=r_{\rm e})=\mu$ (where we recall that $\mu$ is the Fermi energy). In Eq. (\ref{rho_LDA}), $\Omega_d = \pi^{d/2}/\Gamma(1+d/2)$ is the volume of the $d$-dimensional unit sphere. As in the $1d$ case discussed above [see Eqs. (\ref{sine_k}) and (\ref{airy_k})], the correlation kernel $K_\mu({\bf x}, {\bf y})$ displays different scaling behavior in the {bulk}, i.e. for ${\bf x}$ and ${\bf y}$ far from the edge of the density profile $\tilde \rho({\bf x})$, and at the edge, i.e. for ${\bf x}$ and ${\bf y}$ both close to the edge.
\begin{itemize}
\item[$\bullet$]{In the bulk and for ${\bf x}$ and ${\bf y}$ close by, the kernel takes the scaling form
\begin{align}
& K_\mu({\bf x},{\bf y})\approx \tilde k({\bf x})^d K_d^{\rm b}(\tilde k({\bf x})|{\bf x}-{\bf y}|)\;\;{\rm with}\;\;K_d^{\rm b}(r)=\frac{\J_{d/2}(r)}{(2\pi r)^{d/2}}\;,\label{K_LDA}\\
&{\rm and}\;\;\tilde k({\bf x})=\frac{\sqrt{2m\left[\mu-V(|{\bf x}|)\right]}}{\hbar}\;,\nn
\end{align}
where ${\J}_{\nu}(x)$ denotes the Bessel function of index $\nu$. {The quantity $\tilde k({\bf x})$ can be interpreted as the local Fermi wavevector.}
As expected, the scaling function $K_d^{\rm b}(r)$ in (\ref{K_LDA}) is the same as the one describing the bulk behaviour in the hard box (\ref{bulk_kernel_d_intro}). In particular, evaluating this formula (\ref{K_LDA}) for ${\bf x} = {\bf y}$ gives the expression for the density in~(\ref{rho_LDA}). }

\item[$\bullet$]{Close to an edge point ${\bf x}_{\rm e}$, such that $V(|{\bf x}_{\rm e}|=r_{\rm e})=\mu$, and using the coordinates ${\bf x}=(x_n={\bf x}_{\rm e}\cdot{\bf x},{\bf x}_t)$, with ${\bf x}_{\rm e}\cdot{\bf x}_t=0$, it has been shown \cite{DPMS:2015,fermions_review} that the kernel takes the scaling form 
\begin{align}
&K_{\mu}({\bf x},{\bf y})=\frac{1}{w_N^d}K_d^{\rm soft}\left(\frac{{\bf x}-{\bf x}_{\rm e}}{w_N},\frac{{\bf y}-{\bf x}_{\rm e}}{w_N}\right)\;,\;\;{\rm with}\;\;w_N=\left(\frac{\hbar^2}{2m V'(r_{\rm e})}\right)^{\frac{1}{3}}\;,\label{K_d_soft}\\
&{\rm and}\;\;K_d^{\rm soft}({\bf u},{\bf v})=\int\frac{d^{d-1}{\bf l}}{(2\pi)^{d-1}}\e^{i{\bf l}\cdot({\bf u}_t-{\bf v}_t)}\int_{{\bf l}^2}^{\infty}dz \Ai(u_n+z)\Ai(v_n+z)\;.\nn
\end{align}
The superscript 'soft' stands for a soft edge, as opposed to the `hard edge' studied below. In fact, for $d\geq 2$, the angular part of the 
$(d-1)$ dimensional integral over ${\bf l}$ in (\ref{K_d_soft}) can be computed explicitly using Eqs. (\ref{angular_part_2}) and (\ref{angular_part_3}) to obtain   
\be\label{K_d_soft_2}
K_d^{\rm soft}({\bf u},{\bf v})=\int_0^{\infty} dl \left(\frac{l}{2\pi}\right)^{\frac{d-1}{2}}\frac{\J_{\frac{d-3}{2}}(l|{\bf u}_t-{\bf v}_t|)}{|{\bf u}_t-{\bf v}_t|^{\frac{d-3}{2}}}\int_{l^2}^{\infty}dz \Ai(u_n+z)\Ai(v_n+z)\;.
\ee
}

\end{itemize}

Let us now turn to the derivation of the results for potentials inducing hard edges to the density, starting from the simplest cases in dimension one at zero temperature.

\section{One-dimensional hard box at zero temperature} \label{hb_1d_T0}

We focus on the case $d=1$ and $T=0$ and first consider the simple case of a hard box potential where the potential is uniformly equal to zero in the box. Then, we study the case of non-uniform box-potentials. The case of singular power-law potentials is discussed in a separate section (see Section \ref{sing_pot}). 

\subsection{Uniform hard box potential} \label{hb1d}
\subsubsection{Finite $N$ solution and connection with the Jacobi Unitary Ensemble}\hfill\\

In the case of a one dimensional hard box potential of the form
\be\label{hb_pot_1d}
V_R(x)=\begin{cases}
        &\displaystyle 0\;,\;\;|x|\leq R\\
        & \displaystyle\infty\;,\;\;|x|>R\;,
       \end{cases}
\ee
the single-particle eigenfunctions and energies are given for $n=1,2,\cdots$ as
\be\label{hb_1d_wf}
\phi_n(x)=\sin\left(\frac{n\pi}{2R}(x+R)\right)\;\;{\rm and}\;\;\epsilon_n=\q k_n^2=\frac{\hbar^2\pi^2}{8 m R^2}n^2\;.
\ee
We set in the following $R=1$, which amounts to rescale all positions by $R$. The $N$-body ground state wave function is given by the 
Slater determinant constructed from the single-particle eigenfunctions in Eq. \eqref{hb_1d_wf},
\be\label{Slater_det}
\Psi_0({x}_1,\cdots,{x}_N)=\frac{1}{\sqrt{N!}}\det_{1\leq i,j\leq N}\phi_j(x_i)=\frac{1}{\sqrt{N!}}\det_{1\leq i,j\leq N}\sin\left(\frac{j \pi}{2R}(x_i+R)\right)\;.
\ee
This Salter determinant can be written in a more convenient way by using the identity $\sin(nx)=\sin(x)U_{n-1}(\cos(x))$ where $U_n(t)$ is the Chebychev polynomial of second kind of degree $n$. By rearrangements of rows and columns, the joint quantum PDF of the positions reads
\be\label{joint_PDF_hb_1d}
|\Psi_0({x}_1,\cdots,{x}_N)|^2=\frac{1}{Z_N}\prod_{k=1}^N \cos^2\left(\frac{\pi x_k}{2}\right)\prod_{i<j}^N \left|\sin\left(\frac{\pi x_i}{2}\right)-\sin\left(\frac{\pi x_j}{2}\right)\right|^2\;.
\ee
Introducing the new variables $u_i=(1+\sin(\pi x_i/2))/2$, the joint PDF of $u_1,\cdots,u_N$ can be worked out from \eqref{joint_PDF_hb_1d}. 
It coincides with the joint PDF of the eigenvalues of a matrix belonging to the Jacobi Unitary Ensemble of Random Matrix Theory (RMT) \cite{mehta,For10,Cunden1D,FFGW03}
\be\label{JUE}
P(u_1,\cdots,u_N)=\frac{1}{Z_N'}\prod_{k=1}^N \sqrt{u_k(1-u_k)}\prod_{i<j}^N \left|u_i-u_j\right|^2\;,
\ee
as announced in the Introduction in Eqs. (\ref{joint_PDF_hb_1d_intro}) and (\ref{JUE_intro}). 
%

The correlation kernel $K_{\mu}(x,y)$ for this case can be worked out exactly from the expression in Eq. \eqref{Kernel}, and using the single-particle wave-function in Eq. \eqref{hb_1d_wf}. This yields for any finite value of $N$
\be\label{khb_1d_N}
K_\mu({x}, {y}) = \sum_{n=1}^N \phi_{n}^*(x) \phi_{n}(y)  = \frac{\sin\left(\frac{(2N+1)\pi}{4}(x-y)\right)}{4\sin\left(\frac{\pi}{4}(x-y)\right)}-\frac{\sin\left(\frac{(2N+1)\pi}{4}(2+x+y)\right)}{4\sin\left(\frac{\pi}{4}(2+x+y)\right)} \;.
\ee 
We deduce the finite $N$ density $\tilde \rho(x)$ by inserting Eq. \eqref{khb_1d_N} in Eq. \eqref{dens},
\be\label{dens_1_N}
\tilde \rho(x)=\frac{2N+1}{4}-(-1)^N\frac{\cos\left(\frac{(2N+1)\pi x}{2}\right)}{4\cos\left(\frac{\pi x}{2}\right)}\;.
\ee
One can easily check that $\tilde \rho(x)$ in Eq. \eqref{dens_1_N} is normalized to $N$, using
\be
\int_{-1}^{1}\tilde \rho(x)dx=\frac{2N+1}{2}-\frac{(-1)^N}{4}\int_{-1}^{1}\frac{\cos\left(\frac{(2N+1)\pi x}{2}\right)}{\cos\left(\frac{\pi x}{2}\right)}dx=\frac{2N+1}{2}-\frac{1}{2}=N\;.
\ee
The exact expression in Eq. (\ref{dens_1_N}) shows that the density vanishes at the edges for $x=\pm 1$ on a typical scale $\sim N^{-1}$ and oscillates in the bulk around a constant value $N/2+1/4$.
For large $N$, we thus distinguish two different regimes:
\begin{itemize}
 \item A bulk regime for $|x\pm 1|\gg N^{-1}$ with a nearly constant density $\rho_{\rm b}\approx N/2$, obtained by neglecting the oscillating term in Eq. \eqref{dens_1_N}.
Note that we obtain exactly the bulk density in Eq. \eqref{rho_LDA} inserting $V(x)=0$,
\be\label{k_F_hb}
\tilde \rho(x)\approx\tilde \rho_1^{\rm b}(x)=\frac{\sqrt{2m\mu}}{\pi\hbar}\Theta(1-|x|)=\frac{k_F}{\pi}\Theta(1-|x|)\;\;{\rm with}\;\;k_F=\frac{\sqrt{2m\mu}}{\hbar}=\frac{N\pi}{2}\;.
\ee
 \item An edge regime at a distance $\sim 1/k_F=2/(N\pi)$ of the edges in $x=\pm 1$ where the density takes the scaling form
\be\label{edge_dens_hb_1d}
\tilde \rho(x)\approx \rho_{\rm b}F_1(k_F(1\pm x))\;,\;\;k_F(1\pm x)=O(1)\;\;{\rm with}\;\;
F_1(z)=1-\frac{\sin 2z}{2z}\;.
\ee
The boundary conditions impose that for any $n$, the single particle eigenfunctions defined in Eq. \eqref{hb_1d_wf} satisfy $\phi_n(\pm 1)=0$. Therefore, the density has to vanish at $x=\pm 1$ and the scaling function $F_1(z)$ in Eq. \eqref{edge_dens_hb_1d} describes the cross-over from the constant value in the bulk to $0$ exactly at $x=\pm 1$. This function is plotted in Fig. \ref{Fig_box} along with its finite $N$ counterpart obtained from Eq. \eqref{dens}. Its asymptotic behaviours read
\be\label{dens_edge_hb_1d_as}
F_1(z)\approx\begin{cases}
                   &\displaystyle \frac{2}{3}z^2+O(z^4)\;,\;\;z \ll 1\\
                   &\\
                   &\displaystyle 1+O(z^{-1})\;,\;\;z \gg 1 \;.\\
                  \end{cases}
\ee
In particular the large $z\gg 1$ behavior of $F_1(z)$ in Eq. \eqref{dens_edge_hb_1d_as} together with \eqref{edge_dens_hb_1d} shows a smooth matching between the edge and bulk densities.
\end{itemize}
We now analyze the correlation kernel in the large $N$ limit.

\subsubsection{Correlation kernel for large $N$}\hfill\\

In the limit $N\gg 1$, the kernel in Eq. \eqref{khb_1d_N} exhibits the same two regimes as the density (bulk and edge regimes):
\begin{itemize}
\item  In the bulk, where both $x$ and $y$ are far from the walls, we set $x-y = r/k_F = 2 r/(N \pi)$, where $r$ is a dimensionless variable, in the exact
expression for $K_\mu(x,y)$ in Eq. (\ref{khb_1d_N}). In the large $N$ limit, the second term is of order $O(1)$ and is highly oscillating, while the first
term is of order $O(N)$, and is thus the leading contribution for large $N$. Hence we find that the kernel $K_\mu(x,y)$ takes the scaling form
\be\label{khb_1d_b}
K_\mu({x}, {y})=k_F K_1^{\rm b}(k_F(x-y)),\;\;{\rm with}\;\;K_1^{\rm b}(r)=\frac{\sin r}{\pi r}\;,
\ee
which is the sine-kernel, as announced in Eq. (\ref{sine_k}). 
%
\item At the edge, i.e. when both $x$ and $y$ are close to $1$, we set $k_F(1-x)=O(1)$ and $k_F(1-y)=O(1)$ in the expression for $K_\mu(x,y)$ in Eq. (\ref{khb_1d_N}). In this case, both terms in Eq. (\ref{khb_1d_N}) are now of order $O(N)$ and they equally contribute to the limiting form of the kernel. We thus find that $K_\mu(x,y)$ takes the scaling form~\cite{LLMS17,CMV2011}
\be\label{khb_1d_e}
K_\mu({x}, {y})=k_F K_1^{\rm e}(k_F(1-x),k_F(1-y)),\;\;{\rm with}\;\;K_1^{\rm e}(u,v)=\frac{\sin(u-v)}{\pi(u-v)}-\frac{\sin(u+v)}{\pi(u+v)}\;.
\ee
The first term of Eq. \eqref{khb_1d_e} is exactly the bulk term $K_1^{\rm b}(u-v)$. The position $y=1-v/k_F$ has a mirror image $y^{T}=1+v/k_F$ by reflexion with respect to the wall at $x=1$. Therefore, the rescaled distance between $x=1+u/k_F$ and $y^{T}$ is $k_F(x-y^{T})=u+v$. This is precisely the argument of the second term in Eq. \eqref{khb_1d_e}. In particular, it ensures that the kernel vanishes at the wall for $u=0$ or $v=0$, such that $y=y^{T}$. One can check that when taking the position $x$ far from the edge $k_F(1-x)=u\gg 1$ but with a fixed distance with respect to the position $y$, such that $k_F(x-y)=u-v=O(1)$, only the second term in Eq.~\eqref{khb_1d_e} vanishes and the edge scaling function reduces to the bulk scaling function $K_1^{\rm b}(u-v)$.
\end{itemize}
These limiting kernels allow in particular via Eq. \eqref{correlation} the study of all $p$-points correlation functions, both in the bulk and at the edge. 

To further extend our study to the case of a non-zero potential inside the box, it is convenient to use a different method, which is quite
versatile, and relies on the use of the quantum propagator \cite{fermions_review}. This method is reviewed in the next section for the pure hard box potential (\ref{hb_pot_1d}).


\subsubsection{Quantum propagator}\label{prop_1d}\hfill\\

Let us introduce the quantum imaginary time propagator $G(x,y,t)$ \cite{fermions_review} defined as
\be\label{propagator_1d}
G({x},{y},t)=\langle {y}|e^{-\frac{t H}{\hbar}}|{x}\rangle=\sum_{k=1}^{\infty}\phi_k^*({ x})\phi_{k}({ y})e^{-\frac{t E_{ k}}{\hbar}}\;,
\ee
where the $\phi_k$'s are the (single-particle) eigenfunctions of $H$. This propagator is simply related to the correlation kernel via Laplace transformation~\cite{fermions_review} 
\be\label{kernel_propagator_1d}
K_{\mu}({x},{y})=\int_{\cal C} \frac{dt}{2i\pi t}e^{\frac{\mu t}{\hbar}}G({x},{y},t)\;,
\ee
where ${\cal C}$ indicates a Bromwich contour in the complex $t$-plane. For the hard-wall potential in Eq. (\ref{hb_pot_1d}) with $R=1$, this propagator satisfies the free diffusion equation
\be\label{free_diffusion}
-\hbar\partial_t G(x,y,t)=-\q \partial_y^2 G(x,y,t)\;\;{\rm with}\;\;G(x,y,0)=\delta(x-y),
\ee
together with the boundary condition at the wall, $G(x,y=\pm 1,t) = 0$, for all $x \in [-1,1]$ and $t$. From the linearity of Eq. \eqref{free_diffusion}, we can build a linear combination of solutions that satisfy the boundary conditions. This solution can be built from the free propagator 
\be\label{free_prop_1d}
G_1^{\rm free}(r,t)=\sqrt{\frac{m}{2\pi \hbar t}}\exp\left(-\frac{m r^2}{2\hbar t}\right)\;,
\ee
using the so-called {\it method of images}. The superscript 'free' stands for the solution of the free diffusion equation \eqref{free_diffusion} with free boundary conditions. It is convenient to rewrite this free propagator under the scaling form
\be\label{free_prop_1d_scaled}
G_1^{\rm free}(r,t)=k_F\tilde G_1(k_F r,\mu t/\hbar)\;\;{\rm with}\;\;\tilde G_{1}(s,\tau)=\frac{e^{-\frac{s^2}{4\tau}}}{\sqrt{4\pi\tau}}\;.
\ee 
with $\mu=\frac{\hbar^2\pi^2}{8m}N^2$ and $k_F = \sqrt{2 m \mu}/\hbar$. In terms of the free propagator, the full propagator reads~\cite{LLMS17}
\be\label{image_prop_1d}
G(x,y,t)=\sum_{n=-\infty}^{\infty}\left[G_1^{\rm free}\left(4n+x-y,t \right)-G_1^{\rm free}\left((4n+2)-x-y,t\right)\right]\;.
\ee
Substituting Eq. \eqref{image_prop_1d} in Eq. \eqref{kernel_propagator_1d}, we obtain
\be\label{K_sum}
K_{\mu}(x,y)=\int_{\cal C} \frac{dt}{2i\pi t}e^{\frac{\mu t}{\hbar}}\sum_{n=-\infty}^{\infty}\left[G_1^{\rm free}\left(4n+x-y,t \right)-G_1^{\rm free}\left((4n+2)-x-y,t\right)\right]\;.
\ee
The Bromwich integral over $t$ can be performed using the inversion formula
\be\label{inversion_prop}
\int_{\cal C}\frac{d\tau}{2i\pi \tau^{d/2+1}}\exp\left(z\tau-\frac{a}{\tau}\right)=\left(\frac{z}{a}\right)^{\frac{d}{4}}\J_{\frac{d}{2}}(2\sqrt{az})\;,
\ee
where $\J_\alpha(x)$ is the Bessel function of order $\alpha$. Indeed, one has
\be\label{free_scaling}
\int_{\cal C}\frac{dt}{2\pi i t}e^{\frac{\mu t}{\hbar}}G_1^{\rm free}(r,t)=k_F\int_{\cal C}\frac{d\tau}{2i\pi \tau}e^{\tau}\tilde G_1(k_F r,\tau)=k_F\frac{{\rm J}_{\frac{1}{2}}(k_F r)}{\sqrt{2\pi k_F r}}\;.
\ee
where we have used the scaling form in Eq. (\ref{free_prop_1d_scaled}) together with the formula (\ref{inversion_prop}) specialised to $d=1$, $z=1$ and $a=(k_F r/2)^2$. 
Finally using $\J_{1/2}(x)=\sqrt{2/(\pi x)}\sin(x)$, and integrating term by term in Eq. \eqref{K_sum} using Eq. \eqref{free_scaling}, this yields
\be\label{image_Ker_1d}
K_\mu(x,y)=k_F\sum_{n=-\infty}^{\infty}\left[K_1^{\rm b}\left(k_F(4n+x-y)\right)-K_1^{\rm b}\left(k_F(4n+2-x-y)\right)\right]\;,
\ee
where $K_1^{\rm b}(r)=\sin(r)/(\pi r)$ is the bulk scaling function of the kernel and $k_F = N \pi/2$. Note that one can check that this expression (\ref{image_Ker_1d}) coincides with the formula given in Eq. (\ref{khb_1d_N}). From this formula (\ref{image_Ker_1d}) one can see that in the bulk regime, where $k_F(1-x)\gg 1$ and $k_F(x-y)=O(1)$, only the term for $n=0$, and more specifically the first term in the square brackets -- which is a function of $x-y$ only -- in Eq. \eqref{image_Ker_1d} is dominant. 
One recovers the bulk scaling function of Eq. \eqref{khb_1d_b}. On the other hand, in the edge regime where $k_F(1-x)=O(1)$ and $k_F(1-y)=O(1)$, for $n=0$, both terms in the square brackets in Eq. \eqref{image_Ker_1d} contribute to the same order while all terms for $n\neq 0$ vanish. One then recovers the edge scaling function of Eq. \eqref{khb_1d_e}. 

This propagator method can be extended to an arbitrary hard-box potential $V(x)$, which is non-zero inside the box. In that case, the propagator $G(x,y,t)$ satisfies the imaginary time Schr\"odinger equation which reads 
\be\label{constant_pot}
-\hbar\partial_t G(x,y,t)=-\q \partial_y^2\, G(x,y,t) + V(y) G(x,y,t)\;\;{\rm with}\;\;G(x,y,0)=\delta(x-y)\;,
\ee
together with the boundary condition at the wall, $G(x,y=\pm 1,t) = 0$, for all $x$ and $t$. Of course, if the potential $V(x)$ is sufficiently confining
such that it imposes an edge at $x=r_{\rm e}$ (with $V(r_{\rm e}) = \mu$) with $r_{\rm e} - 1 \gg  w_N$ -- where $w_N$ is the associated scale of fluctuations at this edge [see Eq. (\ref{K_d_soft})] --  
then the density at the wall is essentially zero. In that case, {and if $V(x)$ is smooth,} the limiting kernel at the edge $r_{\rm e}$ is given by the Airy kernel (\ref{airy_k}).  
On the other hand, if $r_{\rm e} > 1$, { i.e. if $V(1) > \mu$ (more precisely $\frac{V(1)}{\mu}>1$),}
and if the potential is sufficiently smooth close to the wall at $x=1$, then the limiting kernel is the hard-wall
kernel in (\ref{khb_1d_e_intro}). In Ref. \cite{LLMS17}, it was actually shown that the precise condition for this to hold reads 
\be\label{condition_hb}
\tilde k(1)  = \frac{\sqrt{2m[\mu - V(1)]}}{\hbar}  \gg \frac{1}{w_N} \;,
\ee 
where $w_N$ is the width of the smooth edge regime {and $\tilde k(1)$ is the local Fermi wavector at the position of the wall, defined
in \eqref{K_LDA}.} If this condition (\ref{condition_hb}) is satisfied, then the limiting kernel takes the scaling form
\be\label{LDA_images}
K_\mu(x,y)\approx \tilde k(1) K_1^{\rm e}\left(\tilde k(1)(1-x),\tilde k(1)(1-y)\right)\;, 
\ee
where $K_1^{\rm e}(u,v)$ is the standard hard-wall kernel (\ref{khb_1d_e_intro}). In the next section, we study the situation where the condition (\ref{condition_hb}) is not satisfied.

\subsection{Non-uniform hard box potentials}\label{trunc_1d}

Let us consider a hard box potential between $[-1,1]$ and imagine that we switch on progressively a non-uniform linear potential $V(x)=\mu |x|/r_{\rm e}$ within the box, such that the potential reads
\be\label{trunc}
V_{\rm tr}(x)=\begin{cases}
      &\mu \dfrac{|x|}{r_{\rm e}}\;,\;\;|x|\leq 1\\
      &\\
      &+\infty\;,\;\;|x|>1\;,
     \end{cases}
\ee
where the subscript `tr' stands for `truncated' potential. The study of the linear potential (\ref{trunc}) is particularly interesting, first because it is exactly solvable and second because, as we will see later, the more general potentials can be mapped, in the large $\mu$ limit, onto this linear case. In the absence of a hard wall at $x = \pm 1$, the linear potential $V(x) = \mu \frac{|x|}{r_{\rm e}}$ would
create two edges at $x = \pm r_{\rm e}$, where the corresponding scale of fluctuations would be $w_N=k_F^{-2/3}r_{\rm e}^{1/3}$ [see Eq. (\ref{K_d_soft})], with $k_F = \sqrt{2 m \mu}/\hbar$. 
The situation of the previous sections corresponds to the case $r_{\rm e}\to\infty$ for which $V(x)=0$. Now let us increase progressively the potential by decreasing $r_{\rm e}$. We anticipate that the various regimes are controlled by the dimensionless parameter $\ell$ defined here as
\be
\ell=\frac{1-r_{\rm e}}{w_N}\;.\label{ell}
\ee

\begin{figure}[h]
 \centering
 \includegraphics[width=1.3\textwidth]{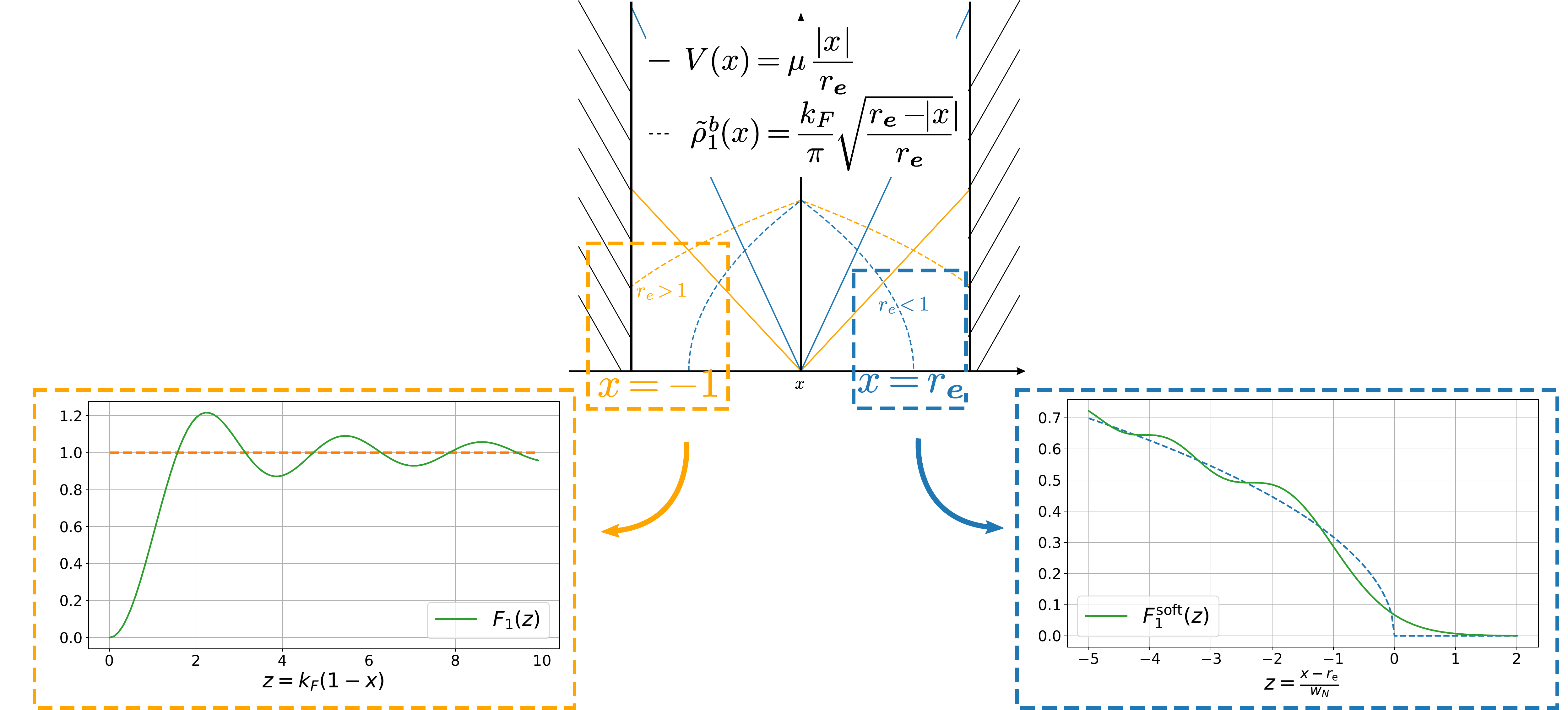}
 \caption{Sketch of two extreme situations for a linear potential of increasing slope in the bulk. In the first situation -- in orange -- the potential $V(x)$ is not strong enough to create an edge $r_e$ of the density $\tilde \rho_1^{\rm b}(x)=\tilde k(x)/\pi$ that lies in the box ($|x|\leq 1$). On a scale $s=k_F(1-x)=O(1)$ close to the hard edge, the density takes the hard edge scaling form, with the scaling function $F_1(s)$ given in Eq. \eqref{edge_dens_hb_1d}. In the second situation -- in blue -- the potential creates an edge $r_{\rm e}$ that lies in the box, but far enough from the walls for them to have an impact. On a scale $s=(r_{\rm e}-x)/w_N=O(1)$ -- where $w_N=k_F^{-2/3}r_{\rm e}^{1/3}$ -- the density takes the soft edge scaling form with the scaling function $F_{1}^{\rm soft}(s)$ given in Eq. \eqref{dens_oh_edge}. Note that the solid orange and blue lines in the top
 figure are the plots of $V(x)$ vs. $x$ and that their dotted counterparts are the corresponding bulk density bulk densities $\tilde \rho_1^b(x)$ vs. $x$.}\label{Fig_distance}
 
\end{figure}
There are indeed two extreme cases depending on the value of $\ell$:
\begin{itemize}
\item For $r_{\rm e}\gg 1$, i.e., {$\ell\approx - \frac{r_e}{w_N} =-(k_F r_{\rm e})^{2/3}<0$} with $|\ell| \gg 1$, the soft edge lies outside the box, it is therefore not encountered. Moreover, in that case {the local Fermi wavector 
defined in \eqref{K_LDA}, at the position of the wall, $\tilde k(1)=k_F \sqrt{1-\frac{1}{r_e}}$, satisfies $\tilde k(1)\,w_N = \sqrt{|\ell |} \gg 1$ and the condition in Ref. (\ref{condition_hb}) is obeyed}. Hence, close to the wall, the kernel takes the hard edge scaling form described in Eq. \eqref{LDA_images}. This situation is described graphically by the  orange case in Fig. \ref{Fig_distance}.

\item For $r_{\rm e}\ll 1$, i.e., $\ell\approx 1/w_N =   k_F^{2/3} r_{\rm e}^{-1/3}>0$ with $\ell\gg1 $, the soft edge lies inside the bulk and the density drops to zero for $|x|\geq r_{\rm e}$. Because the density is already negligible at the position of the wall, the presence of the wall has no impact on the system. Close to the soft edge, the kernel takes the soft edge scaling form described in Eq. \eqref{K_d_soft}. This situation is described graphically by the blue case in Fig. \ref{Fig_distance}. 
\end{itemize}
In the intermediate region where $r_{\rm e}\sim 1$, i.e. $\ell=O(1)$, we will need to consider both the presence of the wall and the spatial variation of the potential. We now turn to the analysis of this intermediate regime for the linear truncated potential described in Eq. \eqref{trunc}.

\subsubsection{Analytical solution for the linear potential}\hfill \\

In the case of this linear potential~\eqref{trunc}, the quantum propagator in Eq. \eqref{propagator_1d} satisfies the equation
\be\label{prop_lin}
-\hbar\partial_t G(x,y,t)=- {\frac{\hbar^2}{2 m}}
\partial_y^2 G(x,y,t)+\mu\frac{|y|}{r_{\rm e}} G(x,y,t)\;,\;\;{\rm with}\;\; G(x,y,0)=\delta(x-y)\;,
\ee
and with vanishing boundary conditions for $x,y=\pm 1$.
In the interesting case where $r_e=O(1)$ and $|r_{\rm e}-1|\ll 1$, we investigate the behavior of the propagator close to the hard edge by introducing the rescaled propagator
\be\label{re_lin_prop}
G(x,y,t)=\frac{1}{w_N}G_0\left(\frac{x-1}{w_N},\frac{y-1}{w_N},t\right)\;,
\ee
where we recall that $w_N=\hbar^{2/3}/(2m V'(r_{\rm e}))^{1/3}=k_F^{-2/3}r_{\rm e}^{1/3}$ in our case. The potential at the hard edge reads $V(1)=\frac{\mu}{r_e} = \mu+ \hbar^{2}\ell/(2m w_N^2)$. From Eq. \eqref{prop_lin}, the rescaled propagator in Eq. \eqref{re_lin_prop} is a solution of the equation
\be\label{eq_re_lin_prop}
\left(\frac{2m w_N^2}{\hbar^2}(\hbar\partial_t+\mu)+\ell\right)G_0(u,v,t)=(\partial_v^2-v)G_0(u,v,t)\;,\;\;{\rm with}\;\; G_0(u,v,0)=\delta(u-v)\;,
\ee
for $u,v \leq 0$ and with vanishing boundary conditions for $u,v=0$. It is convenient to write the propagator as
\be\label{G_1}
G_0(u,v,t)=G_1(u,v,t)\exp\left(-\frac{\mu t}{\hbar}-\frac{\hbar\ell t}{2m w_N^2}\right)\;,
\ee 
where the propagator $G_1(u,v,t)$ satisfies the equation
\be\label{prop_FF}
\frac{2m w_N^2}{\hbar}\partial_t G_1(u,v,t)=(\partial_v^2-v)G_1(u,v,t)\;,\;\;{\rm with}\;\; G_1(u,v,0)=\delta(u-v)\;,
\ee
and with vanishing boundary conditions at $u,v=0$. This equation (\ref{prop_FF}) appears in the study of the first passage time of the Brownian motion to a parabolic barrier \cite{Airywall}. It also appears in the study of the persistence properties of the Airy$_2$ process \cite{FerFri}. Its solution is given by~\cite{FerFri,Airywall}
\begin{align}
&G_1(u,v,t)=\int_{-\infty}^{\infty}\sigma(s,s+u)\sigma(s,s+v)\exp\left(\frac{\hbar s t}{2m w_N^2}\right)ds\;,\;\;u,v\leq0\;,\label{prop_sol_FF}\\
&{\rm with}\;\;\sigma(r,s)=\frac{\Ai(r)\Bi(s)-\Bi(r)\Ai(s)}{\sqrt{\Ai^2(r)+\Bi^2(r)}}\;.\label{sigma}
\end{align}
where ${\rm Ai}(x)$ and ${\rm Bi}(x)$ are the two linearly independent solutions of the Airy differential equation $y''-x y = 0$. Combining Eqs. (\ref{prop_sol_FF}) and (\ref{G_1}), we obtain that the propagator takes the scaling form at the edge
\begin{align}
G(x,y,t)&=\frac{1}{w_N}G_0\left(\frac{x-1}{w_N},\frac{y-1}{w_N},t\right)\;\;{\rm with}\nn\\
G_0(u,v,t)&=e^{-\frac{\mu t}{\hbar}}\int_{-\infty}^{\infty}\sigma(s,s+u)\sigma(s,s+v)e^{\frac{\hbar(s-\ell)t}{2m w_N^2}}ds\;.\label{final_scaling_prop}
\end{align}
Inserting Eq. \eqref{final_scaling_prop} in Eq. \eqref{kernel_propagator_1d}, we find that the kernel takes the exact form at the edge (which is close to the wall)
\begin{align}
&K_{\mu}(x,y)=\frac{1}{w_N}K_1^{\ell}\left(\frac{x-1}{w_N},\frac{y-1}{w_N}\right)\;,\;\;w_N=\left(\frac{\hbar^2}{2m V'(r_{\rm e})}\right)^{\frac{1}{3}}\;,\label{scaling_K_1_l}\\
&{\rm with}\;\;K_1^{\ell}(u,v)=\int_{-\infty}^{\infty}ds\int_{\cal C}\frac{dt}{2i\pi t}\exp\left(\frac{\hbar(s-\ell)t}{2m w_N^2}\right)\sigma(s,s+u)\sigma(s,s+v)\label{K_1_l_1}\;,\;\;u,v\leq0\;.
\end{align}
We note that the integral over $t$ in Eq.~\eqref{K_1_l_1}  can be explicitly performed, using 
\be\label{Heaviside}
\int_{\cal C}\frac{dt}{2i\pi t}e^{xt}=\Theta(x)\;,
\ee
with $\Theta(x)$ the Heaviside step function. This ultimately leads to
\begin{align}\label{K_1_l_final}
K_1^{\ell}(u,v)=\int_{\ell}^{\infty}\sigma(s,s+u)\sigma(s,s+v)ds\;,\;\;u,v\leq0\;,\;\;\ell=\frac{1-r_{\rm e}}{w_N}\;,
\end{align}
where we recall that $\sigma(r,s)$ is given in Eq. \eqref{sigma}.
One can check that this kernel gives the expected asymptotic limits (see Appendix \ref{limits} for details):
\begin{itemize}
\item For $\ell\to+\infty$, rescaling the position close to the soft edge (in $u=-\ell$), we obtain the soft edge scaling function of Eq. \eqref{K_d_soft} 
\be
K_1^{\ell}(u-\ell,v-\ell) \to K_{1}^{\rm soft}(u,v)\;,\;\;\ell\to +\infty\label{l+}
\ee
\item For $\ell\to -\infty$, rescaling further at the hard edge with $\sqrt{|\ell|}$, we obtain for $\ell \to -\infty$ the hard edge scaling function of Eq. \eqref{LDA_images}
\be
\frac{1}{\sqrt{|\ell|}}K_{1}^{\ell}\left(\frac{u}{\sqrt{|\ell|}},\frac{v}{\sqrt{|\ell|}}\right)\to K_1^{\rm e}(u,v)\;,\;\;\ell\to -\infty\;\label{l-}.
\ee
In this case, the total rescaling parameter $\sqrt{-\ell}/w_N=k_F\sqrt{1-1/r_{\rm e}}$ is precisely identical to $\tilde k(1)$, the rescaling parameter in Eq. \eqref{LDA_images}.
\end{itemize}

\begin{figure}[h]
 \centering
 \includegraphics[width=0.8\textwidth]{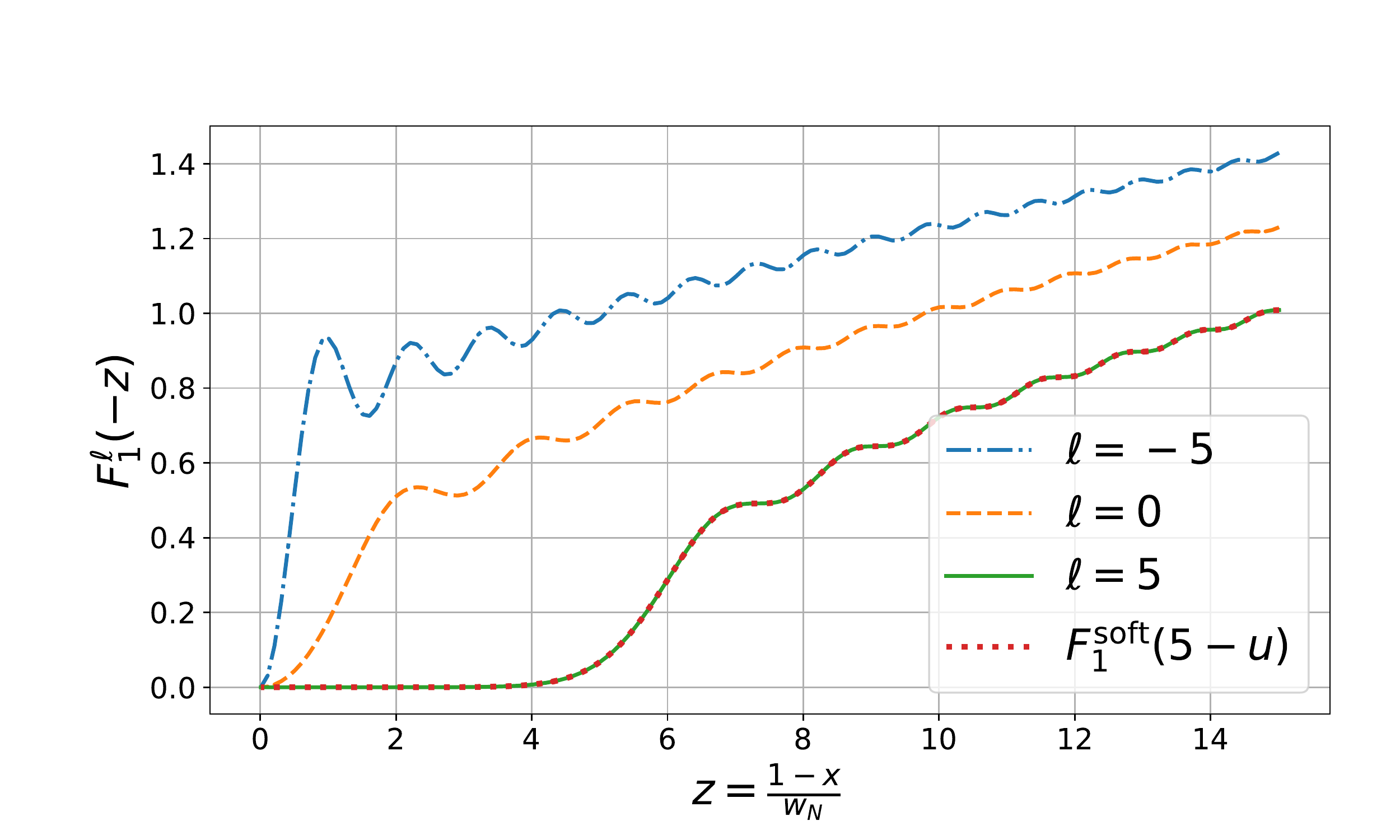}
 \caption{Plot of the density $F_1^{\ell}(-z)$ in Eq. \eqref{rho_1_l} as a function of the rescaled distance with the wall $z=(1-x)/w_N\geq 0$ for $\ell=-5,0,5$ respectively in dashed-dotted blue, dashed orange and solid green lines. In this plot, the wall lies at $z=0$ with the bulk to its right while the soft edge lies at $z=\ell$. For $\ell=5$, the density already coincides with the formula obtained taking the Airy kernel (dotted red line) corresponding to the limit $\ell\to\infty$ in Eq. \eqref{l+}}\label{Fig_rho_ell}
 
\end{figure}

The scaling function for the density at the edge of the Fermi gas is obtained by evaluating the kernel in (\ref{K_1_l_final}) at coinciding points $u=v$. This yields
\be\label{rho_1_l}
\tilde \rho(x)\approx\frac{1}{w_N}F_1^{\ell}\left(\frac{x-1}{w_N}\right)\;,\;\;{\rm with}\;\;F_1^{\ell}(z)=\int_{\ell}^{\infty}\sigma(s,s+z)^2 ds\;.
\ee
This density scaling function is plotted in Fig. \ref{Fig_rho_ell} for different values of $\ell=-5,0,5$. For positive values of $\ell$, it converges very rapidly to the asymptotic limit for $\ell\to+\infty$ given by $F_{1}^{\rm soft}(s=z-\ell)=\Ai'(s)^2-s\Ai(s)^2$, while the convergence for $\ell<0$ is much slower.


%
%
%

As shown below, this analysis can actually be generalized straightforwardly to a more general class of non-uniform (smooth) potentials $V(x)$ inside the box $[-1,1]$.

\subsubsection{General case of a truncated potential}\hfill \\

Let us consider the general case of a truncated potential of the form
\be\label{gen_trunc_pot}
V_{\rm tr}(x)=\begin{cases}
      &V(x)\;,\;\;|x|\leq 1\\
      &\\
      &\infty\;,\;\;|x|>1\;,
     \end{cases}
\ee
where $V(x)$ is a smooth potential, e.g. $V(x) \propto |x|^p$ with $p>0$, that would create an edge at position $x=r_{\rm e}$ in the absence of the wall.
We have seen in the previous section that the extreme cases $r_{\rm e}\gg 1$, described by Eq. \eqref{LDA_images} and $r_{\rm e}\ll 1$, described by Eq. \eqref{K_d_soft} are universal with respect to  the confining potential $V(x)$.
In this general case, the kernel $K_{\mu}(x,y)$ can once again be obtained through the imaginary time propagator $G(x,y,t)$ which satisfies 
\be
-\hbar\partial_t G(x,y,t)=-\q \partial_y^2 G(x,y,t)+V(y)G(x,y,t)\;\;{\rm with}\;\; G(x,y,0)=\delta(x-y)\;,\label{gen_diff}
\ee
and with vanishing boundary conditions at $x,y=\pm 1$. Close to the soft edge on a scale $|x-r_{\rm e}|\sim w_N$, we expect $w_N$ to be small and  
we thus expand the potential in Taylor series close to $x=r_{\rm e}$ in~(\ref{gen_diff}). This yields
\be\label{lin}
V(r_{\rm e}+w_N z)=\mu+\frac{\hbar^2}{2m w_N^2}\left[z+O\left(\frac{w_N}{r_{\rm e}}\right)\right]\;.
\ee
It appears linear on this scale and the situation is then similar to the previous section where we considered a linear potential from the start.
Introducing the rescaled propagator $G_0(u,v,t)$ close to the hard wall as in Eq. \eqref{re_lin_prop}, it will be solution (at leading order) of Eq. \eqref{eq_re_lin_prop} as $V(y=1+w_N v)\approx\mu+\hbar^2(\ell+v)/(2m w_N^2)$. The solution for the propagator will be identical to the case of the linear potential and therefore, the kernel will have the same scaling function at the edge described in Eq. \eqref{K_1_l_final} with parameters given by
\be\label{rescaling_factors}
V(r_{\rm e})=\mu\;\;,\;\;w_N=\left(\frac{\hbar^2}{2mV'(r_{\rm e})}\right)^{\frac{1}{3}}\;\;{\rm and}\;\;\ell=\frac{1-r_{\rm e}}{w_N}\;.
\ee
Note that the same result can be obtained by using the general short time expansion developed in Section VII of Ref. \cite{fermions_review}. 

To summarize this section, we have found a new kernel $K_1^{\ell}(u,v)$  \eqref{K_1_l_final}, parametrised by $\ell=(1-r_{\rm e})/w_N$, which interpolates between the Airy kernel for $\ell\to+\infty$ and the hard box kernel in the limit $\ell\to -\infty$. Furthermore, we have shown that this potential is universal with respect to a large class of potentials $V(x)$ that are smooth at the box edge.

\section{Higher dimensional spherical hard box at zero temperature}\label{hb_d_T0}

So far, we only considered the case of dimension one. 
Even though in higher dimensions $d>1$ the connection with RMT is lost, we can still use the determinantal properties of the process to obtain all correlation functions (\ref{correlation}). In higher dimensions, the ground state energy $E_0=\sum_{k=1}^N \epsilon_k$ can be degenerate. In such cases, the situation is a priori more complicated to handle. However, in the large $N$ limit, the effects of the degeneracy are sub-leading~\cite{fermions_review} and we can therefore restrict our study to the case where the ground state is not degenerate. Let us first focus on the case of the spherical hard box potential with zero potential inside, before treating more complicated cases, where the potential within the box is non-uniform.

\subsection{Uniform hard box potential}\label{shb_t0}

\begin{figure}[h]
 \centering
 \includegraphics[width=0.5\textwidth]{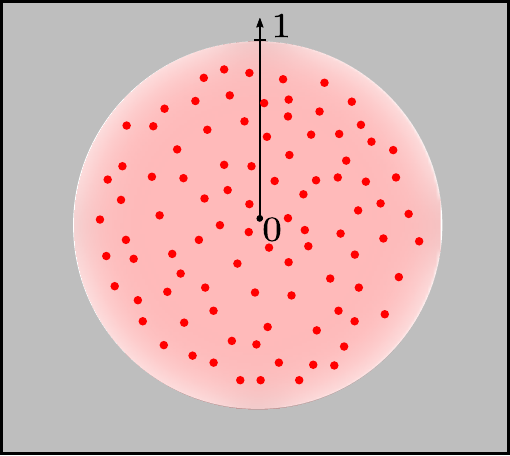}
 \caption{Sketch of a hard box in dimension $d=2$. The potential is $V({\bf x})=0$ inside the disk $|{\bf x}|<1$ and $V({\bf x})=+\infty$ outside (in gray).}\label{box2d}
\end{figure}

We consider now the generalization of \eqref{hb_pot_1d} to higher dimensions $d>1$ (see Fig. \ref{box2d})
\be\label{hb_pot_d}
V_R({\bf x})=\begin{cases}
        & 0\;,\;\;|{\bf x}|\leq R\\
				&\\
        & \infty\;,\;\;|{\bf x}|>R\;.
       \end{cases}
\ee

As in the one-dimensional case, the only length scale of the problem is $R$ such that we can rescale all lengths and set the wall at $R=1$.
We will first obtain the kernel for any finite number $N$ of particles and then analyze the large $N$ asymptotic behavior. We will then consider the generalisation to more complicated non-uniform potentials.

\subsubsection{Finite $N$ results}\hfill\\

Using the spherical symmetry of the potential in Eq.~\eqref{hb_pot_d}, it is natural to introduce the spherical coordinates ${\bf x}=(r,{\bm \theta})$ where ${\bm \theta}$ is a $d-1$ dimensional vector encoding the angular variables. We then rewrite the single-particle Hamiltonian in these coordinates 
\be\label{hamiltonian_radial}
H=-\q\Delta_{\bf x}+V_R({\bf x})=-\q r^{\frac{1-d}{2}}\partial_r^2 r^{\frac{d-1}{2}}+\frac{\hbar^2}{8m r^2}(d-1)(d-3)+\frac{\hat{\bf  L}^2}{2m r^2}+V_R(r)\;,
\ee
where $\hat{\bf L} \equiv \hat{\bf L}({\bm \theta})$ is the angular momentum operator. In terms of the spherical coordinates, the single-particle eigenfunction reads
\be\label{sg_decomp_d}
\phi_{\bf k}({\bf x})=r^{\frac{1-d}{2}}\chi_{n,l}(r)Y_{l,{\bm m}}({\bm \theta})\;,
\ee
where the quantum numbers are also decomposed as ${\bf k}=(n,l,{\bm m})$, where ${\bm m}$ is a $(d-2)$ dimensional vector. In Eq. (\ref{sg_decomp_d}), the functions $Y_{l,{\bm m}}({\bm \theta})$ denote the eigenfunctions of~$\hat{\bf  L}^2$
\be\label{spherical_harmonics}
\hat{\bf  L}^2 Y_{l,{\bm m}}({\bm \theta})=\hbar^2 l(l+d-2)Y_{l,{\bm m}}({\bm \theta})\;,\;\;{\rm with}\;\;l\in \mathbb{N}\;.
\ee
These functions are the $d$-dimensional spherical harmonics. For each value of $l$ (the orbital quantum number), the corresponding degeneracy is \cite{harmonic_osc} 
\be\label{degeneracy}
g_d(l)=\frac{(2l+d-2)(l+d-3)!}{l!(d-2)!}\;.
\ee
It corresponds to all possible distinct values of ${\bm m}$ (the quantum number ${\bm m}$ being a $d-2$ dimensional $\mathbb{Z}$ vector), that is ${\bm m}=\pm 1$ in $d=2$ (thus $g_2(0)=1$ and $g_2(l)=2$ for $l \geq 1$) and ${\bm m}=m\in \mathbb{Z}$ between $-l$ and $+l$ in $d=3$. 
Inserting the exact form of the single particle eigenfunctions in Eq.~\eqref{sg_decomp_d} in the Hamiltonian in Eq. \eqref{hamiltonian_radial}, one obtains an effective Schr\"odinger equation for $\chi_{n,l}(r)$. For $r>0$, this yields
\be\label{effective_hamiltonian}
H_{\rm eff}\chi_{n,l}(r)=\epsilon_{n,l}\chi_{n,l}(r)\;\;{\rm with}\;\;H_{\rm eff}=-\q\partial_r^2+V_{\rm eff}^l(r)\;,
\ee
where the effective potential $V_{\rm eff}^l(r)$ depends on the orbital quantum number $l$ and reads
\be\label{effective_potential}
V_{\rm eff}^l(r)=\begin{cases}
                  &\displaystyle\frac{\hbar^2}{2mr^2}\left(l+\frac{d-3}{2}\right)\left(l+\frac{d-1}{2}\right)\;,\;\;r<1\\
                  &\\
                  &\displaystyle\infty\;,\;\;r>1\;.
                 \end{cases}
\ee
The radial part of the eigenfunctions $\chi_{n,l}(r)$ reads for $r\geq 0$
\be\label{radial_wf}
\chi_{n,l}(r)=\begin{cases}
                  &\displaystyle\frac{\sqrt{2r}}{\J_{\nu-1}(k_{n,l})}\J_{\nu}(k_{n,l}\,r)\;,\;\;r \leq 1\\
                  &\\
                  &\displaystyle 0\;,\;\;r>1\;,
                 \end{cases}\;\;{\rm where}\;\;\nu=l+\frac{d-2}{2}\;.
\ee
The single-particle energies $\epsilon_{n,l}$ are fixed by the boundary condition $\chi_{n,l}(r=1)=0$,
\be\label{eigenE}
\epsilon_{n,l}=\q k_{n,l}^2\;,\;\;{\rm where}\;\;k_{n,l}=j_{\nu,n}\;,\;\;{\rm and}\;\;\J_\nu(j_{\nu,n})=0\;,
\ee
that is $j_{\nu,n}$ is the $n^{\rm th}$ real zero of the Bessel function of order $\nu$ \cite{zero_Bessel}.

To compute the correlation kernel $K_{\mu}({\bf x},{\bf y})$, we use the general formula in Eq.~\eqref{Kernel} 
where the single particle eigenfunctions $\phi_{\bf k}({\bf x})$ in spherical coordinates are given in (\ref{sg_decomp_d}). Parameterizing ${\bf x} = (r,{\bm \theta})$ and ${\bf y} = (r',{\bm \theta'})$, this
yields
\be\label{kernel_spherical}
K_{\mu}({\bf x},{\bf y})=(r\,r')^{\frac{1-d}{2}}\sum_l\left(\sum_{\bm m} Y_{l,{\bm m}}^*({\bm \theta})Y_{l,{\bm m}}({\bm \theta'})\right)K^l_{\rm eff}(r,r')\;,
\ee
where $K^l_{\rm eff}(r,r')$ is an effective one dimensional kernel given by
\be\label{effective_kernel}
K^l_{\rm eff}(r,r')=\sum_n \chi_{n,l}(r)\chi_{n,l}(r')\Theta(k_F-k_{n,l}) \;,\;\;{\rm with}\;\;k_F=\frac{\sqrt{2m\mu}}{\hbar}\;.
\ee
The sum over ${\bm m}$ in Eq. \eqref{kernel_spherical} can be worked out exactly \cite{harmonics}
\be\label{sum_m}
\sum_{\bm m} Y_{l,{\bm m}}^*({\bm \theta})Y_{l,{\bm m}}({\bm \theta'})=\frac{g_d(l)}{S_d}{\sf P}_{l,d}(\eta)\;,{\rm where}\;\;\eta=\frac{{\bf x}\cdot{\bf y}}{|{\bf x}||{\bf y}|}\;,
\ee
where $S_d=2\pi^{d/2}/\Gamma(d/2)$ is the surface of the $d$-dimensional sphere and $g_d(l)$ is the degeneracy given in Eq. \eqref{degeneracy}. In Eq. (\ref{sum_m}), ${\sf P}_{l,d}(\eta)$ is the solution of the differential equation
\be
(1-\eta^2){\sf P}_{l,d}''(\eta)+(1-d)\eta\,{\sf P}_{l,d}'(\eta)+l(l+d-2){\sf P}_{l,d}(\eta)=0 \;, \label{eq:Legendre} 
\ee
with the conditions ${\sf P}_{l,d}(-\eta)=(-1)^l {\sf P}_{l,d}(\eta)$ and ${\sf P}_{l,d}(1)=1$. Note that this function ${\sf P}_{l,d}(\eta)$ can be expressed in terms of the Gegenbauer polynomials $C_l^{m}(\eta)$ \cite{Grad} as
\be\label{Gegenbauer}
g_d(l){\sf P}_{l,d}(\eta)=\frac{2l+d-2}{d-2}C_l^{\frac{d-2}{2}}(\eta)\;.
\ee
Note additionally that for $d=2$, we have ${\bm m}=\pm 1$ and $Y_{l,m}(\theta)=e^{\pm il\theta}/\sqrt{2\pi}$ such that
\be\label{spher_harm_2d}
\sum_{\bm m} Y_{l,{ m}}^*({ \theta})Y_{l,{ m}}({\theta'})=\frac{1}{\pi}\cos\left(l(\theta-\theta')\right)=\frac{1}{\pi}T_l(\cos(\theta-\theta'))\;,
\ee
with $T_l(x)$ the Chebychev polynomial of first kind of degree $l$. This Eq. \eqref{spher_harm_2d} is in full agreement with Eqs. \eqref{sum_m} and \eqref{Gegenbauer} as $S_2=2\pi$ and from Ref. \cite{Gegen}
\be\label{GegentoCheby}
\frac{l C_l^m(x)}{2m}\to T_l(x)\;,\;\;m\to 0\;.
\ee

We have thus obtained an exact formula for the kernel $K_{\mu}({\bf x},{\bf y})$ for finite value of $N$ (equivalently finite $\mu$). Indeed, inserting Eqs. \eqref{effective_kernel} and \eqref{sum_m} into Eq. \eqref{kernel_spherical}, we obtain (setting ${\bf x} = (r,{\bm \theta})$ and ${\bf y} = (r',{\bm \theta'})$)
\be
K_{\mu}({\bf x},{\bf y})=\frac{2(r r')^\frac{2-d}{2}}{S_d}\sum_{l=0}^\infty\sum_{n=0}^\infty g_d(l){\sf P}_{l,d}(\eta) \frac{\J_\nu(j_{\nu,n}r)\J_\nu(j_{\nu,n}r')}{\J_{\nu-1}(j_{\nu,n})^2}\Theta(k_F-j_{\nu,n})\;,\label{kernel_exact_d_t0}
\ee
where $\nu = l + (d-2)/2$ and $k_F = \sqrt{2 m \mu}/\hbar$ while $g_d(l)$ and ${\sf P}_{l,d}(\eta)$ are given respectively in Eqs. (\ref{degeneracy}) and (\ref{Gegenbauer}). From Eq. \eqref{kernel_exact_d_t0}, we obtain that we only need to consider values of $j_{\nu,n}\leq k_F$ to compute the kernel at zero temperature.
\begin{figure}[h]
 \centering
\includegraphics[width=0.8\textwidth]{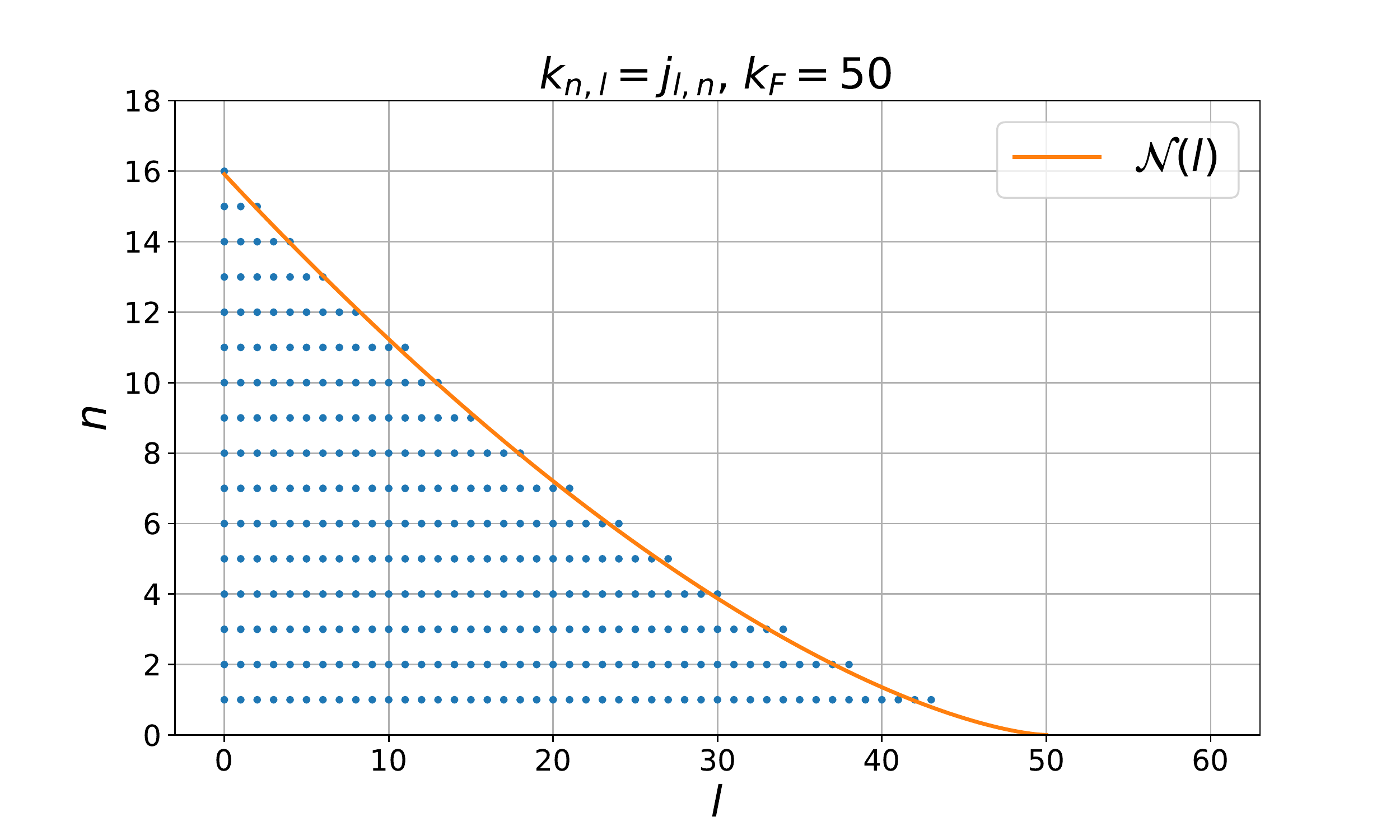}
\caption{Values of $n$ such that the $n^{\rm th}$ zero $j_{l,n}$ of the Bessel function $J_{l}(j_{l,n})=0$ verifies $j_{l,n}\leq k_F=50$ as a function of $l$ (blue dots). The orange curve represents $k_F{\cal N}(l/k_F)$, that is the maximum value $m_l$ of $n$ such that $j_{l,n}\leq k_F$ in the limit of large $k_F$. Note also that there is a maximum value of $l$, $l^*=44$ such that $j_{l^*,1}\leq k_F$.}\label{Fig_filling}
 \end{figure}
Furthermore, $j_{\nu,n}$ is an increasing function of both $\nu$ and $n$, it imposes for instance  that at fixed $l$, there must be a maximum value $m_l$ of $n$ such that
\be\label{m_l}
k_{m_l,l}=j_{\nu,m_l}\leq k_F\;\;{\rm and}\;\;k_{m_l+1,l}=j_{\nu,m_l+1}> k_F\;,
\ee
On the other hand, there must also be a maximal value $l^*$ of $l$ such that 
\be\label{l_star}
j_{l^*+\frac{d-2}{2},m_l}\leq k_F\;\;{\rm with}\;\;m_{l^*}\geq 1\;\;{\rm and}\;\;j_{l^*+\frac{d}{2},1}>k_F\;.
\ee
An example of filling of the different energy levels is given in Fig. \ref{Fig_filling}. Let us specify the formula for the kernel (\ref{kernel_exact_d_t0}) in $d=2$, which reads more explicitly, with ${\bf x} = (r,\theta)$ and ${\bf y} = (r',\theta')$
\be
K_{\mu}({\bf x},{\bf y})=\frac{2}{\pi}\sum_{l=0}^\infty\sum_{n=0}^\infty \cos{l(\theta-\theta')} \frac{\J_l(j_{l,n}r)\J_l(j_{l,n}r')}{\J_{l-1}(j_{l,n})^2}\Theta(k_F-j_{l,n})\;.\label{kernel_exact_2d}
\ee

From the exact formula for the kernel (\ref{kernel_exact_d_t0}) at coinciding point, 
we obtain the exact density as
\be\label{dens_d_N}
\tilde \rho({\bf x})=\frac{2r^{2-d}}{S_d}\sum_{l=0}^{\infty}g_d(l)\sum_{n=0}^{\infty}\left(\frac{\J_{l+\frac{d-2}{2}}(k_{n,l}r)}{\J_{l+\frac{d}{2}}(k_{n,l}r)}\right)^2\Theta(k_F-k_{n,l})\;,
\ee
with $r = |{\bf x}|$. 
\begin{figure}[t]
 \centering
 \includegraphics[width=0.8\textwidth]{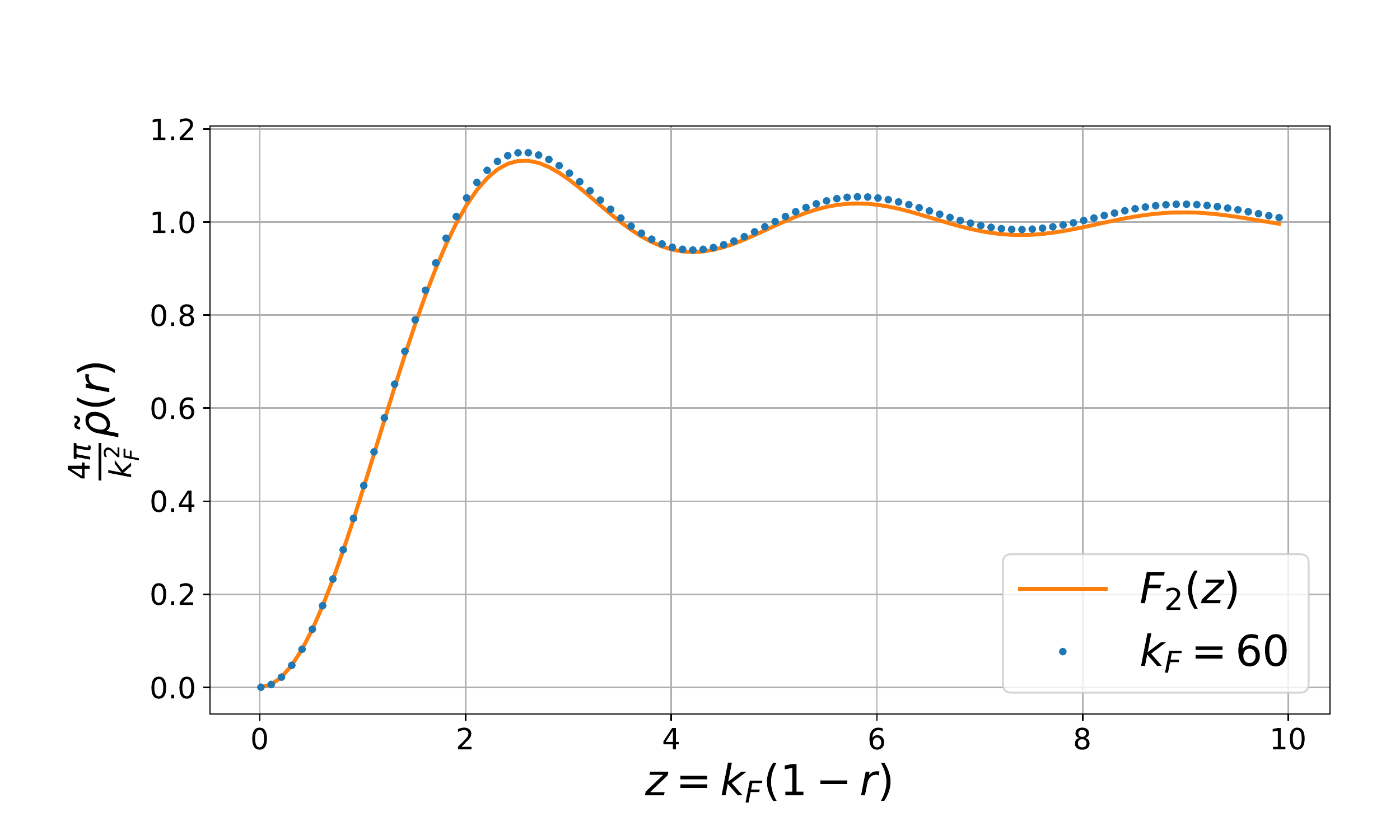}
 \caption{Plot of the rescaled density for a two dimensional hard box $(4\pi)/k_F^2\tilde \rho(r)$ given at finite $N$ in Eq. \eqref{dens_d_N} as a function of $z=k_F(1-r)$ for $k_F=60$ (blue dots). This density shows a good agreement with the scaling function $F_2(z)$ (orange line) given in Eq. \eqref{dens_d}}
 \label{Fig F2}
\end{figure}
where we have used that $\eta = 1$ for ${\bf x} = {\bf y}$ [see Eq. (\ref{sum_m})] as well as $P_{l,d}(1) = 1$. Note that, in this formula, the sums are actually bounded $0\leq l\leq l^*$ and $1\leq n\leq m_l$ such that $k_{n,l}\leq k_F$ (see Fig. \ref{Fig_filling}). 
A plot of this density is shown in Fig. \ref{Fig F2}. Extracting the large $N$ limit of this formula is quite hard (see below). But far from the edge, the density is given by the bulk density formula 
in Eq. \eqref{rho_LDA}. Inserting $V(|x|)=0$ in Eq. \eqref{rho_LDA}, we obtain
\be\label{LDA_d}
\tilde \rho_{d}^{\rm b}({\rm x})=\Omega_d\left(\frac{k_F}{2\pi}\right)^d\Theta(1-|{\rm x}|)\;.
\ee
We can now use the normalisation condition to obtain the expression of $k_F$ as a function of $N$ in the large $N$ limit,
\be
\int d^d {\rm x}\, \tilde \rho_d^{\rm b}({\rm x})=N\;\;\Rightarrow\;\;k_F=2\pi\left(\frac{N}{\Omega_d^2}\right)^{\frac{1}{d}}=2\Gamma\left(\frac{d}{2}+1\right)^\frac{2}{d}N^{\frac{1}{d}}\;,\label{k_F_N}
\ee
where we used $\Omega_d=\pi^{d/2}/\Gamma(d/2+1)$. Therefore, the limit $N\gg1$ is a limit where $k_F\sim N^{1/d}\gg 1$ as expected. Note that taking $d=1$ in Eq. \eqref{k_F_N} using $\Gamma(3/2)^2=\pi/4$, we recover the exact relation $k_F=N\pi/2$ for the one-dimensional case in Eq. \eqref{k_F_hb}. We will now analyze the behavior of the kernel $K_{\mu}({\bf x},{\bf y})$ in the limit of large $k_F$ and particularly the behavior at the edge of the Fermi gas.

\subsubsection{Kernel for large $N$}\hfill\\

In this section, we perform the large $N$ asymptotics of the kernel given Eq. (\ref{kernel_spherical}) and obtain the scaling form both at the edge (at $r=1$) and
in the bulk. We start from the exact formula in Eq. (\ref{kernel_spherical}) and evaluate explicitly first the radial part $K_l^{\rm eff}(r,r')$ and then the angular part, i.e. $\sum_{\bm m} Y_{l,{\bm m}}^*({\bm \theta})Y_{l,{\bm m}}({\bm \theta'})$. The result for the radial part is given in Eq. \eqref{edge_effk_hb_d} and the result for the angular is given in Eqs. (\ref{angular_part}) and (\ref{angular_part_3}). These two factors are put together in Eq. (\ref{kernel_ab3}) leading to the final results in \eqref{k_d_final} and \eqref{kernel_edge_hb_d_b_2}.

\noindent{\it Radial part}.
We anticipate that the behavior at the edge of the Fermi gas is dominated by the eigenfunctions of high energy $\epsilon_{n,l}=\q k_{n,l}^2\sim \mu$. In this regime, we expect both $n$ and $l$ to be large. The radial wave function $\chi_{n,l}(r)$ can be evaluated in this limit using the asymptotic form of the Bessel functions \cite{as_Bessel}
\begin{align}\label{bessel_large}
&\J_{\nu}(k_{n,l}r)\approx\left(\frac{2k_{n,l}}{\pi l \sqrt{(k_{n,l}r)^2-l^2}}\right)^{\frac{1}{2}}\cos\left(k_{n,l}\xi\left(r,\frac{l}{k_{n,l}}\right)-\frac{\pi}{4}\right)\;,\\
&{\rm with}\quad \xi(r,\tilde l)=\sqrt{r^2-\tilde l^2}-\tilde l\arccos\left(\frac{\tilde l}{r}\right) \;,\label{xi_def}
\end{align}
where we used that for large $l$, $\nu=l+(d-2)/2\approx l$. As $k_{n,l}=j_{\nu,n}$ is an increasing function of both $n$ and $l$, we expect that in this regime $k_{n,l}\gg 1$. This expression in Eq. \eqref{xi_def} allows to obtain the asymptotic form of these wave-vectors $k_{n,l}$ for large $n\gg 1$ and $l\gg 1$ as
\be\label{asymp_k_nl}
k_{n,l}\xi\left(1,\frac{l}{k_{n,l}}\right)=\sqrt{k_{n,l}^2-l^2}-l\arccos\left(\frac{l}{k_{n,l}}\right)= n\pi+\frac{3\pi}{4}\approx n\pi\;.
\ee
We see that this relation is valid only for $0\leq l/k_{n,l}\leq1$ and gives a value of $n$ such that $0\leq n/k_{n,l}\leq 1$.
Moreover, since the maximal value of these wave-vectors is $k_F$, it is natural to rescale $n=k_F \, \tilde{n}$ and similarly $l=k_F \, \tilde{l}$ with $\tilde{n},\tilde{l}\leq 1$. In these notations, $k_{n,l}$ takes a scaling form 
\be\label{asymp_K}
k_{n,l}=k_F{\cal K}\left(\frac{n}{k_F},\frac{l}{k_F}\right)\;\;{\rm where}\;\;\sqrt{{\cal K}^2-\tilde{l}^2}-\tilde{l}\arccos\left(\frac{\tilde l}{\cal K}\right)=\tilde{n}\pi\;.
\ee
From this relation, we see that for $k_{n,l}=k_F$, ${\cal K}(\tilde{n},\tilde{l})=1$. From Eq. \eqref{m_l}, this situation corresponds to the case where $n=m_l$ is maximum at large but fixed value of $l$ such that
\be\label{N_l}
m_l\approx k_F {\cal N}\left(\frac{l}{k_F}\right)\;,\;\;{\rm with}\;\;{\cal N}(\tilde l)=\frac{\sqrt{1-\tilde{l}^2}-\tilde{l}\arccos{\tilde{l}}}{\pi}\;. 
\ee
Furthermore, as ${\cal N}(1)=0$, we get that $l^*\to k_F$ in the limit $N\gg 1$.
Going back to Eq. \eqref{bessel_large} to investigate the behavior of the radial wave-function close to the edge, we expand Eq. \eqref{xi_def} for large $k_F$, keeping 
$k_F(1-r)=O(1)$. This yields
\be\label{dev_xi_edge}
k_F{\cal K}\xi\left(r,\frac{\tilde l}{\cal K}\right)-\frac{\pi}{4}=n\pi+\sqrt{{\cal K}^2-\tilde{l}^2}k_F(r-1)+O(k_F)^{-1}\;.
\ee
Inserting this relation in Eq. \eqref{bessel_large}, we obtain
\be\label{J_large_edge}
J_{\nu}(k_{n,l}r)\approx\left(\frac{2{\cal K}}{\pi k_F \tilde{l} \sqrt{{\cal K}^2-\tilde{l}^2}}\right)^{\frac{1}{2}}\sin\left[\sqrt{{\cal K}^2-\tilde{l}^2}k_F(r-1)\right]\;.
\ee
Using the identity for Bessel functions $\J_{\nu-1}(x)=\J_{\nu}'(x)+(\nu/x)\J_{\nu}(x)$ with $J_{\nu}(j_{\nu,n})=0$ and deriving Eq. \eqref{J_large_edge} with respect to $r$, we can compute the normalizing term $J_{\nu-1}(k_{n,l})$ in Eq. \eqref{radial_wf}. Finally, inserting this form and Eq. \eqref{J_large_edge} in Eq. \eqref{radial_wf}, we rewrite the radial wave-function $\chi_{n,l}(r)$ at the edge $k_F(1-r)=O(1)$ as
\be\label{xi_edge}
\chi_{n,l}(r)=\sqrt{\frac{2{\cal K}^2}{{\cal K}^2-\tilde{l}^2}}\sin\left[\sqrt{{\cal K}^2-\tilde{l}^2}k_F(r-1)\right]\;.
\ee
For large $k_F$,the effective one-dimensional kernel in Eq. \eqref{effective_kernel} is rewritten as an integral over $\tilde n=n/k_F$. Using this expansion of $\chi_{n,l}(r)$ in Eq. \eqref{xi_edge} close to the edge yields
\be\label{k_eff_int}
K^l_{\rm eff}(r,r')=k_F\int_0^{{\cal N}(\tilde l)}d\tilde n \frac{2{\cal K}^2}{{\cal K}^2-\tilde{l}^2}\sin\left[\sqrt{{\cal K}^2-\tilde{l}^2}k_F(r-1)\right]\sin\left[\sqrt{{\cal K}^2-\tilde{l}^2}k_F(r'-1)\right]\;.
\ee
Finally, introducing the variable $z=\sqrt{{\cal K}^2(\tilde n,\tilde l)-\tilde{l}^2}$ and deriving \eqref{asymp_K} with respect to $\tilde n$, one obtains
\be
\frac{\partial z}{\partial \tilde n}=\frac{\partial {\cal K}}{\partial \tilde n}\frac{\cal K}{\sqrt{{\cal K}^2-\tilde{l}^2}}=\pi \frac{{\cal K}^2}{{\cal K}^2-\tilde{l}^2}\;.
\ee
The integral in Eq. \eqref{k_eff_int} can then be performed explicitly by performing the change of variable $\tilde n \to z$. Using for the integration limits ${\cal K}(0,\tilde{l})=\tilde{l}$ and ${\cal K}({\cal N}(\tilde l),\tilde{l})=1$, we obtain the scaling form for the effective one-dimensional kernel at the edge
\begin{align}\label{edge_effk_hb_d}
&K_{\rm eff}^l(r,r')=\tilde k_l K_{1}^{\rm e}\left(\tilde k_l(1-r),\tilde k_l(1-r')\right)\;,\\
&{\rm with}\;\;\tilde k_l=k_F\cos\omega_l=\sqrt{k_F^2-l^2}\;\;{\rm and}\;\;K_1^{\rm e}(u,v)=\frac{\sin(u-v) }{\pi(u-v)}-\frac{\sin(u+v)}{\pi(u+v)}\;.\nn
\end{align}
Note that this result (\ref{edge_effk_hb_d}) is exactly what would be obtained from Eq. \eqref{LDA_images} with the substitution $V(1) \to V_{\rm eff}^l(1)=\q l^2$, which amounts
to neglect the variation of the effective potential in Eq. \eqref{effective_potential}. 


\noindent{\it Angular part}. We now turn to the analysis of the angular part of the kernel in the limit where $l=k_F\sin\omega_l\gg 1$ is large and the two arguments ${\bf x}$ and ${\bf y}$ form a small angle, i.e. $\eta={\bf x}\cdot{\bf y}/(|{\bf x}||{\bf y}|)\approx 1$. We consider a boundary point ${\bf x}_w$ and two points ${\bf x}$ and ${\bf y}$ close to this boundary point ($k_F|{\bf x}-{\bf x}_w|\sim k_F|{\bf y}-{\bf x}_w|=O(1)$). 
We introduce the set of coordinates in the collinear and perpendicular decomposition
\begin{align}
k_F({\bf x}_w-{\bf x})={\bf u}=\begin{cases}
&u_n=k_F({\bf x}_w-{\bf x})\cdot{\bf x}_w=O(1)\\
&\\
&{\bf u}_t=k_F(({\bf x}\cdot{\bf x}_w){\bf x}_w-{\bf x})=O(1)\\
\end{cases}\label{trans_col_decomp}\\
k_F({\bf x}_w-{\bf y})={\bf v}=\begin{cases}
&v_n=k_F({\bf x}_w-{\bf y})\cdot{\bf x}_w=O(1)\\
&\\
&{\bf v}_t=k_F(({\bf y}\cdot{\bf x}_w){\bf x}_w-{\bf y})=O(1)\\
\end{cases}\;.\nn
\end{align} 
The subscript '$t$' stands for the transverse direction while '$n$' stands for the radial direction.
Expanding $\eta=\frac{{\bf x}{\bf y}}{|{\bf x}||\bf y|}$ up to second order for large $k_F$, we obtain
\be
\eta\approx 1-\frac{|{\bf u}_t-{\bf v}_t|^2}{2k_F^2}\approx \cos\left(k_F^{-1}|{\bf u}_t-{\bf v}_t|\right)\;.
\ee
In the limit of large orbital quantum number $l=k_F\sin \omega_l\gg 1$ and small angle between ${\bf x}$ and ${\bf y}$, such that $\arccos \eta=|{\bf u}_t-{\bf v}_t|/k_F\ll 1$ with the product $|{\bf u}_t-{\bf v}_t|\sin \omega_l=\psi=O(1)$ we rewrite the differential equation for ${\sf P}_{l,d}(t)$ in Eq.~\eqref{eq:Legendre} as a differential equation for $f_d(\psi)={\sf P}_{l,d}(t)/S_d$ using $\partial_\eta\approx -(l^2/\psi) \partial_\psi$ and $1-\eta^2\approx(\psi/l)^2$. We obtain at leading order 
\be
f_d''(\psi)+\frac{d-2}{\psi}f'(\psi)+f(\psi)=0\;.
\ee 
The solution of this equation such that $f_d(0)={\sf P}_{l,d}(1)/S_d=1/S_d$ leads to the scaling form
\be\label{f_d}
\frac{1}{S_d}{\sf P}_{l,d}(\eta)\sim f_d(\psi)=\frac{\Gamma(d-1)}{4\pi}\frac{\J_{\frac{d-3}{2}}(\psi)}{(2\pi\psi)^{\frac{d-3}{2}}}\;,\;\;\psi=l|{\bf x}_t-{\bf y}_t|=|{\bf u}_t-{\bf v}_t|\sin\omega_l\;.
\ee
Finally, to consider the limit of large quantum orbital number $l=k_F \sin \omega_l\gg 1$ for the angular part in Eq. \eqref{sum_m}, we need the behavior of the degeneracy factor
\be\label{degeneracy_large}
g_d(l)=\frac{(2l+d-2)(l+d-3)!}{l!(d-2)!}\approx k_F^{d-2}\frac{2(\sin\omega_l)^{d-2}}{\Gamma(d-1)}\;.
\ee
Inserting Eqs. \eqref{f_d} and \eqref{degeneracy_large} in Eq. \eqref{sum_m} we obtain for large values of $k_F$
\be\label{angular_part}
\sum_{\bm m} Y_{l,{\bm m}}^*({\bm \theta})Y_{l,{\bm m}}({\bm \theta'})=\frac{g_d(l)}{S_d}{\sf P}_{l,d}(\eta) \approx k_F^{d-1}\left(\frac{\sin\omega_l}{2\pi}\right)^{\frac{d-1}{2}}\frac{\J_{\frac{d-3}{2}}(|{\bf u}_t-{\bf v}_t|\sin \omega_l)}{|{\bf u}_t-{\bf v}_t|^{\frac{d-3}{2}}}\;.
\ee
Note at this stage that this expression for the angular part of the kernel in Eq. \eqref{angular_part} can be rewritten as the scaling form
\be\label{angular_part_2}
\frac{g_d(l)}{S_d}{\sf P}_{l,d}\left(\frac{{\bf x}\cdot{\bf y}}{|{\bf x}||{\bf y}|}\right)=l^{d-1} K_d^{\rm ang}(l|{\bf x}_t-{\bf y}_t|)\;\;
{\rm with}\;\;K_d^{\rm ang}(z)=\frac{1}{2\pi}\frac{\J_{\frac{d-3}{2}}(z)}{(2\pi z)^{\frac{d-3}{2}}}\;,
\ee
which is universal with respect to all spherically symmetric potentials $V(|{\bf x}|)$ in the limit of large $l$ and small angle difference $|{\bf x}_t-{\bf y}_t|$ with fixed $l|{\bf x}_t-{\bf y}_t|=O(1)$. This scaling function $K_d^{\rm ang}(z)$ admits an alternative representation which is obtained by deriving the integral representation
\be\label{integral_Bessel}
\int_{|{\bf m}|\leq l}\frac{d^{d-1} {\bf m}}{(2\pi)^{d-1}}e^{i {\bf m}\cdot {\bf u_t}}=\left(\frac{l}{2\pi |{\bf u}_t|}\right)^{\frac{d-1}{2}}\J_{\frac{d-1}{2}}(l|{\bf u}_t|)\;,
\ee
with respect to $l$ using the identity $\partial_x(x^a \J_a(x))=x^a \J_{a-1}(x)$. This yields
\begin{align}\label{angular_part_3}
&\sum_{\bm m} Y_{l,{\bm m}}^*({\bm \theta})Y_{l,{\bm m}}({\bm \theta'})\approx l^{d-1} K_d^{\rm ang}(l|{\bf x}_t-{\bf y}_t|)\\
&{\rm with}\;\;K_d^{\rm ang}(z)=\int_{|{\bf m}|=1}\frac{d^{d-2}{\bf m}}{(2\pi)^{d-1}}e^{i z{\bf m}\cdot {\bf e}}\;,\;\;|{\bf e}|=1\;.\nn
\end{align}

\begin{figure}[h]
\centering
 \includegraphics[width=0.2\textwidth]{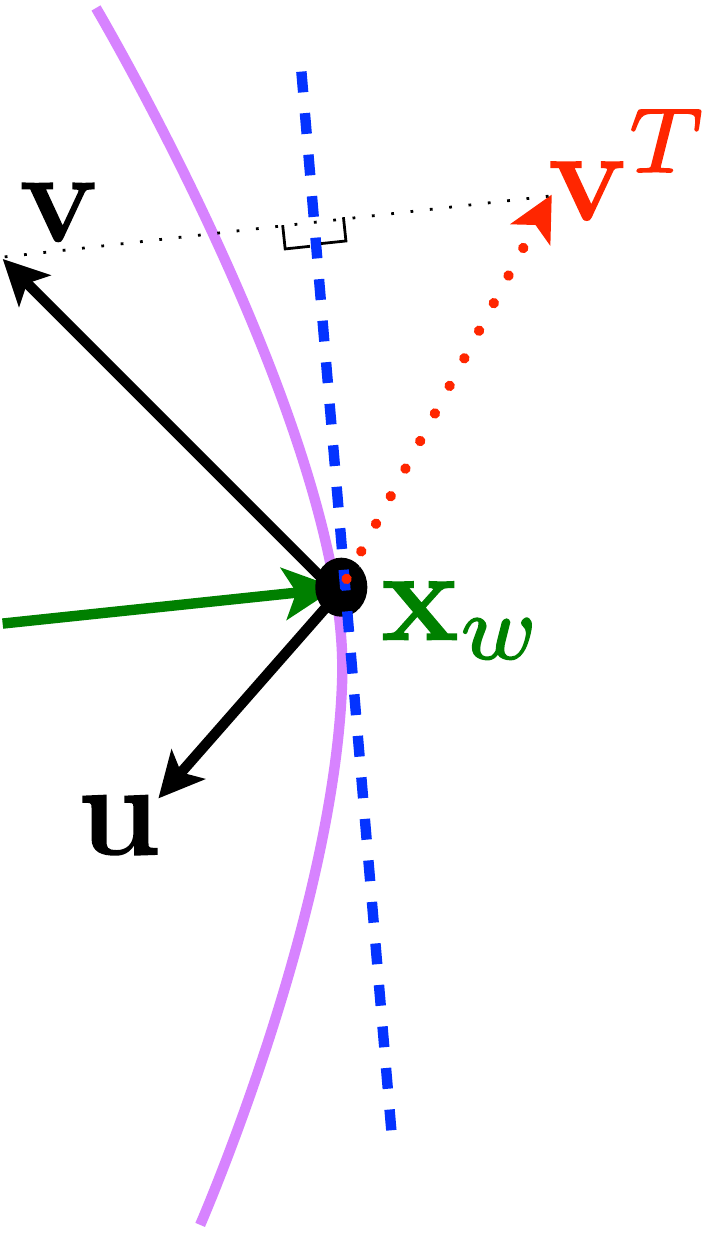}
 \caption{The two points lie respectively in the rescaled positions ${\bf u}$ and ${\bf v}$ with the rescaled origin in the boundary point ${\bf x}_w$. Here ${\bf v}^T$ is the image of ${\bf v}$ by the reflexion with respect to the hyperplane in blue which is orthogonal to ${\bf x}_w$.}
 \label{fig_image}
\end{figure}

\noindent{\it Edge Kernel}. We now put together the radial and the angular parts. Close to a boundary point ${\bf x}_w$, we use the set of coordinates in radial and transverse directions introduced in Eq. \eqref{trans_col_decomp}. Inserting Eqs. \eqref{edge_effk_hb_d} and \eqref{angular_part} in Eq. \eqref{kernel_spherical}, we replace in the limit of large $l=k_F\sin \omega_l$ the discrete sum over $l$ by an integral over $\omega=\omega_l$ such that $l/k_F=\sin\omega$. We obtain the scaling form for the kernel at the edge
\begin{align}\label{kernel_ab3}
&K_{\mu}({\bf x},{\bf y})=k_F^d K_{d}^{\rm e}(k_F({\bf x}_w-{\bf x}),k_F({\bf x}_w-{\bf y}))\;,\\
&K_{d}^{\rm e}({\bf u},{\bf v})=\int_0^{\frac{\pi}{2}}\left(\frac{\sin\omega}{2\pi}\right)^{\frac{d-1}{2}}\frac{\J_{\frac{d-3}{2}}(|{\bf u}_t-{\bf v}_t|\sin \omega)}{|{\bf u}_t-{\bf v}_t|^{\frac{d-3}{2}}}(\cos\omega)^2 K_{1}^{\rm e}\left(u_n\cos\omega,v_n\cos\omega\right)d\omega\nn
\end{align}
As the effective one dimensional kernel given in Eq. \eqref{edge_effk_hb_d} is the difference between two terms, this edge kernel will keep the same structure. We first analyze the bulk contribution, replacing in the effective one dimensional kernel $K_{1}^{\rm e}$ by $K_1^{\rm b}(|u_n-v_n|\cos\omega)$. This yields
\be
\frac{1}{(2\pi)^{\frac{d}{2}}}\int_0^{\frac{\pi}{2}}\left(\sin\omega\right)^{\frac{d-1}{2}}\frac{\J_{\frac{d-3}{2}}(|{\bf u}_t-{\bf v}_t|\sin \omega)}{|{\bf u}_t-{\bf v}_t|^{\frac{d-3}{2}}}(\cos\omega)^{\frac{3}{2}} \frac{\J_{\frac{1}{2}}(|u_n-v_n|\cos\omega)}{\sqrt{|u_n-v_n|}}d\omega\;,\label{bulk_contrib}
\ee  
where we used $K_1^{\rm b}(z)=\J_{1/2}(z)/\sqrt{2\pi z}$. Quite remarkably, this integral over $\omega$ can be performed explicitly using the identity \cite{int_Bessel}
\be\label{integral_bessel}
\int_0^{\frac{\pi}{2}}(\cos\omega)^{\tau+1}\J_{\tau}(y\cos\omega)(\sin\omega)^{\sigma+1}\J_{\sigma}(z\sin\omega)d\omega
=\frac{z^{\sigma}y^{\tau}\J_{\tau+\sigma+1}(\sqrt{z^2+y^2})}{(\sqrt{z^2+y^2})^{\tau+\sigma+1}}\;.
\ee
Specializing to $\tau=1/2$, $y=|u_n-v_n|$, $\sigma=(d-3)/2$ and $z=|{\bf u}_t-{\bf v}_t|$, Eq. \eqref{bulk_contrib} leads to
\begin{eqnarray}\label{bulk_contrib_2}
&&\hspace*{-1cm}\frac{1}{(2\pi)^{\frac{d}{2}}}\int_0^{\frac{\pi}{2}}\left(\sin\omega\right)^{\frac{d-1}{2}}\frac{\J_{\frac{d-3}{2}}(|{\bf u}_t-{\bf v}_t|\sin \omega)}{|{\bf u}_t-{\bf v}_t|^{\frac{d-3}{2}}}(\cos\omega)^{\frac{3}{2}} \frac{\J_{\frac{1}{2}}(|u_n-v_n|\cos\omega)}{\sqrt{|u_n-v_n|}}d\omega \nonumber \\
&&\hspace*{-1cm}=
\frac{\J_{\frac{d}{2}}\left(\sqrt{|u_n-v_n|^2+|{\bf u}_t-{\bf v}_t|^2}\right)}{\left(2\pi\sqrt{|u_n-v_n|^2+|{\bf u}_t-{\bf v}_t|^2}\right)^{\frac{d}{2}}}=\frac{\J_{\frac{d}{2}}\left(|{\bf u}-{\bf v}|\right)}{\left(2\pi|{\bf u}-{\bf v}|\right)^{\frac{d}{2}}}\;.
\end{eqnarray}
This part of the scaling function $K_d^{\rm e}({\bf u},{\bf v})$ reproduces exactly the bulk contribution $K_d^{\rm b}(|{\bf u}-{\bf v}|)$ given in Eq. \eqref{K_LDA}.
The second term appearing in the scaling function in Eq. \eqref{kernel_ab3} is very similar to this first term but one needs now to replace $K_1^{\rm e}$ by $K_1^{\rm b}(|u_n+v_n|\cos\omega)$. This term is analyzed similarly and leads to
\be\label{image_d}
\frac{\J_{\frac{d}{2}}\left(\sqrt{|u_n+v_n|^2+|{\bf u}_t-{\bf v}_t|^2}\right)}{\left(2\pi\sqrt{|u_n+v_n|^2+|{\bf u}_t-{\bf v}_t|^2}\right)^{\frac{d}{2}}}=\frac{\J_{\frac{d}{2}}\left(|{\bf u}-{\bf v}^{T}|\right)}{\left(2\pi|{\bf u}-{\bf v}^{T}|\right)^{\frac{d}{2}}}\;,\;\;{\rm with}\;\;v^T=(-v_n=-{\bf v}\cdot{\bf x}_w,{\bf v_t})\;.
\ee
From \eqref{image_d}, one can realize that ${\bf v}^{T}$ is the mirror image of ${\bf v}=(v_n,{\bf v}_t)$ with respect to the hyperplane orthogonal to ${\bf x}_w$ (see Fig. \ref{fig_image}). The structure of this edge scaling function is reminiscent of a method of images, that is still present in this higher dimensional case. Finally, collecting the two terms in Eqs. \eqref{bulk_contrib_2} and \eqref{image_d}, we obtain the final scaling form for the kernel at the edge
\begin{align}\label{k_d_final}
&K_\mu({\bf x},{\bf y})=k_F^d K_{d}^{\rm e}(k_F({\bf x}_{w}-{\bf x}),k_F({\bf x}_{w}-{\bf y}))\;,\;\;K_d^{\rm e}({\bf u},{\bf v})=K_d^{\rm b}(|{\bf u}-{\bf v}|)-K_d^{\rm b}(|{\bf u}-{\bf v}^T|)\;,\\
&{\rm with}\;K_d^{\rm b}(z)=\frac{\J_{\frac{d}{2}}(z)}{(2\pi z)^{\frac{d}{2}}}\;,{\bf u}=(u_n,{\bf u}_t)\;,\;{\bf v}=(v_n,{\bf v}_t)\;{\rm and}\;\;{\bf v}^T=(-v_n,{\bf v}_t)\;.\nn
\end{align}

Alternatively, another representation for this scaling function $K_{d}^{\rm e}$ is obtained by inserting Eqs. \eqref{khb_1d_e} together with \eqref{angular_part_3} in Eq. \eqref{kernel_spherical}. In the limit of large $l= k_F\tilde l$, we replace the sum on $l$ by an integral over $\tilde l$. Finally, the one dimensional integral over $\tilde l$ and the $d-2$ dimensional integral over ${\bf m}$ (coming from Eq. \eqref{angular_part_3}) is replaced by a $d-1$ dimensional integral over ${\bf l}=(\tilde l,{\bf m})$ such that $|{\bf l}|=\tilde l$. This second representation of the scaling function reads
\be
K_{d}^{\rm e}({\bf u},{\bf v})=\int_{|{\bf l}|<1}\frac{d^{d-1}{\bf l}}{(2\pi)^{d-1}}e^{i {\bf l}\cdot ({\bf u}_t-{\bf v}_t)}\sqrt{1-{\bf l}^2}K^{\rm e}_{1}\left(u_n\sqrt{1-{\bf l}^2},v_n\sqrt{1-{\bf l}^2}\right)\;.\label{kernel_edge_hb_d_b_2}
\ee
In particular, this formula establishes a nice relation between the $d$-dimensional kernel, with $d\geq 1$ and the one dimensional kernel, and it will be useful in the following. 
A similar structure was actually obtained in the case of a soft edge [see Eq. (288) of~\cite{fermions_review}]. 

As an application of this formula \eqref{k_d_final} for the kernel we compute the density at the edge, which is simply obtained by evaluating it at coinciding points ${\bf y}={\bf x}$ near the boundary located at ${\bf x}_w$. It reads 
\be\label{dens_d}
\tilde \rho({\bf x}) = K_\mu({\bf x}, {\bf x}) \approx \frac{N}{\Omega_d}F_d(k_F{\sf d})\;\;{\rm with}\;\;F_d(z)=1-\Gamma\left(\frac{d}{2}+1\right)\frac{\J_{\frac{d}{2}}(2z)}{z^{\frac{d}{2}}}\;,
\ee
where ${\sf d}$ is the distance of ${\bf x}$ from the boundary. The asymptotic behaviors of $F_d(z)$ read
\be\label{F_d_as}
F_d(z)\approx\begin{cases}
                   \frac{2}{2+d} z^2+O(z^4)\;&,\;\; z\to 0\\
                   &\\
                   1+O(z^{-\frac{d+1}{2}})\;&,\;\; z\to +\infty\;.
                  \end{cases}
\ee
In particular, using the second line of Eq. \eqref{F_d_as}, the constant bulk density is recovered far from the edge. 
Taking $d=1$ in Eq. \eqref{dens_d}, we obtain the same scaling function that was obtained previously in Eq. \eqref{edge_dens_hb_1d}.
This scaling function for $d=2$ is plotted in Fig. \ref{Fig F2} and it is compared with a numerical evaluation of the exact formula for the density \eqref{dens_d_N} for finite $N$. This shows that the scaling form (\ref{dens_d}) describes very well the density profile near the edge.  


To address the question of the universality of these results \eqref{k_d_final}, \eqref{kernel_edge_hb_d_b_2} and \eqref{dens_d} with respect to the potential inside the spherical hard box, it is convenient, as was done for $d=1$ in section~\ref{prop_1d}, to introduce the quantum propagator in imaginary time.

\subsubsection{Quantum propagator in $d>1$}\hfill\\

Let us first present an alternative derivation of this result for the kernel in Eqs. \eqref{k_d_final} and \eqref{kernel_edge_hb_d_b_2} using the 
$d$-dimensional imaginary time quantum propagator    
\be\label{propagator_d}
G({\bf x},{\bf y},t)=\langle {\bf y}|e^{-\frac{t H}{\hbar}}|{\bf x}\rangle=\sum_{\bf k}\phi_{\bf k}^*({\bf x})\phi_{\bf k}({\bf y})e^{-\frac{t E_{\bf k}}{\hbar}}\;.
\ee
It is related to the correlation kernel via \cite{fermions_review}
\be\label{kernel_propagator_d}
K_{\mu}({\bf x},{\bf y})=\int_{\cal C} \frac{dt}{2i\pi t}e^{\frac{\mu t}{\hbar}}G({\bf x},{\bf y},t)\;,
\ee
where ${\cal C}$ indicates the Bromwich integration contour in the complex $t$-plane. As in one dimension \eqref{free_diffusion}, this propagator satisfies the free diffusion equation
\be\label{eq_free_propagator_d}
-\hbar\partial_t  G({\bf x},{\bf y},t)=-\q\Delta_{{\bf y} } G({\bf x},{\bf y},t),\;\;{\rm with}\;\; G({\bf x},{\bf y},0)=\delta^d({\bf x}-{\bf y})\;,
\ee
and with vanishing boundary conditions $G({\bf x}, {\bf y},t) = 0$ for all ${\bf x}$, $t$ and $|{\bf y}|=1$. This equation \eqref{eq_free_propagator_d} is invariant by translation of the coordinates within the boundaries. 
Therefore we place the origin at a position ${\bf x}_w$ of the hard edge ($|{\bf x}_w|=1$).  We want to obtain the correlation kernel in the edge regime $k_F({\bf x}_w-{\bf x})=O(1)$ for large $\mu=\hbar^2k_F^2/(2m)$. Therefore, from Eq. \eqref{kernel_propagator_d} it will correspond to a regime where $t\sim \hbar/\mu\ll 1$ for the propagator. It is convenient to work with dimensionless space and time variables \cite{LLMS17}   
\be\label{scaling_prop_free_d}
G({\bf x},{\bf y},t)=k_F^d G_0\left(k_F({\bf x}_w-{\bf x}),k_F({\bf x}_w-{\bf y}),\frac{\mu t}{\hbar}\right)\;,
\ee
where, according to \eqref{eq_free_propagator_d}, the rescaled propagator $G_0$ is solution of the dimensionless equation
\be \label{eq_free_propagator_d_scaled}
\partial_{\tau}G_0({\bf u},{\bf v},\tau)=\Delta_{{\bf v}} G_0({\bf u},{\bf v},\tau)\;,\;\;{\rm with}\;\;G_0({\bf u},{\bf v},0)=\delta^{d}({\bf u}-{\bf v})\;,
\ee
and with $G_0({\bf u},{\bf v}, \tau) = 0$ for all ${\bf u}$, $t$ and ${\bf v}$ such that the original variable ${\bf y} = {\bf x}_w- k_F^{-1} {\bf v}$ is on the boundary, i.e. for $({\bf x}_w-k_F^{-1}{\bf v})^2=1$. Since ${\bf x}_w^2 = 1$, this condition simply reads $2{\bf v}\cdot{\bf x}_w=k_F^{-1}{\bf v}^2$. To leading order for large $k_F$ (and ${\bf v}^2=O(1)$), this boundary condition reduces to
\be\label{rescaled_vanishing}
G_0({\bf u},{\bf v},\tau)=0\;,\;\;{\rm for}\;\; {\bf v}\cdot{\bf x}_w=v_n=0\;.
\ee
The solution of the (scaled) diffusion equation \eqref{eq_free_propagator_d_scaled} with the boundary condition (\ref{rescaled_vanishing}) on the hyperplane orthogonal to ${\bf x}_w$ can be obtained by the same method of images as in the one-dimensional case (see section \ref{prop_1d}). It reads
\be\label{G_0_image}
G_0({\bf u},{\bf v},\tau)\approx \frac{1}{(4\pi \tau)^{\frac{d}{2}}} \left(e^{-\frac{({\bf u}-{\bf v})^2}{4\tau}}-e^{-\frac{({\bf u}-{\bf v}^T)^2}{4\tau}}\right)\;,
\ee
where ${\bf v}^T={\bf v}-2({\bf v}\cdot{\bf x}_w){\bf x}_w=(-v_n,{\bf v}_t)$ is the reflexion of ${\bf v}$ by the hyperplane orthogonal to ${\bf x}_w$ (see Fig. \ref{fig_image}).
Inserting the scaling form \eqref{scaling_prop_free_d}-\eqref{G_0_image} in the inversion formula of Eq. \eqref{kernel_propagator_d} and performing the change of variable $t \to \tau=\mu t/\hbar$, we obtain the scaling form valid for large $k_F$
\begin{align}
&K_{\mu}({\bf x},{\bf y}) \approx k_F^d K_d^{\rm e}(k_F({\bf x}_w-{\bf x}),k_F({\bf x}_w-{\bf y}))\;,\label{k_d_prop_scal}\\
&K_d^{\rm e}({\bf u},{\bf v})=\frac{1}{(4\pi)^{\frac{d}{2}}}\int_{\cal C} \frac{d\tau}{2i\pi\tau^{\frac{d}{2}+1}}e^{\tau}\left(e^{-\frac{({\bf u}-{\bf v})^2}{4\tau}}-e^{-\frac{({\bf u}-{\bf v}^T)^2}{4\tau}}\right)\;.\nn
\end{align}
Finally, using the inversion formula in Eq. \eqref{inversion_prop}
\be
\int_{\cal C}\frac{d\tau}{2i\pi \tau^{d/2+1}}\exp\left(z\tau-\frac{a}{\tau}\right)=\left(\frac{z}{a}\right)^{\frac{d}{4}}\J_{\frac{d}{2}}(2\sqrt{az})\;,
\ee
specialized to $z=1$, $a=({\bf u}-{\bf v})^2/4$ for the first term or $a=({\bf u}-{\bf v}^T)^2/4$ for the second one, we obtain the final expression of $K_d^{\rm e}({\bf u},{\bf v})$ given in Eq. \eqref{k_d_final}. 

In Ref. \cite{LLMS17}, this method relying on the propagator was used to extend this result in Eq. \eqref{k_d_final} to any domain ${\cal D}$ with a smooth (twice differentiable) boundary $\partial{\cal D}$. In this case, the formula in Eq. \eqref{k_d_final} holds where ${\bf v}^T$ is now the image of ${\bf v}$ with respect to the tangent plane to the boundary at ${\bf x}_w$ \cite{LLMS17} (see also Fig. \ref{fig_image}). It is then natural to wonder what happens in the presence of a non-zero smooth potential $V({\bf x}) \neq 0$ inside the box. In the absence of the wall, this potential would create an edge in the density at ${\bf x} = {\bf x}_e$ such that $V({\bf x}_e) = \mu$, with an associated width $w_N$ given in Eq.~\eqref{K_d_soft}. Obviously if $|{\bf x}_e|\ll|{\bf x}_w|$ then the density at the wall is essentially zero and the kernel at the edge ${\bf x}_e$ is given by the soft edge kernel given in Eq. \eqref{K_d_soft}. On the other hand, if $|{\bf x}_e| \gg |{\bf x}_w|$, and if the potential does not vary too fast near the boundary $\partial{\cal D}$, it was shown in Ref. \cite{LLMS17}, that the limiting kernel near the wall is given by the result  in \eqref{k_d_final}. More precisely, as in the one-dimensional case (\ref{condition_hb}), if the following condition holds
\be \label{condition_hb_d}
\tilde k({\bf x}_w) = \frac{\sqrt{2m(\mu - V({\bf x}_w))}}{\hbar} \gg \frac{1}{w_N} 
\ee
where $w_N$ is the width of the smooth edge regime close to ${\bf x}_e$ \eqref{K_d_soft}, then the limiting kernel near the wall takes the following universal scaling form \cite{LLMS17}
\be\label{LDA_images_d}
K_\mu({\bf x}, {\bf y}) \approx \left[\tilde k({\bf x}_w)\right]^d K_d^{\rm e}\left(\tilde k({\bf x}_w)({\bf x}_w-{\bf x}),  \tilde k({\bf x}_w)({\bf x}_w-{\bf y})\right) \;,
\ee
where $K_d^{\rm e}({\bf u}, {\bf v})$ is given in Eq. \eqref{k_d_final}. 

Let us now investigate the intermediate regime where the condition (\ref{condition_hb_d}) does not hold and where we expect
to obtain the generalisation to higher dimensions $d>1$ of the interpolating kernel found in $d=1$ in \eqref{K_1_l_final} and \eqref{sigma}.

%
\subsection{Non-uniform hard box potentials}

To study the transition from soft to hard edge behavior, we consider the spherically symmetric truncated potential of the form
\be\label{trunc_pot_d}
V_{\rm tr}({\bf x})=\begin{cases}
      &\mu \left(\frac{|{\bf x}|}{r_{\rm e}}\right)^{p}\;,\;\;|{\bf x}|\leq 1\\
      &\\
      &\infty\;,\;\;|{\bf x}|>1\;,
     \end{cases}
\ee
with $p>0$ which is a generalization to $d$-dimensions of the one-dimensional truncated potential studied in section \ref{trunc_1d} (thus we set the wall at $|{\bf x}_w| = 1$). In the absence of the wall, the potential creates an edge in the density at $r_{\rm e}$ such that $V(r_{\rm e})=\mu$ and here we consider the case where $r_{\rm e}\simeq 1$. As in the one-dimensional case, we anticipate that the crossover between the hard edge and the soft edge kernel is governed by the rescaled distance $\ell$ 
\be
\ell=\frac{1-r_{\rm e}}{w_N}\;\;{\rm with}\;\;w_N=\left(\frac{\hbar^2}{2mV'(r_{\rm e})}\right)^{\frac{1}{3}}=p^{-1/3}k_F^{-2/3}r_{\rm e}^{1/3}\;,
\ee
where $w_N$ is the characteristic length at the soft edge \eqref{K_d_soft}. 
In the presence of a non-zero potential inside the box (\ref{trunc_pot_d}) the $d$-dimensional quantum propagator $G({\bf x},{\bf y},t)$ in (\ref{propagator_d}) satisfies 
\be\label{eq_propagator_d}
-\hbar\partial_t  G({\bf x},{\bf y},t)=-\q\Delta_{{\bf y} } G({\bf x},{\bf y},t)+V(|{\bf y}|)G({\bf x},{\bf y},t)\;,\;\;{\rm with}\;\; G({\bf x},{\bf y},0)=\delta^d({\bf x}-{\bf y})\;,
\ee
with the boundary condition $ G({\bf x},{\bf y},t) = 0$ for all ${\bf x}$, $t$ and $|{\bf y}|=1$. To study the behavior of this propagator close to a boundary point ${\bf x}_{w}$, such that $|{\bf x}_{w}|=1$, it is useful, as done before (\ref{scaling_prop_free_d}), to introduce a rescaled propagator $G_d({\bf u},{\bf v},t)$ defined as
\be\label{prop_scal_d}
G({\bf x},{\bf y},t)=\frac{1}{w_N^d} G_0\left(\frac{{\bf x}-{\bf x}_{w}}{w_N},\frac{{\bf y}-{\bf x}_{w}}{w_N},t\right)\;.
\ee
We introduce at this stage the decomposition of coordinates along radial and transverse directions
\begin{align}\label{def_trans_norm}
&\frac{{\bf x}-{\bf x}_{w}}{w_N}={\bf u}=\begin{cases}
&u_n={\bf x}\cdot{\bf x}_{w}-1\leq 0\\
&\\
&{\bf u}_t={\bf x}-({\bf x}\cdot{\bf x}_{w}){\bf x}_{w}
\end{cases}\\
&{\rm and}\;\;
\frac{{\bf y}-{\bf x}_{w}}{w_N}={\bf v}=\begin{cases}
&v_n={\bf y}\cdot{\bf x}_{w}-1\leq 0\\
&\\
&{\bf v}_t={\bf y}-({\bf y}\cdot{\bf x}_{w}){\bf x}_{w}
\end{cases}\;.\nn
\end{align}
In this new set of coordinates, we develop up to first order in $w_N\sim k_F^{-2/3}\ll 1$ the expression of $|{\bf y}|$,
\be\label{Taylor_y}
|{\bf y}|=|{\bf x}_w+w_N{\bf v}|=\sqrt{|1+w_N v_n|^2+w_N^2|{\bf v}_t|^2}\approx 1+w_N v_n +O(w_N^2)\;.
\ee
We then expand in Taylor series at first order the potential $V(|{\bf y}|)$ using (\ref{Taylor_y}) as well as $1-r_{\rm e}=w_N\ell$
\be\label{lin_pot_d}
V(|{\bf y}|)\approx V(r_{\rm e})+V'(r_{\rm e})w_N(\ell+v_n)=\mu+\frac{\hbar^2}{2m w_N^2}\left[\ell+v_n+O\left(\frac{w_N}{r_{\rm e}}\right)\right]\;,
\ee
where we used $\mu=V(r_{\rm e})$ and $w_N=\hbar^{2/3}/(2m V'(r_{\rm e}))^{1/3}$. Thus we see on (\ref{lin_pot_d}) that, to leading order, this linearized potential has spatial variations only along the direction collinear to the boundary point ${\bf x}_w$. By substituting the potential $V(|{\bf y}|)$ in Eq. \eqref{eq_propagator_d} by its linearized approximation \eqref{lin_pot_d}, we find that the rescaled propagator in Eq. \eqref{prop_scal_d} satisfies 
\begin{eqnarray}\label{rescaled_prop_d}
&&\left(\frac{2m w_N^2}{\hbar^2}(\hbar\partial_t+\mu)+\ell\right)G_0({\bf u},{\bf v},t)=\left(\Delta_{{\bf v} }-v_n\right) G_0({\bf u},{\bf v},t)\;,\\
&&\hspace*{2.6 cm}\;\;{\rm with}\;\;G_0({\bf u},{\bf v},0)=\delta^d({\bf u}-{\bf v}) \nonumber
\end{eqnarray}
and boundary conditions $G_0({\bf u},{\bf v},t) = 0$ for all ${\bf u}$, $t$ and ${\bf v}$ such that $v_n=0$. On the rhs of Eq. (\ref{rescaled_prop_d}) it is useful to write the Laplacian as 
$\Delta_{\bf v}=\Delta_{{\bf v}_t}+\partial_{v_n}^2$ in terms of transverse and radial ${\bf v}$-coordinates (\ref{def_trans_norm}). This equation (\ref{rescaled_prop_d}) can then be solved via the use of Fourier transform with respect to ${\bf u}_t$ and ${\bf v}_t$, i.e.  
\be\label{FT_prop}
G_0({\bf u},{\bf v},t)=\int\frac{d^{d-1}{\bf l}_1}{(2\pi)^{d-1}}\int\frac{d^{d-1}{\bf l}_2}{(2\pi)^{d-1}}e^{i ({\bf l}_1\cdot{\bf u}_t+{\bf l}_2\cdot{\bf v}_t)}\tilde G_0(u_n,{\bf l}_1,v_n,{\bf l}_2,t)\;.
\ee
Inserting this form \eqref{FT_prop} in Eq. \eqref{rescaled_prop_d}, we obtain an equation for $\tilde G_0(u_n,{\bf l}_1,v_n,{\bf l}_2,t)$
\be
\left(\frac{2m w_N^2}{\hbar^2}(\hbar\partial_t+\mu)+\ell+{\bf l}_2^2\right)\tilde G_0(u_n,{\bf l}_1,v_n,{\bf l}_2,t)=\left({\partial^2_{{v}_n }}-v_n\right)\tilde G_0(u_n,{\bf l}_1,v_n,{\bf l}_2,t)\;,\label{prop_trans}
\ee
with initial condition $\tilde G_0(u_n,{\bf l}_1,v_n,{\bf l}_2,0)=\delta(u_n-v_n)(2 \pi)^{d-1}\delta^{d-1}({\bf l}_1+{\bf l}_2)$ and $\tilde G_0(u_n,{\bf l}_1,v_n,{\bf l}_2,t) = 0$ for all $u_n, {\bf l}_1, {\bf l}_2, t$ and $v_n = 0$. We find that the solution of this equation \eqref{prop_trans} with these initial and boundary conditions is of the form  
\be\label{ansatz}
\tilde G_0(u_n,{\bf l}_1,v_n,{\bf l}_2,\tau)=\exp\left(-\frac{\mu t}{\hbar}-\frac{\hbar}{2m w_N^2}({\bf l}_2^2+\ell)t\right)(2 \pi)^{d-1}\delta^{d-1}({\bf l}_1+{\bf l}_2) G_1(u_n,v_n,t)\;,
\ee
where $G_1(u_n,v_n,t)$ is the one-dimensional propagator given in Eq. \eqref{prop_sol_FF}. It satisfies 
\be\label{eff_1d_prop}
\frac{2m w_N^2}{\hbar}\partial_t G_1(u_n,v_n,t)=(\partial_{v_n}^2-v_n)G_1(u_n,v_n,t)\;,\;\;{\rm with}\;\; G_1(u_n,v_n,0)=\delta(u_n-v_n)\;,
\ee
with vanishing boundary conditions for $u_n,v_n=0$.
Inserting this expression \eqref{ansatz} in  \eqref{FT_prop}, we obtain 
\be\label{d_prop}
G_0({\bf u},{\bf v},t)=\int\frac{d^{d-1}{\bf l}}{(2\pi)^{d-1}}\e^{i {\bf l}\cdot({\bf u}_t-{\bf v}_t)}\exp\left(-\frac{\mu t}{\hbar}-\frac{\hbar}{2m w_N^2}({\bf l}^2+\ell)t\right) G_1(u_n,v_n,t)\;.
\ee
The kernel is obtained by inserting Eq. \eqref{prop_scal_d} in the inversion formula \eqref{kernel_propagator_d}. Therefore, we anticipate that the kernel takes the scaling form
\be
K_{\mu}({\bf x},{\bf y})\approx \frac{1}{w_N^d}K_{d}^{\ell}\left(\frac{{\bf x}-{\bf x}_{w}}{w_N},\frac{{\bf y}-{\bf x}_{w}}{w_N}\right)\;.
\ee
The scaling function $K_{d}^{\ell}({\bf u},{\bf v})$ is itself obtained by inserting in Eq. \eqref{kernel_propagator_d} the propagator scaling function in Eq. \eqref{d_prop}, yielding
\be\label{K_d_l_int}
K_{d}^{\ell}({\bf u},{\bf v})=\int_{\cal C}\frac{dt}{2i\pi t}\int\frac{d^{d-1}{\bf l}}{(2\pi)^{d-1}}\e^{i {\bf l}\cdot({\bf u}_t-{\bf v}_t)}\int_{-\infty}^{\infty}dse^{\frac{\hbar(s-{\bf l}^2-\ell)}{2m w_N^2}t}\sigma(s,s+u)\sigma(s,s+v)\;.
\ee
The Bromwich integral on $t$ in Eq. \eqref{K_d_l_int} is performed using the identity
\be
\int_{\cal C}\frac{dt}{2i\pi t}e^{x t}=\Theta(x)\;.
\ee
Finally, the scaling function $K_{d}^{\ell}({\bf u},{\bf v})$ can be reexpressed in terms of the one-dimensional scaling function $K_1^{\ell}(u,v)$ in Eq. \eqref{K_1_l_final}. This yields
\begin{align}
&K_{\mu}({\bf x},{\bf y})\approx \frac{1}{w_N^d}K_{d}^{\ell}\left(\frac{{\bf x}-{\bf x}_{w}}{w_N},\frac{{\bf y}-{\bf x}_{w}}{w_N}\right)\;,\label{K_ell_d}\\
&{\rm with}\;\;K_d^{\ell}({\bf u},{\bf v})=\int\frac{d^{d-1}{\bf l}}{(2\pi)^{d-1}}\e^{i{\bf l}\cdot({\bf u}_t-{\bf v}_t)} K_1^{\ell +{\bf l}^2}(u_n,v_n)\;.\nn
\end{align}
A similar expression was already obtained in the case of a soft edge [see Eq. (288) of Ref. \cite{fermions_review}] where the scaling function in $d>1$ dimension is expressed in term of the one dimensional scaling function
\be
K_d^{\rm soft}({\bf u},{\bf v})=\int \frac{d^{d-1}{\bf l}}{(2\pi)^{d-1}}\e^{i{\bf l}\cdot({\bf u}_t-{\bf v}_t)} K_1^{\rm soft}(u_n+{\bf l}^2,v_n+{\bf l}^2)\;.
\ee
As seen in Eq. \eqref{kernel_edge_hb_d_b_2}, it is also the case for the uniform hard box potential.
In the limit $\ell \to +\infty$, using the asymptotic behavior of $K_1^{\ell}(u_n,v_n)$ displayed in Eq. \eqref{l+}, we obtain the convergence to the soft edge scaling function 
\be
K_d^{\ell}({\bf u}-\ell{\bf x}_w,{\bf v}-\ell{\bf x}_w)\to K_d^{\rm soft}\left({\bf u},{\bf v}\right)\;,\;\;\ell\to+\infty\;.
\ee
On the other hand for $\ell\to -\infty$, rescaling further the positions close to the hard edge as in the one dimensional case, and using the $\ell\to -\infty$ limit for $K_1^{\ell}$ in Eq. \eqref{l-}, we obtain
\be
\frac{1}{|\ell|^{\frac{d}{2}}}K_d^{\ell}\left(\frac{{\bf u}}{\sqrt{|\ell|}},\frac{{\bf v}}{\sqrt{|\ell|}}\right)\to K_d^{\rm e}({\bf u},{\bf v})\;,\;\;\ell\to -\infty\;.
\ee
Note that the total rescaling parameter in this case $\sqrt{|\ell|}/w_N=\sqrt{2m V'(r_{\rm e})(r_{\rm e}-1)}/\hbar$ matches smoothly the limit for $r_{\rm e}\to|{\bf x}_w|=1$ of the rescaling parameter $\tilde k({\bf x}_{w})=\sqrt{2m\left[\mu-V(|{\bf x}_w|)\right]}/\hbar$ appearing in Eq. \eqref{LDA_images_d}. The correlation kernel takes at the edge of the Fermi gas a scaling form in Eq. \eqref{K_ell_d} that does not depend on the details of the potential. It interpolates from the soft to hard edge scaling form depending smoothly on the rescaled  $\ell=(1-r_{\rm e})/w_N$. 
%

Note that there is an alternative form for the kernel $K_d^{\ell}(\bf u,\bf v)$ where one integrates first Eq. (\ref{K_d_l_int}) over ${\bf l}$. We separate the $d-1$ dimensional integral over ${\bf l}$ into a spherical integral over the radial variable $l=|{\bf l}|$ and the $d-2$ dimensional angular variable ${\bf m}$. We recognise the $d-2$ angular integral in Eq. \eqref{angular_part_3} with $z=l|{\bf u}_t-{\bf v}_t|$ and ${\bf e}=\frac{{\bf u}_t-{\bf v}_t}{|{\bf u}_t-{\bf v}_t|}$ and replace its expression by Eq. \eqref{angular_part_2}. This finally yields
\be
K_d^{\ell}({\bf u},{\bf v})=\int_0^{\infty}dl \left(\frac{l}{2\pi}\right)^{\frac{d-1}{2}}\frac{\J_{\frac{d-3}{2}}(l|{\bf u}_t-{\bf v}_t|)}{|{\bf u}_t-{\bf v}_t|^{\frac{d-3}{2}}} K_1^{\ell +l^2}(u_n,v_n)\;.
\ee
This result concludes our study of the non-interacting Fermi gas at zero temperature. We will now study at finite temperature the interplay between quantum and thermal fluctuations in any dimension $d\geq 1$.

\section{Uniform hard box potential at finite temperature $T>0$}\label{hb_d_T}

So far, we focused on the case of zero temperature ($T=0$), where the positions of the fermions constitute a determininantal
point process, in any dimension $d$. However, experiments are usually performed at finite temperature $T>0$ and it is thus important
to characterise the effect of temperature on the correlations of trapped non-interacting fermions. A priori, it is rather natural to work in the canonical
ensemble, where the number of fermions $N$ is fixed. However, in the canonical ensemble, the system is not anymore determinantal \cite{fermions_review}, and the analysis is thus much more complicated (see however \cite{Liechty17}). To circumvent this technical problem, it is useful to consider instead the same system of non-interacting trapped fermions but in the grand-canonical ensemble, where the (finite temperature) chemical potential $\tilde \mu$ is fixed while the total number of fermions fluctuates. The main advantage of the grand-canonical ensemble is that the positions of the fermions constitute a determinantal point process, even at finite temperature $T>0$ \cite{fermions_review,J07,G60,HKPV06}. 

Coming back to the canonical ensemble with a fixed number $N$ of fermions, one can actually show \cite{fermions_review} that the local correlations coincide, in the large $N$ limit, with the predictions obtained in the grand-canonical ensemble, where they have a determinantal structure (note however that this equivalence between the two ensembles can not be used to compute the fluctuations of global observables \cite{GMS17,GMST18}). Hence, in the following we will focus on the grand-canonical ensemble where the positions of the fermions form a determinantal point process with a correlation kernel given  by
\be\label{K_finite_b}
K_{\tilde \mu}({\bf x},{\bf y})=\sum_{\bf k} n_{\rm FD}(\epsilon_{\bf k})\phi^*_{\bf k}({\bf x})\phi_{\bf k}({\bf y})\;\;{\rm with}\;\;n_{\rm FD}(\epsilon)=\frac{\zeta}{e^{\beta \epsilon}+\zeta}\;,
\ee
where $n_{\rm FD}$ is the Fermi-Dirac distribution and $\zeta$ is the fugacity defined as
\be\label{zeta}
\zeta=e^{\beta\tilde\mu}
\ee
with $\tilde\mu$ the finite temperature chemical potential. It is related to $N$, the number of fermions, via 
\be\label{rel_mu_N}
N = \sum_{\bf k} \frac{\zeta}{\zeta+ e^{\beta \epsilon_{\bf k}}} \;.
\ee
In Eq. (\ref{K_finite_b}), we recall that the functions $\phi_{\bf k}(\bf x)$ are the single particle eigenfunctions, with associated
eigenvalue $\epsilon_k$  (\ref{spwf}). By comparing the formula for the finite $T>0$ kernel $K_{\tilde \mu}({\bf x},{\bf y})$ in (\ref{K_finite_b}) and the $T=0$ kernel $K_{\mu}({\bf x},{\bf y})$ given in Eq. (\ref{Kernel}), one easily obtains the useful identity \cite{fermions_review}

\be\label{K_beta}
K_{\tilde \mu}({\bf x},{\bf y})=\int_{0}^{\infty}\frac{\zeta d\mu}{\zeta +e^{\beta\mu}}  \partial_{\mu} K_{\mu}({\bf x},{\bf y})\;,
\ee
where the fugacity $\zeta$ (\ref{zeta}) is independent of the dummy integration variable $\mu$. Hence, with the help of this formula (\ref{K_beta}), 
the limiting scaling form of the finite $T>0$ kernel $K_{\tilde \mu}({\bf x},{\bf y})$ can be obtained rather straightforwardly from the $T=0$ results for $K_{\mu}({\bf x},{\bf y})$  that we have established in the previous sections.

For the case of smooth potentials, we have obtained a number of results at finite temperature and we refer the reader to \cite{us_finiteT,fermions_review,us_Wigner,farthest_f,periodic_airy} for the details. In the case of the one-dimensional harmonic potential, $d=1$, we showed that the soft edge kernel at finite temperature takes a universal scaling function depending continuously on the scaled inverse temperature parameter, denoted by $b$. For the one-dimensional harmonic potential $b \sim N^{1/3}/T$, which shows that the relevant temperature scale at the edge is, in that case, $T \sim N^{1/3}$, much smaller than the $T \sim T_F \sim N$ which is the relevant temperature scale in the bulk. We now study the effect of the temperature for the hard box potential.

\subsection{Finite temperature chemical potential}

For simplicity, we consider the $d$-dimensional hard box potential, studied at $T=0$ in section \ref{shb_t0}, defined as
\be\label{def_box_T}
V_R({\bf x})=\begin{cases}
        & 0\;,\;\;|{\bf x}|\leq R\\
				&\\
        & \infty\;,\;\;|{\bf x}|>R\;,
       \end{cases}
\ee
and we set $R=1$ in the following. It is convenient to introduce the rescaled inverse temperature $b$ defined as 
\be\label{b_temp}
b=\beta\mu=\frac{\beta\hbar^2 k_F^2}{2m}=\frac{T_F}{T}\;,
\ee
where $T_F=\mu/k_B$ is the Fermi temperature (and we recall that $\mu$ is the Fermi energy). In the present case, at variance
with the soft edge case discussed above, the bulk and edge relevant temperature scales are identical and equal to $T_F$. 
Note also that this parameter $b$ can be defined in terms of the De Broglie thermal wave-length
\be\label{lambda_T}
\lambda_T=\sqrt{\frac{2\pi\beta\hbar^2}{m}}\;,
\ee
which is the typical scale of thermal fluctuations and the Fermi wavevector $k_F$ as 
\be\label{b_length}
b=\frac{T_F}{T}=\frac{(k_F\lambda_T)^2}{4\pi}\;.
\ee
The de Broglie wave length $\lambda_T$ plays a very important role as it controls the classical to quantum crossover in the system. Indeed, if $\lambda_T$ is much larger (respectively much smaller) than the inter-particle distance $\propto 1/k_F$, then the system exhibits a quantum (respectively classical) behaviour. Therefore, in the following, to describe the quantum to classical crossover, we will consider the case where $\lambda_T\, k_F$, or equivalently $b$ in  (\ref{b_length}), is fixed, i.e. $b=O(1)$. In this limit, the discrete sum in Eq.~(\ref{rel_mu_N}) can be replaced by an integral, and the relation between $N$ and $\zeta$ can be written as
%
%
\be\label{N_b}
N \approx-\frac{\Omega_d}{\lambda_T^{d}} \Li_{\frac{d}{2}}(-\zeta)\;,
\ee
where ${\Li}_s(x) = \sum_{k\geq 1} x^k/k^s$ is the polylogarithm function of index $s$ and we remind that $\Omega_d=\pi^{d/2}\Gamma(1+d/2)$. On the other hand, at $T=0$, $N$ is related to the Fermi energy $\mu$ via the relation (\ref{k_F_N}) which can be written as
\be\label{N_zero}
N\approx\frac{1}{\Gamma(1+d/2)^2}\left(\frac{k_F}{2}\right)^d=\frac{1}{\Gamma(1+d/2)^2}\left(\frac{m\mu}{2\hbar^2}\right)^{\frac{d}{2}}\;,
\ee
where, in the second equality, we have used that $k_F = \sqrt{2 m \mu}/\hbar$. Hence, by combining Eqs. (\ref{N_b}) and (\ref{N_zero}), we obtain an implicit equation for $\zeta$ as a function of the rescaled inverse temperature $b=\beta \mu= T_F/T$ (\ref{b_temp}),
\be\label{zeta_T}
-\Li_{\frac{d}{2}}(-\zeta)=\frac{b^\frac{d}{2}}{\Gamma\left(\frac{d}{2}+1\right)}\;.
\ee
We can now analyze the asymptotic behaviors of $\zeta$ for $b\to 0$ and $b\to \infty$ using the asymptotic behavior of $-\Li_s(-x)$,
\be\label{Li_as}
-\Li_s(-x)\approx\begin{cases}
\displaystyle x-\frac{x^2}{2^s}&\;,\;\;x\to 0\;,\\
&\\
\displaystyle\frac{(\ln x)^{\frac{d}{2}}}{\Gamma(s+1)}&\;,\;\;x\to +\infty\;.
\end{cases}
\ee
As $b\to 0$, i.e. $T \gg T_F=\mu/k_B$, using the first line of Eq. (\ref{Li_as}), we obtain 
\be
\zeta\approx \frac{b^\frac{d}{2}}{\Gamma\left(\frac{d}{2}+1\right)}\;,\;{\rm for}\;\;b\to 0\;.
\ee
From this relation, we see in particular that the chemical potential $\tilde\mu = \ln \zeta/\beta$ becomes negative at high temperature and we recover the well known expression for the classical ideal monoatomic gas
\be
\tilde \mu\approx\frac{1}{\beta}\ln(\rho \lambda_T^d)\;,\;\;{\rm for}\;\;T\gg T_F\;\;{\rm with}\;\;\rho=\frac{N}{\Omega_d}\;.
\ee
On the other hand, from the second line of Eq. \eqref{Li_as}, as $b\to\infty$, that is very low temperature $T\ll T_F$, we check that the finite temperature chemical potential $\tilde \mu=(\ln\zeta)/\beta$ goes to the Fermi energy $\mu$, as expected.

\subsection{Finite temperature correlation kernel}

Let us first start with the correlation kernel $K_{\tilde \mu}({\bf x}, {\bf y})$ in the bulk, i.e. for ${\bf x}$ and ${\bf y}$ far from the boundary of the box (\ref{def_box_T}). The finite $T$ bulk kernel, for any $d$-dimensional potential $V({\bf x})$ was obtained in Ref. \cite{fermions_review}, using the representation given in Eq. (\ref{K_beta}). In the case of a box potential as in Eq. (\ref{def_box_T}), it reads for ${\bf x}$ and ${\bf y}$ close to the center of the trap, in terms of our notation (see Appendix \ref{fin_T}) 
\begin{equation}\label{k_bulk_b_finite_2}
K_{\tilde \mu}({\bf x},{\bf y})\approx\frac{1}{\lambda_T^d} K_{d,b}^{\rm b}\left(\frac{|{\bf x}-{\bf y}|}{\lambda_T}\right)\;, \; K_{d,b}^{\rm b}({r})=\int_0^{\infty}\frac{\zeta dk}{\zeta+e^{\frac{k^2}{4\pi}}}\left(\frac{k}{2\pi}\right)^{\frac{d}{2}}\frac{\J_{\frac{d}{2}-1}\left(k\,r\right)}{{r}^{\frac{d}{2}-1}}\;,
\end{equation}
where we recall that $\zeta = e^{\beta \tilde \mu}$ and where $\lambda_T$ is the de Broglie wave length given in Eq.~(\ref{lambda_T}). In the limit $T \to 0$, one can show that this expression (\ref{k_bulk_b_finite_2}) crosses over the zero temperature bulk kernel in Eq. (\ref{bulk_kernel_d_intro}).

\begin{figure}[h]
\centering
\includegraphics[width=0.8\textwidth]{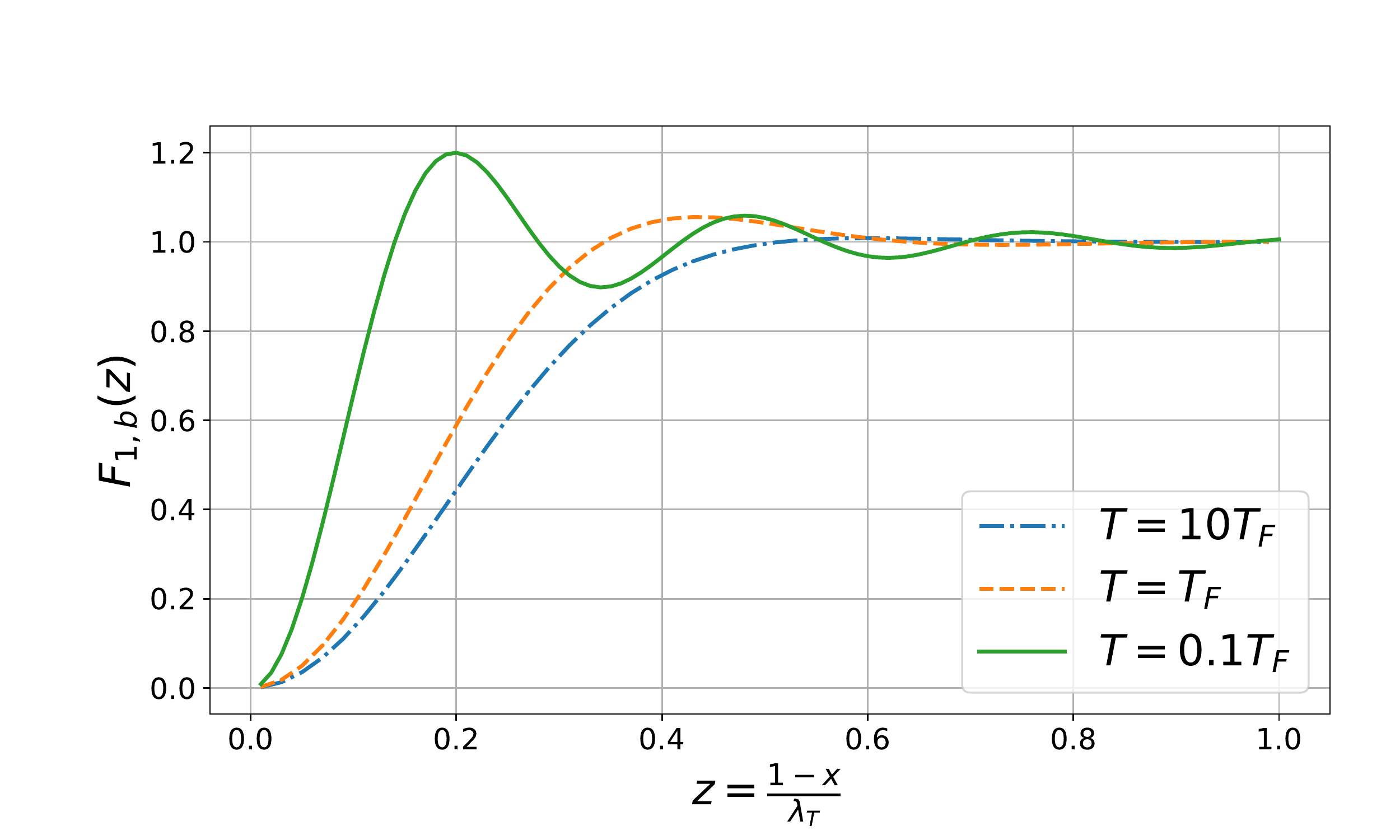}
\includegraphics[width=0.8\textwidth]{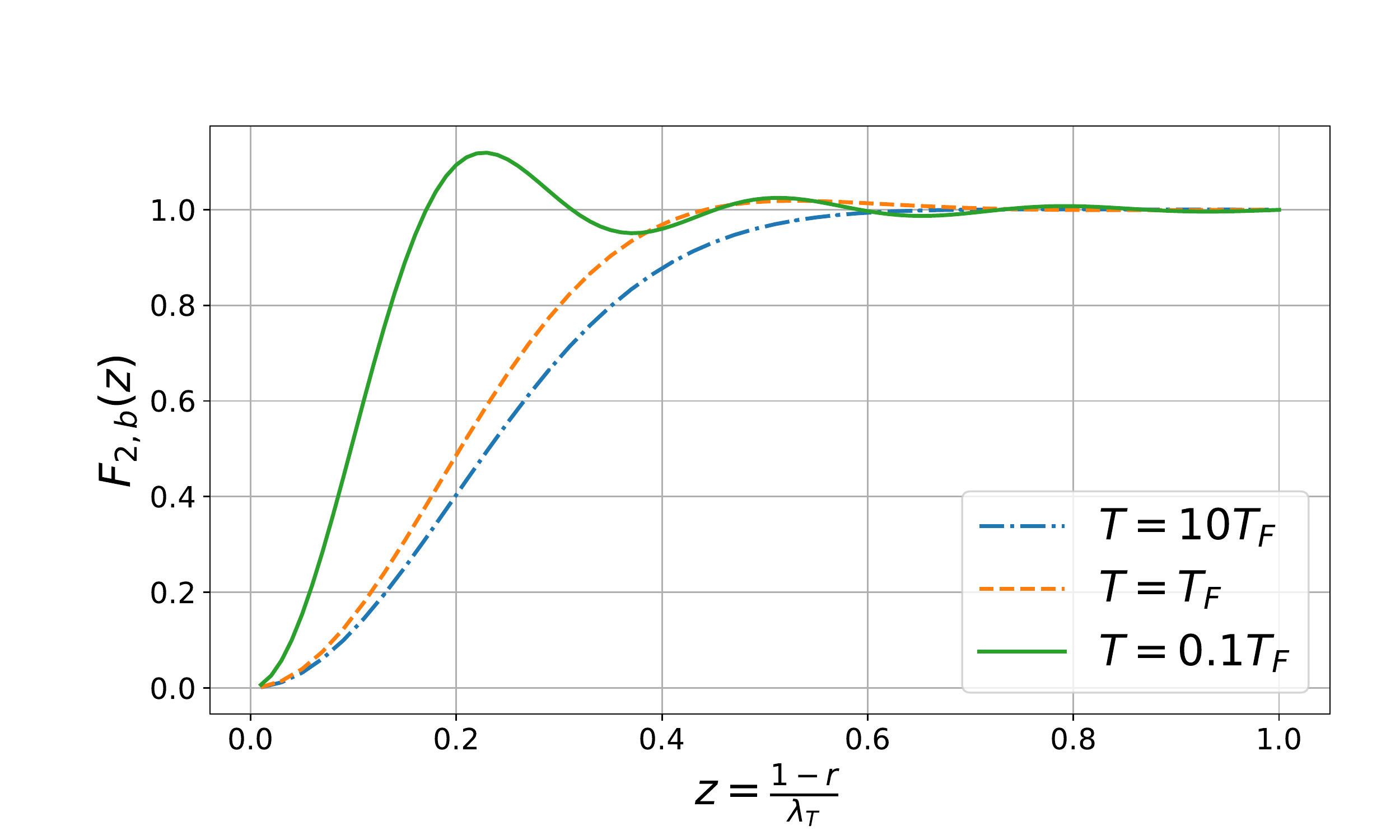}
\caption{Plot of the scaling function $F_{d,b}$ given in Eq. \eqref{F_d_b} for a $d$ dimensional hard box at finite temperature. It is plotted for $d=1$ (top) and $d=2$ (bottom) and for $T/T_F=0.1,1,10$ respectively in dashed-dotted blue, dashed orange and solid green lines. The oscillations, characteristic of the quantum behavior, get damped when increasing the temperature.}\label{fig_dens_t}
\end{figure}

We can now easily obtain the scaling form of the edge kernel $K_{\tilde \mu}({\bf x}, {\bf y})$ for ${\bf x}$ and ${\bf y}$ close to a point on the boundary ${\bf x}_w$ with $|{\bf x}_w| = R =1$ (\ref{def_box_T}). Indeed, at $T=0$, we have shown that the full edge kernel can be obtained from the bulk kernel combined with the image method [see \eqref{k_d_final} and Fig. \ref{fig_image}]. Now the linear relation between $K_{\tilde \mu}$ and $K_{\mu}$
in Eq. (\ref{K_beta}) shows that the image method also holds at finite $T$, which means that $K_{\tilde \mu}$ has the same structure as in Eq. (\ref{k_d_final}) where the $T=0$ bulk kernel $K_d^{\rm b}(z)$ is replaced by its finite $T>0$ generalisation $K_{d,b}^{\rm b}(z)$ given in (\ref{k_bulk_b_finite_2}). Therefore, at the edge, $K_{\tilde \mu}({\bf x}, {\bf y})$ takes the scaling form
\begin{align}\label{k_edge_b_finite_2}
K_{\tilde \mu}({\bf x},{\bf y})&\approx \frac{1}{\lambda_T^d} K_{d,b}^{\rm e}\left(\frac{{\bf x}_w-{\bf x}}{\lambda_T},\frac{{\bf x}_w-{\bf y}}{\lambda_T}\right)\;\\
 \;{\rm with}\;\; K_{d,b}^{\rm e}({\bf u},{\bf v}) &= K_{d,b}^{\rm b}(|{\bf u}-{\bf v}|) -   K_{d,b}^{\rm b}(|{\bf u}-{\bf v}^T|) \nn.
\end{align}  
where $K_{d,b}^{\rm b}(r)$ is given in Eq. (\ref{k_bulk_b_finite_2}). 
From the edge kernel (\ref{k_edge_b_finite_2}) at coinciding point, we find that the density profile $\tilde \rho({\bf x}) = K_{\tilde \mu}({\bf x}, {\bf x})$, close to a point on the boundary at ${\bf x}_w$, takes the scaling form (in the large $N$ limit)
\begin{align}
&\tilde \rho({\bf x})\approx\frac{N}{\Omega_d}F_{d,b}\left(\frac{{\sf d}}{\lambda_T}\right)\;,\label{F_d_b}\nn\\
&{\rm with}\;\;F_{d,b}(z)=1+\frac{1}{\Li_{\frac{d}{2}}(-\zeta)}\int_0^{\infty}\frac{\zeta dk}{\zeta+e^{\frac{k^2}{4\pi}}}\left(\frac{k}{2\pi}\right)^{\frac{d}{2}}\frac{\J_{\frac{d}{2}-1}(2k\, z)}{(2z)^{\frac{d}{2}-1}}\;,
\end{align}
where ${\sf d}$ is the distance of ${\bf x}$ from the boundary (and $z = {\sf d}/\lambda_T$ is the scaled distance). In Eq. (\ref{F_d_b}), we have used the expression of $\lambda_T$ in Eq. (\ref{b_length}) as well as well the relation in Eq. (\ref{zeta_T}). The asymptotic behaviors of this scaling function $F_{d,b}(z)$ in (\ref{F_d_b}) are
\be
F_{d,b}(z)\sim\begin{cases}\label{F_d_b_as}
& \frac{\displaystyle\Li_{\frac{d}{2}+1}(-\zeta)}{\displaystyle\Li_{\frac{d}{2}}(-\zeta)}4\pi z^2+O(z^4)\;,\;\;z\ll 1\\
&\\
&1+O(z^{-\frac{d-1}{2}})\;,\;\; z\gg 1\;.
\end{cases}
\ee
{where $\zeta$ is determined from \eqref{zeta_T}.}
From the first line of Eq. \eqref{F_d_b_as}, we obtain that the density vanishes quadratically near the wall for finite $T$, as in the $T=0$ case [see Eq. \eqref{F_d_as}]. At high temperature $T\gg T_F$, using the first line of Eq. \eqref{Li_as} as $\zeta\ll1$, the scaling function reads $F_{d,b}(z)\approx 4\pi z^2$ for $z\ll 1$. As $\lambda_T\to 0$ this region gets very narrow and the density is constant in nearly all the domain. On the other hand, for $T\ll T_F$, using the second line of Eq. \eqref{Li_as} with $\zeta\approx e^{b\tilde \mu}\gg 1$, one obtains $F_{d,b}(z)\approx 8\pi b z^2/(d+2)$ for $z\ll 1$. Using additionally that $b=(k_F \lambda_T)^2/(4\pi)$ from Eq. (\ref{b_length}) and $z = {\sf d}/\lambda_T$, we recover the zero-temperature scaling form of the density [see the first line of Eq. (\ref{F_d_as}) where $z = {\sf d}\, k_F$. The scaling function $F_{d,b}(z)$ is plotted in Fig. \ref{fig_dens_t} for $d=1,2$ and for different values of $b=T_F/T$.

It would be interesting to study the crossover from hard to soft edges, in the presence of a non-uniform smooth potential, at finite temperature and investigate how the relevant temperature scale crosses over from $T \sim T_F\sim N^{2/d}$ to $T \sim N^{1/(3d)}$ (in the case of the harmonic potential). It can be done rather straightforwardly using the formula in Eq. (\ref{K_beta}) together with the $T=0$ crossover kernel for $d=1$ given in Eqs. (\ref{scaling_K_1_l}) and \eqref{K_1_l_final} and for $d\geq 1$ in Eq. \eqref{K_ell_d}.

 This chapter concludes our study of the non-interacting Fermi gas in a spherical hard box potential. We will now turn to the study of power law potentials in one dimension $V(x)\sim |x|^{-\gamma}$, with $\gamma>0$. At variance with the hard box potential, they are continuous but have an algebraic diverging singularity at the origin.   





\section{Power law potentials in one dimension}\label{sing_pot}

Until now we have studied potentials which explicitly contain a (infinite) hard wall.
This hard wall is an ideal model, leading to the new universality class, which we have studied in detail. But 
one may ask whether this class can be realized in a smooth potential mimicking
a hard wall, e.g. with a divergent singularity at $x=0$. We consider here the potentials
of the form $V(x) \sim 1/|x|^\gamma$ for $\gamma>0$. 

\subsection{$V(x)\sim 1/|x|^{\gamma}$ potentials in one dimension with $0< \gamma<2$.}

Here we study potentials of the form 
\be\label{def_V_singular}
V(x) = \q\frac{\alpha(\alpha+1)}{|x|^\gamma}  \quad , \quad x \neq 0 \;,
\ee 
with $0< \gamma<2$ and $\alpha > 0$.
For a single particle the properties of the eigenstates $\phi(x)$ have been studied in 
\cite{Andrews}. They are solutions of the Schr\"odinger equation $-\q\partial_x^2 \phi(x) + V(x)\,\phi(x) = \varepsilon \phi(x)$ with $\varepsilon \geq 0$ and finite. The main idea is thus that the expectation value of $V(x)$,
$\int dx |\phi(x)|^2 V(x)$, must be finite for any eigenstate $\phi(x)$. As a result 
two different cases can occur:
\begin{enumerate}

\item $0<\gamma<1$: the
barrier is penetrable, i.e. there exists eigenstates which do not vanish at
$x=0$. This is confirmed by an exact solution for $\gamma=1/2$
\cite{Ishkhanyan,Ishkhanyan2}.

\item $1 \leq \gamma<2$: the
barrier is impenetrable, i.e. the eigenfunctions vanish at $x=0$. 
This is confirmed by an exact solution for $\gamma=1$ \cite{Perret}, which we also study below.

\end{enumerate} 
In the first case, $0<\gamma<1$, the potential cannot act as a trap for
a system of $N$ fermions, hence we will not study it here. From now on, we specialize to $1 \leq \gamma<2$. In this case with no loss of 
generality, one
can restrict to $x>0$ and assume that $V(x)=+\infty$ for
$x<0$. 

Let us first study the case $\gamma=1$, for which exact solutions exist.
We set for simplicity $\alpha(\alpha+1)=1$, and study the potential 
\be
V(x) = \frac{\hbar^2}{2m x}  \quad , \quad x >0  \;.
\ee 
Consider first the one particle problem. We study the Schr\"odinger equation for an eigenstate
at energy $E=\q k^2$
\bea \label{schrod1} 
\partial_x^2\phi_k(x) + \left(k^2-\frac{1}{x}\right) \phi_k(x) = 0\;.
\eea 
The spectrum is continuous with $E>0$ and the eigenfunctions with a {\it finite} average potential energy read \cite{Perret}
\be\label{phi_k}
\phi_k(x) = c_k \, x \, e^{- i k x}  \, _1F_1(1- \frac{i}{2k},2,2 ik x ) \;,
\ee
where $_1F_1(a,b,z)$ is the confluent hypergeometric function. One can check that $\phi_k(x)$ in (\ref{phi_k}) is real for $E=\q k^2>0$. 
Note that there is a second solution to the Schr\"odinger equation (\ref{schrod1}), however it has infinite average potential energy and hence it must be discarded \cite{Andrews}. The normalisation factor in (\ref{phi_k}) reads \cite{Perret} 
\be
|c_k|^2 = \frac{2k}{e^{\frac{\pi}{k}}-1} \;,
\ee
which ensures the orthonormality of the wave functions in the continuum
\be
\int_0^{+\infty} dx \phi_k^*(x) \phi_{k'}(x) = \delta(k-k')  \;.
\ee 

Consider now the problem of non-interacting fermions at zero temperature
and Fermi energy $\mu=\q k_F^2$. The exact formula for the kernel reads
\bea
\fl   K_\mu(x,y) &=& \int_0^{k_F} dk \, \phi_k^*(x) \phi_{k}(y)  = \int_0^{k_F} dk \, \phi_k(x) \phi_{k}(y) = k_F K(k_F x ,k_F y;k_F) \\
\fl  K(u, v;k_F) &=&  \frac{2u v}{k_F} \int_0^{1} \frac{t dt}{e^{\frac{\pi}{t k_F}}-1} 
e^{-it(u+v)} \nn \\
\fl && \times  \, _1F_1(1- \frac{i}{2 t k_F },2,2 i t u)
_1F_1(1- \frac{i}{2 t k_F},2,2 i t v) \label{exact_1overx}
\eea
obtained by performing the change of variables $k \to t=k/k_F$. From this expression for the kernel (\ref{exact_1overx}) one obtains an exact expression for the density $\tilde \rho(x) = K_{\mu}(x,x)$.  
\begin{figure}[t]
\centering\includegraphics[width = 0.7 \linewidth]{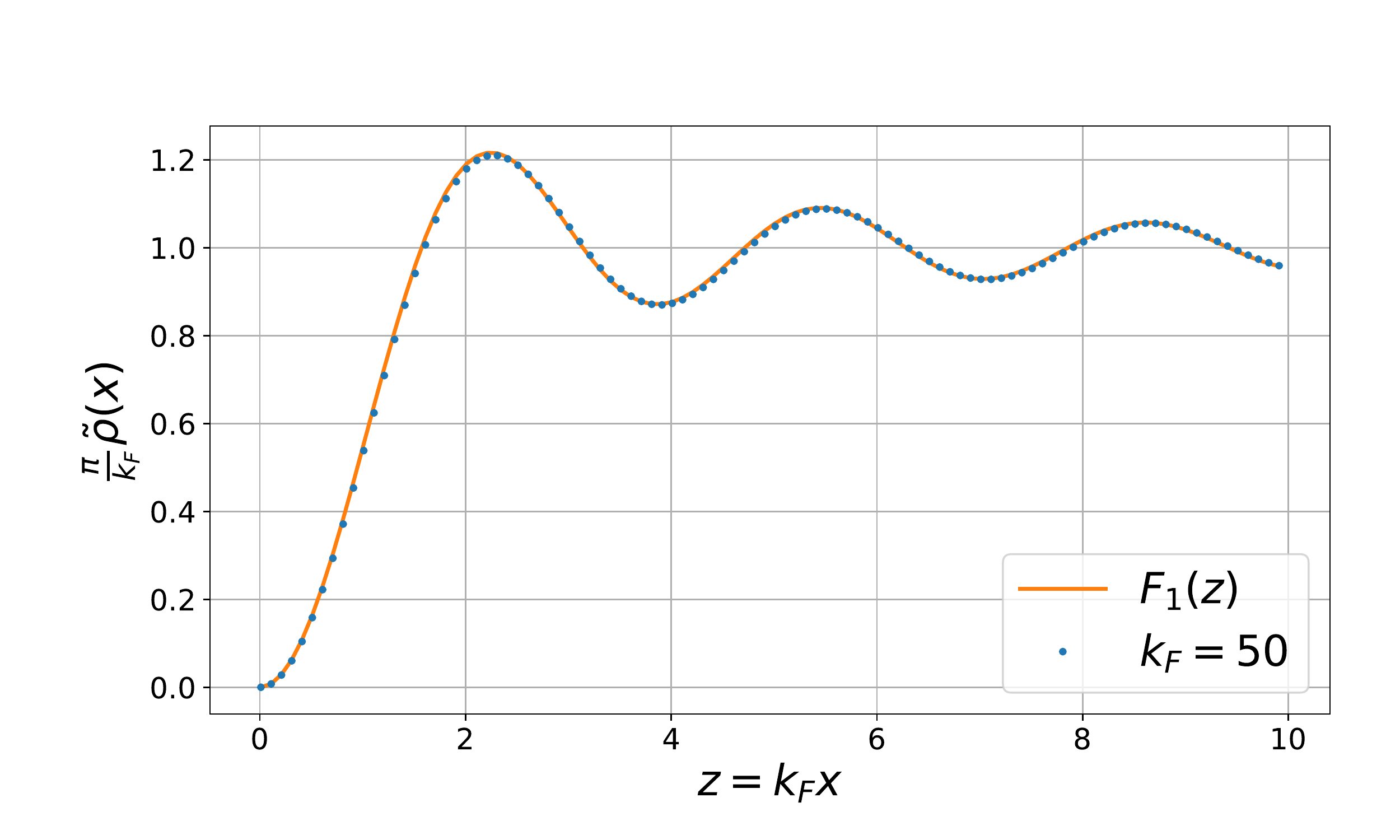}
\caption{Plot of the density $\tilde \rho(x)$ for the potential $V(x) \sim1/x$ as a function of $z=k_F\,x$. The dots correspond to a numerical evaluation of the exact formula obtained from (\ref{exact_1overx}) with $k_F=50$, while the solid line is the limiting hard wall density with $F_1(z)$ given in Eq. (\ref{density_1d_hwall}).}\label{plot:dens_1ox}
\end{figure}
In Fig. \ref{plot:dens_1ox} we show a plot of the exact density for $k_F = 50$. 
Its behavior at small $x$ and fixed $\mu$ is given by
\be \label{smallx}
\tilde \rho(x)= K_\mu(x,x) = C_\mu  \left(x^2 +x^3\right) + O(x^4) 
\ee
where 
\be
C_\mu= \int_0^{k_F} \frac{2t dt}{e^{\frac{\pi}{t}}-1} \underset{k_F\gg 1} {\simeq}\frac{2}{3 \pi} k_F^3 \;.
\ee 
The result \eqref{smallx} matches exactly the small $x$ result given in \eqref{edge_dens_hb_1d}-\eqref{dens_edge_hb_1d_as}, i.e.,
\be
\tilde \rho(x) \simeq \frac{k_F}{\pi}F_1(k_F x)\;\;{\rm with}\;\;F_1(z)\sim\frac{2}{3}z^2\;,\;\;z\ll1\;. 
\ee 

We now analyse this formula for the kernel (\ref{exact_1overx}) in the limit of large $\mu=\q k_F^2$. Using the behavior of the hypergeometric function 
\be
_1F_1(1- \frac{i}{\sqrt{2t k_F }},2,2 i u t) \underset{k_F \to \infty}{\to}  \, _1F_1(1,2,2 i u t) = \frac{1}{2 i u t} (e^{2  i u t}-1)
\ee
in Eq. (\ref{exact_1overx}), we obtain straightforwardly	
\be
K(u, v;\infty) =
\frac{2}{\pi} \int_0^1 dt \sin(u t) \sin(v t) = \frac{\sin (u-v)}{\pi(u-v)} - \frac{\sin (u+v)}{\pi(u+v)} =  K_1^{\rm e}(u,v) \;,
\ee 
where, in the last equality, we have used the expression of the hard box kernel $K_1^{\rm e}(u,v)$ given in \eqref{khb_1d_e}. In particular, as in the pure hard box case, the fermion density close to the origin is given, for large $\mu$, by $\tilde \rho(x) \approx (k_F/\pi) F_1(k_F\,x)$ where $F_1(z)$ is given in  Eq.~(\ref{density_1d_hwall}) (see Fig.~\ref{plot:dens_1ox}).

Let us now turn to the more general case of $V(x) \sim x^{-\gamma}$ (\ref{def_V_singular}) with $1 \leq \gamma < 2$.
As it can be seen from the formula for the fermion density (\ref{rho_LDA}), the position of the edge density is at 
$r_{\rm e}= k_F^{-2/\gamma}$, with $\mu=\q k_F^2$. Hence for $\gamma <2$ the scaled position of the edge, $k_F r_{\rm e}$
tends to zero at large $\mu$. This is consistent with our exact result for $\gamma=1$ 
which recovers the hard box with a wall at $x=0$, in this scaled variable. 
It strongly suggests that the same conclusion will hold for any $1 \leq \gamma < 2$.
A further argument is to consider again the Schr\"odinger equation \eqref{schrod1} for an eigenstate
at energy $E=\q k^2$, in the rescaled coordinate $u=k_F x$, with $q=k/k_F$
($0<q<1$) and for arbitrary $\gamma$. Denoting $ \phi_k(x)= \psi_{q}(u)$ the Schr\"odinger equation \eqref{schrod1}, written
in terms of rescaled variables, reads
\bea\label{schrod_rescaled}
\partial_u^2 \psi_q(u)+\left(q^2- 
\frac{ k_F^{\gamma-2}}{u^\gamma}\right) \psi_q(u) =0 \;.
\eea 
Again we see that, for large $k_F$, if $\gamma<2$ we can neglect the potential term for 
any fixed $u$. However for any $k_F$, one should remember that
the wave-function must vanish at $u=0$. Hence the problem
of non-interacting fermions with large $\mu=\q k_F^2$ near the origin becomes equivalent to
the problem of non-interacting fermions near a hard wall, as in the pure hard box potential. 

From these arguments we also see that we expect some new behavior for 
$\gamma \geq 2$. We now turn to this case, distinguishing the marginal case 
$\gamma=2$ (for which there is a continuously varying family of scaling forms
for the kernel, related to the hard-edge universality class in RMT), and
$\gamma>2$. 

\subsection{$V(x)\sim 1/x^2$ potential in one dimension and the Bessel kernel}\label{sec:bessel}

\subsubsection{Edge kernel at zero temperature}  \hfill\\

Let us now consider the one-dimensional case of a repulsive $1/x^2$ potential of the form
\be\label{defValpha}
V_\alpha(x)=\begin{cases}
&+\infty\;,\;\;x<0\;,\\
&\\
&\q \frac{\alpha(\alpha+1)}{x^2}\;,\;\; x \geq 0
\end{cases}
\ee
with $\alpha>0$ (note that one can also consider an additional quadratic potential $V(x) = V_\alpha(x) + c\, x^2$ with $c>0$, which is exactly solvable~\cite{Nadal_Majumdar}, see also Appendix \ref{App:Jacobi}). For such a potential (\ref{defValpha}), the spectrum is continuous. Denoting the eigenvalues by 
$E= \frac{\hbar^2 k^2}{2 m}$, the corresponding eigenfunctions, denoted $\phi_k(x)$
are given by 
\bea
\phi_k(x) = \sqrt{k x} \J_{\alpha + \frac{1}{2}}(k x) 
\eea 
where $k>0$ is a continuum quantum number. These eigenfunctions satisfy the 
continuum orthonormality condition \cite{Bessel_delta}
\bea \label{phi_alpha}
\int_0^{+\infty} dx \, \phi_k(x)  \phi_{k'}(x)  
= \delta(k-k') \;.
\eea 
In the case $\alpha=0$ one recovers, using $\J_{1/2}(x) = \sqrt{\frac{2}{\pi x}} \sin x$,
the single particle eigenfunctions for the hard wall potential
\bea
\phi_k(x) = \sqrt{\frac{2}{\pi}} \sin(k x) \;.
\eea 
Note also that similar eigenfunctions appear in \eqref{radial_wf} for the problem of the spherical hard box in
$d$ dimensions which reduces to a one-dimensional radial Schr\"odinger equation with an effective
$1/r^2$ potential \eqref{effective_potential} and an amplitude corresponding to $\alpha=l+(d-3)/2$, where $l$ is the orbital quantum number. 

We now obtain the correlation kernel $K_\mu(x,y)$ for noninteracting fermions at zero temperature with Fermi energy $\mu$, by filling all eigenstates with one fermion up to energy $\mu= \frac{\hbar^2}{2 m} k_F^2$. It turns out that it can be obtained exactly for arbitrary value of $\mu$. Indeed we have
\be
K_\mu(x,y)= \int_0^{+\infty} dk \Theta(k_F-k) \phi_k(x) \phi_k(y) 
= \sqrt{x y} \int_0^{k_F} dk \,k \, \J_{\alpha + \frac{1}{2}}(k x) 
\J_{\alpha + \frac{1}{2}}(k y)  \;.
\ee
Computing explicitly this integral \cite{Bessel_k} we obtain
\be\label{Kmu_bessel}
K_\mu(x,y)= \frac{ k_F \sqrt{x y} }{x^2-y^2} \left(x \J_{\alpha + \frac{3}{2}}(k_F x) \J_{\alpha + \frac{1}{2}}(k_F y) 
- y \J_{\alpha + \frac{1}{2}}(k_F x) \J_{\alpha + \frac{3}{2}}(k_F y) \right)
\ee 
which can also be expressed in terms of the so-called Bessel kernel \cite{TW94_Bessel}, defined as
\be\label{K_bessel}
K_{\rm Be,\nu}(u,v)= \frac{ \sqrt{v} \J'_{\nu}(\sqrt{v})\J_{\nu}(\sqrt{u}) 
- \sqrt{u} \J'_{\nu}(\sqrt{u}) \J_{\nu}(\sqrt{v}) }{2 (u-v)} 
\ee
with the value at coinciding points
\be
K_{\rm Be,\nu}(u,u)= \frac{1}{4} \left(  \J_{\nu}(\sqrt{u})^2 - \J_{\nu+1}(\sqrt{u}) \J_{\nu-1}(\sqrt{u}) \right)
\ee 
Note that the Bessel kernel can also be written as
\be
K_{\rm Be,\nu}(u,v)= \frac{1}{4} \int_0^1 dz \,  \J_{\nu+1}(\sqrt{u z}) \J_{\nu+1}(\sqrt{v z})  \;.
\ee
By comparing (\ref{Kmu_bessel}) and (\ref{K_bessel}), we obtain
\bea \label{KmuBe} 
K_\mu(x,y)= 2 k_F^2 \sqrt{x y} K_{\rm Be,\alpha+1/2}(k_F^2 x^2 , k_F^2 y^2) \;.
\eea
Note that the Bessel kernel is self-reproducing
\be\label{reprod_bessel}
\int db K_{\rm Be,\nu}(a,b) K_{\rm Be,\nu}(b,c) = K_{\rm Be,\nu}(a,c)
\ee 
which immediately implies from \eqref{KmuBe} that the kernel $K_\mu(x,y)$ in Eq. (\ref{KmuBe}) is
also self-reproducing as required. This Bessel kernel is characteristic of hard edges in RMT. An example of occurrence of this kernel is the case of Laguerre Unitary Ensemble where the joint probability of eigenvalues reads
\be
P_{\rm joint}(\lambda_1,\cdots,\lambda_N)=\frac{1}{Z_N}\prod_{i<j}|\lambda_i-\lambda_j|^{2}\prod_{k=1}^{N}\lambda_k^{\nu} e^{-N \lambda_k}\;.
\ee
The mean density of eigenvalues converges for large $N$ to the Mar\v cenko-Pastur distribution \cite{MP}
\be
\rho(\lambda)=\frac{1}{N}\moy{\sum_{k=1}^N \delta(\lambda-\lambda_k)}\to\rho_{MP}(\lambda)=\frac{1}{2\pi}\sqrt{\frac{4-\lambda}{\lambda}}
\ee 
which diverges  close to the origin, creating a so called 'hard edge'. Rescaling the kernel close to this hard edge we obtain precisely the Bessel kernel in Eq. \eqref{K_bessel}.

Note that it is also possible to use the representation of the kernel in terms of the Euclidean quantum propagator
$G$ [see Eq. \eqref{kernel_propagator_1d}]. For the present Schr\"odinger problem with a $1/x^2$ potential it reads 
\bea\label{exact_propag}
\fl && G(x,y,t)= \int_0^{+\infty} dk \, \phi_k(x) \phi_k(y) e^{- \frac{\hbar^2 k^2}{2 m} t} = 
\sqrt{x y} \int_0^{+\infty} dk \, k \, \J_{\alpha + \frac{1}{2}}(k x) 
\J_{\alpha + \frac{1}{2}}(k y) e^{- \frac{\hbar^2 k^2}{2 m} t} \nonumber \\
\fl && ~~~~~~~~~~ = \frac{\sqrt{x y}}{2 \tilde t} {\rm I}_{\alpha + 1/2}\left( \frac{x y}{2 \tilde t}\right) e^{- \frac{x^2 + y^2}{4 \tilde t}}
\quad , \quad \tilde t = \frac{\hbar^2}{2 m} t \;,
\eea
where ${\rm I}_\nu(x)$ is the modified Bessel function of index ${\nu}$. Using the formula \eqref{kernel_propagator_1d} and this exact expression for the propagator (\ref{exact_propag}) one can also obtain \eqref{KmuBe}.

\subsubsection{Results for the density} \hfill\\

\begin{figure}[h]
\centering
\includegraphics[width=0.8\textwidth]{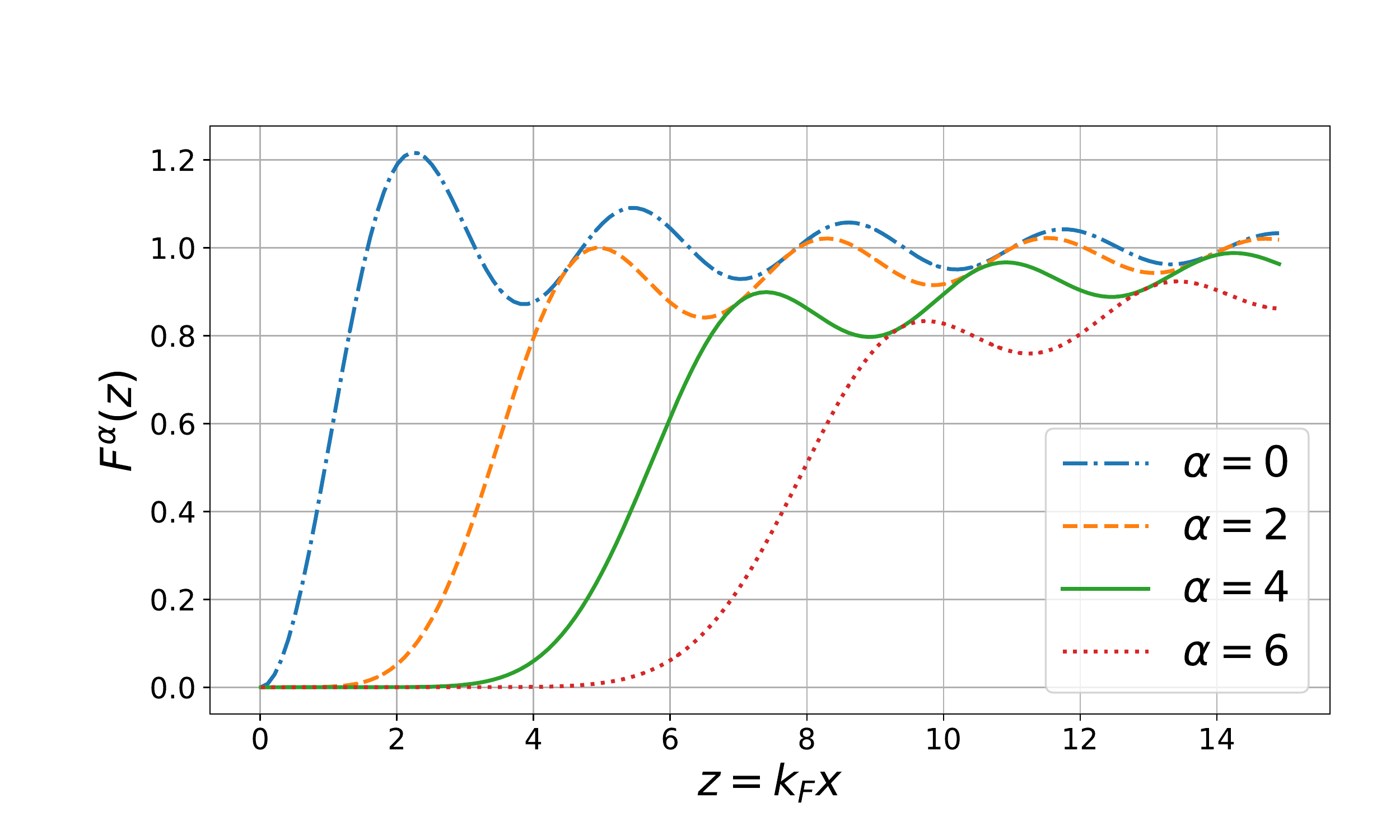}
\caption{Plot of $F^{\alpha}(z)$ as a function of the rescaled position $z=k_F x$ given in Eq. \eqref{rho_alpha} for $\alpha=0,2,4,6$ respectively in dahed-dotted blue, dashed orange, solid green and dotted red lines. As $\alpha$ increases, the density exhibits a more and more pronounced pseudo-gap close to the singularity of the potential at $z=0$.}\label{fig_dens_alpha}
\end{figure}

From the kernel \eqref{KmuBe} we obtain the edge density near the wall as
\be
\tilde \rho(x) 
= \frac{k_F}{\pi} F^{\alpha}(k_F x)\;\;{\rm with}\;\;F^{\alpha}(z)=\frac{\pi z}{2}  \left(  \J_{\alpha+1/2}(z)^2 - \J_{\alpha+3/2}(z) \J_{\alpha-1/2}(z) \right) \;. \label{rho_alpha}
\ee
In particular, for $\alpha =0$, one recovers the density for the hard wall (\ref{density_1d_hwall})
\be
F^{\alpha=0}(z)=F_1(z)=1-\frac{\sin(2z)}{2z} \;.
\ee 
For any $\alpha$ the density converges to its bulk value $k_F/\pi$ for
large $z$ with oscillations 
\be
F^{\alpha}(z) =1  - \frac{\sin (2 z - \pi  \alpha)}{2 z}
-\frac{\alpha (\alpha+1) \cos ^2\left(z-\frac{\pi 
   \alpha}{2}\right)}{z^2} + O\left(\frac{1}{z^4}\right)  \;.
\ee 
Near the origin, the scaling function for the density vanishes as
\be
F^{\alpha}(z) = \frac{2\pi}{4^{1+\alpha} (3 + 2 \alpha) \Gamma(\frac{3}{2} + \alpha)^2} z^{2 + 2 \alpha} 
 -\frac{ 2\pi(2 \alpha +3) (2 \alpha +5)
   z^{2 \alpha +4}}{4^{ \alpha+3}\Gamma \left(\alpha
   +\frac{7}{2}\right)^2} + O(z^{4 + 2 \alpha}) 
\ee 
which agrees with the formula \eqref{dens_edge_hb_1d_as} for $\alpha=0$. In Fig. \ref{fig_dens_alpha} we show a plot of $F^\alpha(z)$ as a function of $z$ and different values of $\alpha$. 
As $\alpha$ increases, we see that a pseudo gap opens more and more near the origin.

We can compare these exact results with the semi-classical (i.e. LDA) formula for the density (\ref{rho_LDA}) 
$\tilde \rho(x)= \frac{1}{\pi \hbar} \sqrt{2 m(\mu- V(x))}$ for $x>r_{\rm e}$
where the edge is by definition
\be
V(r_{\rm e})= \mu \quad \Leftrightarrow \quad r_{\rm e} = k_F^{-1}\sqrt{\alpha(\alpha+1)}  \;.
\ee 
Hence on the scale of the bulk one can consider that $r_{\rm e} \approx 0$ as
$\mu \to +\infty$. All the results above are clearly explicit scaling functions
of the rescaled coordinate $k_F x$.\\

\subsubsection{Large $\alpha$ limit} \hfill \\

In that limit we will use the
convergence of the Bessel kernel towards the Airy kernel, i.e.
\be \label{convergence} 
\lim_{\nu \to +\infty} 2^{2/3} \nu^{4/3} K_{\rm Be,\nu}(\nu^2 + 2^{2/3} \nu^{4/3} \tilde a,
\nu^2 + 2^{2/3} \nu^{4/3} \tilde b) = K_{\Ai}(- \tilde a , - \tilde b)
\ee 
which can be obtained from the relation for Bessel functions $\J_\nu$ of large index $\nu$ \cite{Bessel_Airy}
 \be
 (\nu/2)^{1/3}\J_\nu\left[\nu+(\nu/2)^{1/3}a\right]=\Ai(-a)+O(\nu^{-2/3})\;.
 \ee
For large $\alpha$ we have $k_F r_{\rm e} \simeq \alpha$. Hence we will center
the kernel around this point and define
\be
x= r_{\rm e} + w_N \tilde x \quad , \quad y= r_{\rm e} + w_N \tilde y
\ee 
with $k_F w_N= 2^{-1/3} \alpha^{1/3}$. One can then check that the convergence
property \eqref{convergence} leads to the behaviour of the kernel close to the edge $|x-r_{\rm e}|/w_N\sim |y-r_{\rm e}|/w_N=O(1)$. This yields
 \be\label{airy_k_largea}
 K_{\mu}(x,y)=\frac{1}{w_N}K_{1}^{\rm soft}\left(\frac{r_{\rm e}-x}{w_N},\frac{r_{\rm e}-y}{w_N}\right)\;\;{\rm with}\;\;K_{1}^{\rm soft}(u,v)=\int_0^{\infty}dz \Ai(z+u)\Ai(z+v)\;.
 \ee
This is the celebrated Airy kernel that was already obtained for the edge behavior in smooth potentials (\ref{airy_k}). 
 
\subsubsection{Finite temperature} \hfill \\

To obtain the kernel at finite temperature $K_{\tilde \mu}$ in the grand canonical ensemble with
chemical potential $\tilde \mu$ we use the relation already discussed above 
\eqref{K_beta}. It relates the finite temperature kernel $K_{\tilde \mu}$ with the zero temperature 
kernel $K_\mu$ (\ref{KmuBe}) calculated above and reads
\be
K_{\tilde \mu}(x,y) = \int_0^{+\infty} \frac{d \mu'}{1 + e^{\beta( \mu'-\tilde \mu)}} 
\partial_{\mu'} K_{\mu'}(x,y) \;.
\ee 
Let us define $k_F'$ and $\tilde k_F$ via the relations
$\mu'= \frac{\hbar^2}{2 m} (k_F')^2$ and 
$\tilde \mu= \frac{\hbar^2}{2 m} (\tilde k_F)^2$. One defines the
scaled inverse temperature parameter
\be \label{defbBessel} 
b = \frac{\beta \hbar^2}{2 m} \tilde k_F^2 \quad , \quad k'_F = v \tilde k_F 
\ee 
This leads to 
\be
K_{\tilde \mu}(x,y) =  k_F K_{b}(k_F x, k_F y)
\ee 
with
\be \label{finiteTBessel} 
K_{b}(x, y) = \sqrt{x y} \int_0^{+\infty} \frac{dv}{ 1 + e^{b (v^2-1)}} 
\J_{\alpha + \frac{1}{2}}(v x) 
\J_{\alpha + \frac{1}{2}}(v y) 
\ee 
which converges to $K_\mu(x,y)$ given in \eqref{Kmu_bessel} at zero temperature, i.e. $b \to \infty$. 

Note that if we consider the potential $V(x) \sim 1/x^2$ on the real line (i.e. for all $x \in \mathbb{R}$, $x \neq 0$)
it acts as an impenetrable barrier, and the problem is identical (on each side $x<0$ and $x>0$) to the 
problem studied here. Let us recall that in this section we made no approximation, i.e. all formulae 
are exact for the $1/x^2$ potential.

\subsection{The case $\gamma \geq 2$: Airy universality class}

In the previous section we saw from Eq. \eqref{airy_k_largea} that if the amplitude of the $1/x^2$ potential
becomes large one recovers the Airy universality class, which was obtained for smooth traps \cite{us_finiteT,fermions_review}. In fact, it turns out that the Airy class
is recovered for power law potentials with $\gamma>2$. 
\begin{figure}[t]
\centering
\includegraphics[width = 0.5\linewidth]{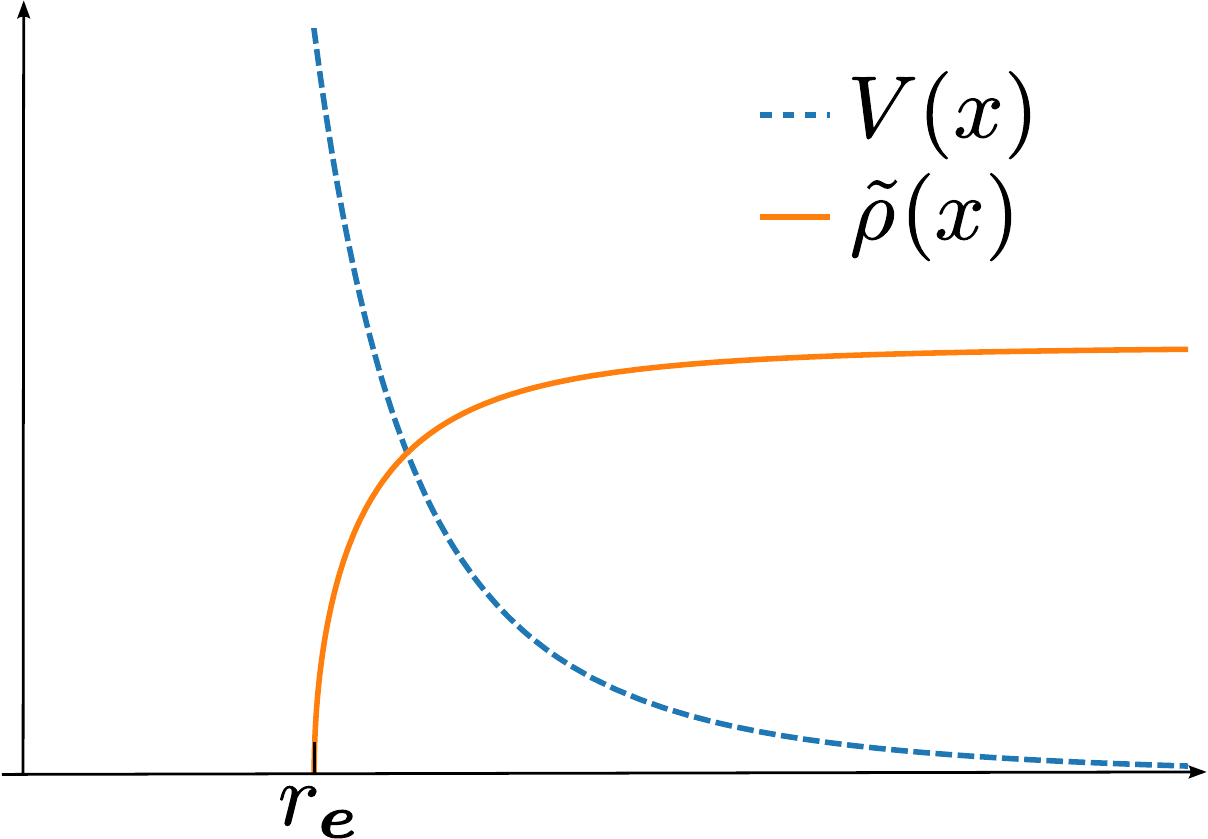}
\caption{Sketch of the density (orange) in the presence of a potential $V(x) \sim |x|^{-\gamma}$ with $\gamma > 2$. In this case, the potential is sufficiently repulsive to push the density away from $x=0$, thus creating a gap on the interval $[0,r_e]$ where the density vanishes. Close to $r_e$, the density vanishes as a square root $\tilde \rho(x) \sim (r_e - x)^{1/2}$ and, as for a soft edge, the liming kernel is given by the Airy kernel (\ref{airy_k}).}\label{fig_1_x3}
\end{figure}

Although it appears
somewhat non intuitive, fast increasing potentials lead to the same 
universality class as smooth traps (see Fig. \ref{fig_1_x3}) ! We will not rederive this property in detail 
here, and refer to the Appendix A in \cite{fermions_review} [see the discussion in the two paragraphs 
below (A. 28)]. We recall these results. The kernel takes the form
\bea \label{Kmu_largeg}
K_\mu(x,y)  \simeq \frac{1}{w_N} K_1^{\rm soft}\left(\frac{x-r_{\rm e}}{w_N},\frac{y-r_{\rm e}}{w_N}\right) \;,
\eea
where $K_1^{\rm soft}(u,v)$ is the Airy kernel (\ref{airy_k}). In Eq. (\ref{Kmu_largeg}), one has 
$V(r_{\rm e})= \mu$, i.e. $r_{\rm e} \sim \mu^{-1/\gamma}$ and 
$w_N \sim \mu^{- \frac{1+\gamma}{3 \gamma}}\ll r_{\rm e}$ is the width of the edge regime. 
It matches correctly the result for large $\alpha$ found in the previous 
section (for $\gamma=2$) in \eqref{airy_k_largea}.


\section{Conclusion}

In this paper, we have computed the spatial correlations of $N$ non-interacting spin-less fermions in any dimension $d$ and at any finite temperature $T$ confined in a hard box potential, i.e. $V({\bf x})=+\infty$ outside of some domain ${\cal D}$ {(the box)}. For any finite $N$, the positions of the fermions form a $d$-dimensional determinantal point process at zero temperature as well as at finite temperature (in the latter case, only in the grand-canonical ensemble). We have mainly focused on the correlation kernel, from which any $n$-point correlation function can be computed. This correlation kernel, and its limiting form in the large $N$ limit, were previously obtained \cite{us_finiteT,DPMS:2015,fermions_review,Eis2013} in the case of spherically symmetric smooth potentials, e.g. $V({\bf x}) \propto |{\bf x}|^{p}$ with $p>0$, which create ``soft edges'' in the density.
{In this paper, we have extended these results to the case of ``hard edges'' where the potential imposes that the density of fermions vanishes exactly on the (smooth) boundary of the domain $\partial {\cal D}$. This can
be achieved in a variety of ways, leading to universal kernels related, in $d=1$, to the
one parameter family of "hard-edge" kernels of RMT, and extending them in higher dimension. 
In the case where the potential is infinite outside the box, and uniform inside, we have 
shown that the correlation kernel takes a universal scaling form (the hard box "hard edge"
kernel) close to the boundary $\partial {\cal D}$ [see e.g. Eqs. (\ref{khb_1d_e_intro}) and (\ref{k_d_final}) at $T=0$], which is well described by an image method.
For the same hard wall box potential, but in the presence of a smoothly varying potential inside the domain ${\cal D}$,
we have obtained a novel universal kernel which interpolates between the hard box "hard edge" kernel and the
"soft edge" Airy kernel [see Eqs. \eqref{K_1_l_final} and \eqref{K_ell_d}]. Another way to realize
a hard edge, which we have investigated only in $d=1$, is through singular potentials of the form $V(x)=x^{-\gamma}$ with $\gamma>1$. In this case, we have shown that the correlation kernel close to the singularity takes universal scaling forms, were the universality class depends on $\gamma$ (see e.g. Eqs. \eqref{K_bessel_intro} and \eqref{K_airy_intro}). In the marginal case $\gamma=2$ we found that the fermion correlations are exactly described by the Bessel kernel, giving thus a complete correspondence
with the full family of hard edge kernels of RMT. Rather suprisingly though, we have shown that if the potential is sufficiently singular, i.e. for $\gamma > 2$, the limiting edge kernel is the Airy kernel, which was actually first found for smooth potential. Finally, in Appendix C, we studied the Jacobi trap, which also
provides interpolations, though of a different kind, between hard and soft edge kernels. 
Thus the present work allows to obtain a rather complete theory for a wide class of ``non-smooth'' potentials that could be engineered in an actual cold atom experiment. In addition,
since most of the results were extended to finite temperature, a comparison with
Fermi gases in current experimental conditions becomes possible.}

It would be interesting to apply these results to the study of extreme value questions for the Fermi gas in non-smooth potentials (see \cite{LLMS17} for preliminary results), such as the position of the farthest fermion from the center of the trap, as recently done for smooth confining potentials \cite{farthest_f}. Another interesting question concerns the Wigner function for fermions in hard-box and how to generalise the recent results obtained for smooth potentials \cite{us_Wigner,wiegman}. More generally, we hope that this work will inspire both theoretical end experimental works to investigate further the spatial properties of Fermi gases.

\ack

We thank D. S. Dean for useful discussions and ongoing collaborations. We also thank F. D. Cunden, N. O'Connell and F. Mezzadri for interesting discussions. 
This research was partially supported by ANR grant ANR-17-CE30-0027-01 RaMaTraF.


\appendix

\section{Limits of $K_1^{\ell}(u,v)$}\label{limits}
In this section we will derive the asymptotic limits of the correlation kernel $K_1^{\ell}(u,v)$ for $\ell=\pm \infty$ in Eqs. \eqref{l+} and \eqref{l-}.

\subsection{Limit for $\ell\to +\infty$}

For $\ell=(1-r_{\rm e})/w_N\gg 1$ and $\ell>0$, i.e. $r_{\rm e}\ll 1$ (see blue case of Fig. \ref{Fig_distance}), the soft edge is encountered first. We change coordinates to rescale close to this soft edge by introducing $u=u'-\ell$ and $v=v'-\ell$. The scaling function at this soft edge is
\be\label{rescaling_+}
K_{1}^{\ell}(u'-\ell,v'-\ell)=\int_\ell^{\infty}\sigma(s,s+u'-\ell)\sigma(s,s+v'-\ell)ds=\int_0^{\infty}\sigma(s+\ell,s+u')\sigma(s+\ell,s+v')ds\;.
\ee
We then introduce the asymptotic expansion of the functions $\Ai(z)$ and $\Bi(z)$ for large positive argument \cite{Ai_large}
 \be\label{AiBi_+}
\Ai(z)\sim\frac{e^{-\frac{2}{3}z^{\frac{3}{2}}}}{2\sqrt{\pi}z^{\frac{1}{4}}}\;\;{\rm and}\;\;
\Bi(z)\sim\frac{e^{\frac{2}{3}z^{\frac{3}{2}}}}{\sqrt{\pi}z^{\frac{1}{4}}}\;\;{\rm for}\;\;z\to+\infty\;.
\ee
Inserting these asymptotic expansions in Eq. \eqref{sigma}, we obtain the asymptotic expansion
\be
\sigma(s+\ell,s+u)=\frac{\Bi(s+\ell)\Ai(s+u)-\Ai(s+\ell)\Bi(s+u)}{\sqrt{\Ai^2(s+\ell)+\Bi^2(s+\ell)}}\approx \Ai(s+u)\;\;{\rm for}\;\;\ell\to +\infty\;.
\ee
It yields for the scaling function 
\be\label{limit_l_+}
K_1^{\ell}(u'-\ell,v'-\ell)\approx \int_0^{\infty}ds \Ai(s+u')\Ai(s+v')=K_{1}^{\rm soft}(u',v')\;.
\ee
We obtain therefore exactly the soft edge scaling function when the wall is far beyond the soft edge.
\subsection{Limit for $\ell\to -\infty$}

For $\ell=(1-r_{\rm e})/w_N\gg 1$ and $\ell<0$, i.e. $r_{\rm e}\gg 1$ (see orange case of Fig. \ref{Fig_distance}), we introduce close to the wall the rescaled coordinates $u'=\sqrt{|\ell|}u$ and $v'=\sqrt{|\ell|}v$. The scaling function reads in these coordinates
\be\label{rescaling_-}
\frac{1}{\sqrt{|\ell|}}K_{1}^{\ell}\left(\frac{u'}{\sqrt{|\ell|}},\frac{v'}{\sqrt{|\ell|}}\right)=\int_{\ell}^{\infty}\frac{ds}{\sqrt{|\ell|}}\sigma\left(s,s+\frac{u'}{\sqrt{|\ell|}}\right)\sigma\left(s,s+\frac{v'}{\sqrt{|\ell|}}\right)\;.
\ee
The integral in Eq. \eqref{rescaling_-} is dominated by the negative contributions of $s\in[\ell,0]$ with $\ell<0$. Therefore we perform the change of variable $t = s/\ell$ and obtain
\be\label{rescaling_-_2}
\frac{1}{\sqrt{|\ell|}}K_{1}^{\ell}\left(\frac{u'}{\sqrt{|\ell|}},\frac{v'}{\sqrt{|\ell|}}\right)\approx\sqrt{|\ell|}\int_0^1 dt \sigma\left(t\ell,t\ell+\frac{u'}{\sqrt{|\ell|}}\right)\sigma\left(t\ell,t\ell+\frac{v'}{\sqrt{|\ell|}}\right)\;.
\ee
Next, we use the asymptotic expansions of the functions $\Ai(z)$ and $\Bi(z)$ for large negative argument \cite{Ai_large}
\be\label{AiBi_-}
\Ai(-z)\sim\frac{1}{\sqrt{\pi}z^{\frac{1}{4}}}\cos\left(\frac{2}{3}z^{\frac{3}{2}}-\frac{\pi}{4}\right)\;\;{\rm and}\;\;
\Bi(-z)\sim\frac{1}{\sqrt{\pi}z^{\frac{1}{4}}}\sin\left(\frac{2}{3}z^{\frac{3}{2}}-\frac{\pi}{4}\right)\;\;{\rm for}\;\;z\to+\infty\;.
\ee
 Inserting this in Eq. \eqref{sigma}, we obtain
\be
|\ell|^{\frac{1}{4}}\sigma\left(t\ell,t\ell+\frac{u'}{\sqrt{|\ell|}}\right)\approx\frac{\Theta(t)}{\sqrt{\pi}t^{\frac{1}{4}}}\sin\left(u'\sqrt{t}\right)\;\;{\rm for}\;\;\ell\to -\infty\;.\label{wf_l_-}
\ee
Finally, inserting Eq. \eqref{wf_l_-} in Eq. \eqref{rescaling_-_2} and performing the change of variable $t \to k=\sqrt{t}$, we obtain
\begin{align}
\frac{1}{\sqrt{|\ell|}}K_{1}^{\ell}\left(\frac{u'}{\sqrt{|\ell|}},\frac{u'}{\sqrt{|\ell|}}\right)\approx\frac{2}{\pi}\int_0^{1}dk\sin(k u')\sin(k v')=K_1^{\rm e}(u',v')\;.\label{limit_l_-}
\end{align}
Therefore we recover the hard edge scaling function (\ref{khb_1d_e_intro}). Note that, in this case, the complete total scaling form reads
\be\label{LDA_kernel}
K_{\mu}(x,y)\approx k_{\rm lin}(1)K_{1}^{\rm e}(k_{\rm lin}(1)(x-y))\;\;{\rm with}\;\;k_{\rm lin}(1)=\frac{\sqrt{|\ell|}}{w_N}=\frac{\sqrt{2mV'(r_{\rm e})(r_{\rm e}-1)}}{\hbar}\;,
\ee
in agreement with the limiting form given in Eq. \eqref{LDA_images} since $\tilde k(1)\to k_{\rm lin}(1)$ for $r_{\rm e}\to 1_+$ where the subscript ${\rm lin}$ refers to the linearized version of the potential. 

\section{Finite temperature bulk kernel}\label{fin_T}

In the case of the hard box potential, the zero temperature kernel takes in the bulk the scaling form 
\be
K_{\mu}({\bf x},{\bf y})\approx k_F^{d}K_{d}^{\rm b}(k_F|{\bf x}-{\bf y}|)=\left(\frac{k_F}{2\pi|{\bf x}-{\bf y}|}\right)^{\frac{d}{2}}\J_{\frac{d}{2}}(k_F|{\bf x}-{\bf y}|)\;.\label{K_bulk_hb}
\ee 
Deriving Eq. \eqref{K_bulk_hb} with respect to $k_F$ using the relation $\partial_x(x^a \J_{a}(x))=x^a \J_{a-1}(x)$ and inserting into Eq. \eqref{K_beta}, we obtain, after performing the change of variable $\mu \to p=\sqrt{2m \mu}/\hbar$
\be
K_{\tilde \mu}({\bf x},{\bf y})\approx \int_0^{\infty}\frac{\zeta dp}{e^{\beta\q p^2}+\zeta}\left(\frac{p}{2\pi}\right)^{\frac{d}{2}}\frac{\J_{\frac{d}{2}-1}(p|{\bf x}-{\bf y}|)}{|{\bf x}-{\bf y}|^{\frac{d}{2}-1}}\;,\label{K_bulk_hb_2}
\ee
where $\zeta=e^{\beta \tilde \mu}$ is the finite temperature chemical potential for the hard box potential. Using $\lambda_T=\sqrt{2\pi \hbar^2\beta/m}$, we perform yet another change of variable $p \to k=\lambda_T p$ in Eq. \eqref{K_bulk_hb_2} and obtain the scaling form in Eq. \eqref{k_bulk_b_finite_2}. Finally, changing from $p$ to $q=\beta\q p^2=\lambda_T^2 p^2/(4\pi)$ in Eq. \eqref{K_bulk_hb_2}, we obtain
\be
K_{\tilde \mu}({\bf x},{\bf y})\approx \frac{1}{\lambda_T^d}\left(\frac{\lambda_T}{|{\bf x}-{\bf y}|}\right)^{\frac{d}{2}-1}\int_0^{\infty}\frac{ \zeta dq}{e^{q}+\zeta}\left(\frac{q}{\pi}\right)^{\frac{d-2}{4}}\J_{\frac{d}{2}-1}\left(\frac{2\sqrt{\pi q}|{\bf x}-{\bf y}|}{\lambda_T}\right)\;,\label{K_bulk_hb_3}
\ee
which is precisely Eq. (274) of Ref. \cite{fermions_review} specialized to the case $V({\bf x})=0$.

\section{Another solvable box potential: the ``Jacobi trap'' at zero temperature}\label{App:Jacobi}

Solvable cases of one dimensional traps studied until now include the harmonic oscillator, associated to the GUE ensemble of RMT, the $1/x^2$ potential associated to the Laguerre ensemble of RMT,
and the hard box associated to one realization of the Jacobi ensemble. It is possible to consider a more
general trapping potential which reduces in various limits to these cases. It is associated to the general 
version of the Jacobi random matrix ensemble \cite{For10,DimitriuEdelman}. 

\subsection{Single particle}

The quantum model is defined by the single particle Hamiltonian $H = - \frac{1}{2} \frac{\partial^2 }{\partial \theta^2} + V(\theta)$ 
(with $\hbar=m=1$) on the interval $\theta \in ]0,\pi[$. The potential is parameterized by two real numbers $a$ and $b$ (not to be confused with the inverse temperature - we work here at zero temperature)
and reads
\be\label{potential_Jacobi}
V(\theta) = \frac{a^2- \frac{1}{4}}{8 \sin^2(\frac{\theta}{2})} + \frac{b^2- \frac{1}{4}}{8 \cos^2(\frac{\theta}{2})}
\ee
and we consider here the repulsive case $a,b \geq 1/2$. The spectrum is discrete with 
eigenvalues
\be
\epsilon_n = \frac{1}{2} (n + \frac{a + b +1}{2})^2   \quad , \quad n =0,1,2.. 
\ee
The normalized eigenfunctions read
\be
\phi_n(\theta) = c_n \sin^{a + \frac{1}{2}}\left(\frac{\theta}{2}\right) \cos^{b + \frac{1}{2}}\left(\frac{\theta}{2}\right) P_n^{a,b}(\cos \theta) 
\ee 
where $P_n^{a,b}(z)$ are the Jacobi polynomials. Recalling 
that for $\theta \in [0,\pi]$, $x = \cos \theta$, $\sin\frac{\theta}{2}=\sqrt{\frac{1-x}{2}}$ 
and $\cos\frac{\theta}{2}=\sqrt{\frac{1+x}{2}}$, $d\theta=\frac{dx}{\sqrt{1-x^2}}$, and 
using the orthogonality relation of the Jacobi polynomials
\be
\int_{-1}^{+1} dx (1-x)^a (1+x)^b P_m^{a,b}(x) P_n^{a,b}(x) = 
\frac{2^{a + b + 1} \delta_{mn} }{2 n + a + b + 1} \frac{\Gamma(n+a+1) \Gamma(n+b+1)}{\Gamma(n+ a + b + 1) n!} 
\ee
we obtain the normalization constant
\be
c_n^2 = \frac{\Gamma(n+ a + b + 1) n!}{(2 n + a + b + 1)
\Gamma(n+a+1) \Gamma(n+b+1)}
\ee

\subsection{$N$ fermions}

Consider now $N$ noninteracting fermions with single particle Hamiltonian $H$
at $T=0$. The ground state $\Psi_0$ is a Slater determinant 
obtained by filling the $N$ lowest energy levels, and the quantum probability $|\Psi_0|^2$
can be written as a determinant
\be
\Psi_0(\theta_1,\cdots,\theta_N) = \frac{1}{\sqrt{N!}} \det_{1 \leq i,j \leq N} \phi_{i-1}(\theta_j) \quad , \quad 
|\Psi_0(\theta_1,\cdots,\theta_N)|^2 = \det_{1 \leq i,j \leq N}
K_\mu(\theta_i,\theta_j)
\ee
with the kernel
\be
K_\mu(\theta,\theta') =   \sin^{a + \frac{1}{2}}(\frac{\theta}{2}) \cos^{b + \frac{1}{2}}(\frac{\theta}{2}) 
\sin^{a + \frac{1}{2}}(\frac{\theta'}{2}) \cos^{b + \frac{1}{2}}(\frac{\theta'}{2}) 
\sum_{n=0}^{N-1} c_n^2 P_n^{a,b}(\cos \theta) P_n^{a,b}(\cos \theta') 
\ee

Hence at $T=0$ the positions of the fermions, $\theta_i$, form a determinantal point process with Kernel $K_\mu$. 
It is convenient to use instead the variables $\lambda_i = \frac{1-\cos(\theta_i)}{2}$. In these
variables, using the Slater determinant form of the wave functions, it is easy to see that
\be
|\Psi_0(\theta_1,\cdots,\theta_N)|^2 d \theta_1 \cdots d \theta_N = P_J(\lambda_1,\cdots,\lambda_N) d \lambda_1 \cdots d \lambda_N
\ee 
where 
\be \label{Jacobi} 
P_J(\lambda_1,\cdots,\lambda_N) \sim \prod_{k=1}^N \lambda_k^{a} (1-\lambda_k)^{b}
\prod_{i<j} |\lambda_i - \lambda_j|^2 \quad , \quad \lambda_i \in  [0,1]
\ee 
is the JPDF of the eigenvalues of the Jacobi ensemble. One random matrix realization
is obtained as follows \cite{DimitriuEdelman,ForresterJacobi,VivoThesis}. Consider the matrix
$J=(W_1+W_2)^{-1} W_1$ 
where $W_j=M_j^{-1} X_j X_j^\dagger$, $j=1,2$, are two 
normalized Wishart matrices 
such that the $X_j$ are two $N \times M_j$ rectangular matrices ($M_{1,2} \geq N$), whose elements are i.i.d
complex random variables with normal distribution $N(0,\frac{1}{\sqrt{2}})+ i N(0,\frac{1}{\sqrt{2}})$. 
The eigenvalues $\lambda_i$, $i=1,..,N$ of $J$ 
belong to the unit interval, $\lambda_i \in [0,1]$, with JPDF given by \eqref{Jacobi}
and $a = M_1-N$ and $b=M_2-N$
(note that this construction yields only integer values). 

We can now calculate the fermion density in the Jacobi trap. In the large $N$ limit and in the bulk,
the Wigner function (see e.g. \cite{us_Wigner}) is given by
\be
W(\theta,p) \simeq \frac{1}{2 \pi} \Theta( \mu - \frac{p^2}{2} - V(\theta) ) 
\ee 
where $\Theta(x)$ is the Heaviside step function and $\mu \simeq \frac{1}{2}N^2$. From 
this formula we obtain the density as $\tilde \rho(\theta)= \int dp W(\theta,p)$, leading to
\be
\tilde \rho(\theta) = \frac{1}{\pi} \sqrt{2 (\mu - V(\theta))}  = \frac{1}{\pi} \sqrt{2} 
\sqrt{\mu - \frac{a^2- \frac{1}{4}}{8 \sin^2(\frac{\theta}{2})} 
- \frac{b^2- \frac{1}{4}}{8 \cos^2(\frac{\theta}{2})} }
\ee 
In the variable $\lambda=\frac{1-\cos(\theta)}{2}$, we have $\tilde \rho_J(\lambda) d\lambda=
\tilde \rho(\theta) d\theta$ and
\be
\tilde \rho_J(\lambda)  = \frac{\sqrt{2}}{\pi \sqrt{\lambda (1-\lambda)} }
\sqrt{\mu - \frac{a^2- \frac{1}{4}}{8 \lambda} 
- \frac{b^2- \frac{1}{4}}{8 (1-\lambda)} } \simeq \frac{N}{\pi} 
\frac{ \sqrt{ (\lambda_+-\lambda)(\lambda-\lambda_-) }}{\lambda (1-\lambda) }
\ee 
with $\lambda_+=1- \frac{b^2}{4 N^2} + O(\frac{1}{N^4})$, $\lambda_-=\frac{a^2}{4 N^2}
+ O(\frac{1}{N^4})$, and in the last equation we dropped subdominant terms in $N$. 
For large $N$, with $a$ and $b$ of order unity, one recovers
the density of eigenvalues of the Jacobi ensemble \cite{DimitriuEdelman}
with $\lambda_-=0$ and $\lambda_+=1$ (see also (2.4.12) in \cite{VivoThesis}).
Note that some results exist for the PDF of the largest eigenvalue
(rightmost fermion) and for the gap probabilities \cite{DimitriuEdelman,ForresterJacobi,Constantine}.


\subsection{From Jacobi to Laguerre and Gaussian ensembles: $1/x^2$ and $x^2$ potentials}

As is well known, the Jacobi ensemble degenerates to the Laguerre ensemble \cite{mehta,For10}. 
The same happens for the fermion problem.
Consider the limit where the Jacobi parameter $b \to +\infty$ and $\theta=\frac{2 z}{\sqrt{b}}$. Then the (scaled and centered) 
single particle
Hamiltonian becomes
\be\label{model_laguerre}
\lim_{b \to +\infty} \frac{4}{b} (H - \frac{b^2}{8}) = - \frac{1}{2} \frac{\partial}{\partial z^2} + V_1(z) \quad , \quad V_1(z) = \frac{z^2}{2} + \frac{a^2-\frac{1}{4}}{2 z^2} 
\ee
a quantum problem now defined for $z>0$. It has eigenvalues $\epsilon_n = 2 n + a + 1$ and eigenvectors
\be
\phi_n(x) = c_n e^{-z^2/2} z^{a+\frac{1}{2}} L_n^{a}(z^2)  
\ee
where the $L_n^{a}(z)$ are the Laguerre polynomials, $c_n^2 =  \frac{2 n!}{\Gamma(n+a+1)}$,
obtained as limits of the Jacobi eigenfunctions. This model is a slight generalization of the
potential $1/x^2$ studied in Section \ref{sec:bessel} in the presence of an additional quadratic trap. In Ref. \cite{Nadal_Majumdar} this model (\ref{model_laguerre})
was studied in the context of non-intersecting interfaces in the presence of a substrate. The $N$-fermion problem now maps to the Laguerre ensemble, i.e. the quantum JPDF becomes the Laguerre measure with weight
$\tilde \lambda^a e^{-\tilde \lambda}$ in the limit, with $\tilde \lambda=z^2$.
This is seen by writing $\lambda=\frac{1-\cos \theta}{2} \simeq \frac{z^2}{b} = \frac{\tilde \lambda}{b}$
and substituting in \eqref{Jacobi}. It is useful to point out that for this model also the quantum propagator is known (using the Poisson Kernel identity)
\bea
G(x,y,t) = \frac{\sqrt{x y}}{\sinh t}  e^{- \frac{1}{2} (x^2+y^2) \coth t}  \,
{\rm I}_a\left(\frac{x \,y}{\sinh t}\right) 
\eea 
which allows to obtain the grand-canonical kernel at arbitrary temperature $T$ and 
chemical potential $\tilde \mu$ as (see Eq. 241 in \cite{fermions_review})
\bea
K_{\tilde \mu}(x,y) =  \int dE \frac{1}{1+ e^{\beta(E-\tilde \mu)}} \int_{\cal C} \frac{dt}{2 i \pi} e^{E t} G(x,y,t) 
\eea 
where ${\cal C}$ is the Bromwich contour. 
Although we will not do it here, it can be used to study the edge behavior for the model
\eqref{model_laguerre} near $z=0$ at any temperature,
and one recovers exactly the same results as in Section \ref{sec:bessel}, i.e. the so-called
hard-edge with the Bessel kernel, as is obtained for the potential $V(z)=\frac{a^2-\frac{1}{4}}{2 z^2}$
(with $a=\alpha+1/2$ in the notations of Section \ref{sec:bessel}). 

Another degeneracy of the Jacobi ensemble is towards the Gaussian model. It can be realized for
$a=1/2$ and $b \to +\infty$. It corresponds to the harmonic oscillator. It can also be realized for $a=b \to +\infty$. This is consistent with
the limit studied in Section \ref{sec:bessel} ($\alpha \to +\infty$), which is shown to
recover the universality of the harmonic potential (soft-edge), see the end of 
the Section \ref{sing_pot}. 

Finally, let us note that for the complete Jacobi model, with $a,b$ fixed and arbitrary, we expect
that the edge behavior will be identical to the one of the $1/x^2$ potential studied
in Section \ref{sec:bessel} (hard edge). The Jacobi trap thus allows to study the crossover
between the hard and soft edge, i.e. between the $1/x^2$ and $x^2$ behaviors. This is an
alternative interpolating family of potentials, different from the box plus quadratic potential
studied in Section \ref{trunc_1d}.

\newpage

{}

\end{document}